\documentclass[aps,prb,superscriptaddress,twocolumn]{revtex4-1}

\pdfoutput=1

\usepackage{graphicx}
\usepackage{amsmath}
\usepackage{amssymb}
\usepackage{verbatim}
\usepackage{braket}
\usepackage{color}
\usepackage{hyperref}
\hypersetup{
	colorlinks=true,
	linkcolor=blue,      
	urlcolor=blue,
	citecolor=blue
}
\date{\today}
\makeatletter
\let\cat@comma@active\@empty
\makeatother
\begin{document}
\title{Topological phase transition between non-high symmetry critical phases and curvature function renormalization group}
	
\author{Ranjith R Kumar}
\email{ranjith.btd6@gmail.com}
\affiliation{Theoretical Sciences Division, Poornaprajna Institute of Scientific Research, Bidalur, Bengaluru-562164, India.}
\affiliation{Graduate Studies, Manipal Academy of Higher Education, Madhava Nagar, Manipal-576104, India.}
\author{Y R Kartik}
\email{yrkartik@gmail.com}
\affiliation{Theoretical Sciences Division, Poornaprajna Institute of Scientific Research, Bidalur, Bengaluru-562164, India.}
\affiliation{Graduate Studies, Manipal Academy of Higher Education, Madhava Nagar, Manipal-576104, India.}
\author{Sujit Sarkar}
\email{Corresponding author: sujit.tifr@gmail.com}
\affiliation{Theoretical Sciences Division, Poornaprajna Institute of Scientific Research, Bidalur, Bengaluru-562164, India.}
	
\begin{abstract}
	The interplay between topology and criticality has been a recent interest of study in condensed matter physics. A unique topological transition between certain critical phases has been observed as a consequence of the edge modes living at criticalities. In this work, we generalize this phenomenon by investigating possible transitions between critical phases which are non-high symmetry in nature. We find the triviality and non-triviality of these critical phases in terms of the decay length of the edge modes and also characterize them using the winding numbers. The distinct non-high symmetry critical phases are separated by multicritical points with linear dispersion at which the winding number exhibits the quantized jump, indicating a change in the topology (number of edge modes) at the critical phases. Moreover, we reframe the scaling theory based on the curvature function, i.e. curvature function renormalization group method to efficiently address the non-high symmetry criticalities and multicriticalities. Using this we identify the conventional topological transition between gapped phases through non-high symmetry critical points, and also the unique topological transition between critical phases through multicritical points. The renormalization group flow, critical exponents and correlation function of Wannier states enable the characterization of non-high symmetry criticalities along with multicriticalities. 
	%Finally, we discuss the future scope of this work and possible experimental investigations.
%\noindent \textbf{Keywords} : {Topological quantum phase transition, Quantum criticality, Dirac equation, Winding number, Curvatuture function renormalization group, Critical exponents, Wannier state correlation function, Entanglement entropy, Entanglement spectra, Central charge, Fidelity succeptibility, Ultracold atoms, Superconducting circuits}
\end{abstract}
\maketitle
\section{Introduction}
\hspace{0.4cm}Topological states of matter have recieved a huge attention from both theoretical and experimental physicists in recent years \cite{haldane1988model,hasan2010colloquium,wang2017topological,goldman2016topological,narang2021topology}. Non-trivial topology of the electronic band structure dictates the formation of localized stable edge modes which are protected by the bulk gap \cite{kitaev2001unpaired,kane2005quantum}. Number of edge modes are counted using topological invariant number, which is defined as the integral of the curvature function (Berry connection, Berry curvature, etc) over the Brillouin zone \cite{thouless1982quantized,berry1984quantal,zak1989berry}. The topological invariant shows quantized jump associated with the bulk gap closing at a critical point. Therefore, a topological transition between distinct gapped phases is characterized by the bulk gap closing and opening along with the quantized jump in the values of invariant numbers~\cite{altland1997symmetry,sarkar2018quantization,rahul2019interplay,kartik2021topological}.

\hspace{0.4cm}Moreover, the quantization signifies the divergence in the curvature function at the critical point, which allows one to frame a scaling theory and correlation factors using the curvature function~\cite{chen2016scaling,chen2016scalinginvariant,chen2017correlation,chen2018weakly,chen2019universality,molignini2018universal,panahiyan2020fidelity,molignini2020generating,abdulla2020curvature,malard2020scaling,molignini2020unifying,kumar2021multi}. A renormalization group (RG) method developed by iteratively finding a parameter space away from the critical point such as to reduce the divergence in the curvature function by driving it to its fixed point configuration. As this procedure does not change the topology of the band structure, eventually, the RG flow lines characterize the topological phase transition. The Lorentzian form of the curvature function near a critical point allows one to obtain the decay length of the edge modes at gapped topological phases\cite{molignini2018universal,continentino2020finite}. The decay length diverges on approaching a critical point, indicating the edge modes decays into the bulk. Therefore, the edge modes were believed to exist only with a finite bulk gap. %The critical points where the bulk gap vanishes were considered to be trivial phases since edge excitations were not observed~\cite{bibid}.  

\hspace{0.4cm}Recently, this conventional understanding has been re-investigated and the edge modes are observed to be localized and stable even at certain critical points~\cite{verresen2018topology,verresen2019gapless,jones2019asymptotic,verresen2020topology,rahul2021majorana,niu2021emergent,PhysRevB.104.075132,PhysRevResearch.3.043048,fraxanet2021topological,keselman2015gapless,scaffidi2017gapless,duque2021topological,kumar2021topological}. Therefore, similar to the gapped topological phases, certain critical phases also possess localized stable edge modes. The critical phases with topological and non-topological characters are identified as the high symmetry (HS) in nature since the gap closing occurs at the HS points in momentum space~\cite{kumar2021topological}. The distinct HS critical phases are separated by the multicritical points and favor an unusual topological transition between them. This transition occurs without gap closing and opening at HS points in contrast to the conventional topological transition~\cite{kumar2021topological,kumar2021multi,rahul2021majorana,verresen2020topology}. The multicritical points which favor the transition are found to have quadratic dispersion. In general, they are the intersection points of the distinct criticalities and are studied in different contexts~\cite{rufo2019multicritical,malard2020multicriticality,malard2020multicriticality,sim2022quench}.
The scaling theory developed to identify the topological transition between gapped phases, are reframed to identify the topological transition between HS critical phases~\cite{kumar2021topological,kumar2021multi}. The RG flow lines, correlation factors and decay length of edge modes at criticality, effectively characterized the topological transition~\cite{kumar2021topological}. 
%between HS critical phases through multicritical points.  
%{\color{red} Can add more details from different paper. JPSJ, PRL, arxiv, JSP etc.}

\hspace{0.4cm}In the topological systems, increasing the nearest-neighbor couplings leads to prominent behavior of non-high symmetry (non-HS) critical points apart from the HS ones \cite{hsu2020topological,niu2012majorana}. The existence of stable localized edge modes at non-HS critical phases has not been explored previously. Furthermore, the possibility of unique topological transition between distinct non-HS critical phases is an interesting open question and require a detailed investigation. 
%also stands topological with stable localized edge modes and the unique topological transition can be generalized to non-HS criticalities.
%However, the possibility of the topological transition between distinct non-HS critical phases has not been expolred previously. 
On the other hand, scaling theory based on the curvature function fails to capture the topological transition at a non-HS critical point between gapped phases \cite{abdulla2020curvature,malard2020scaling}. Therefore, the generalization of this method to identify the possible topological transition between non-HS critical phases is not straightforward.

\hspace{0.4cm}Therefore, in the present work, our motivation is threefold. (i) We generalize the curvature function renormalization group (CRG) theory to characterize the non-HS quantum criticality. We develop and perform the CRG to identify the topological transition, at a non-HS critical point, between gapped phases.
(ii) We identify the topological and trivial non-HS critical phases by investigating the existence of edge modes both analytically and numerically.
(iii) We identify and explore the unique topological phase transition between non-HS critical phases via multicritical points. We reframe the CRG method to capture the topological transition, at a multicritical point, between non-HS critical phases.
%At first, we develop and perform the CRG to identify the topological transition, at a non-HS critical point, between gapped phases. Later, we reframe the method to capture the unique topological transition, at a multicritical point, between non-HS critical phases.

\hspace{0.4cm}The article is laid out as follows: In Section.\ref{II} we introduce the model and the topological phase diagrams. Section.\ref{CRG-non} describes the CRG method for topological transition at non-HS critical point between gapped phases. The diverging property, critical exponents and correlation factors using curvature functions are discussed. In Section.\ref{IV} we demonstrate the existence of trivial and topological non-HS critical phases and the edge mode localizations. These results are supported by the winding number calculations and numerical analysis carried at non-HS criticality. In Section.\ref{V}, the CRG method for topological transition at multicritical point between non-HS critical phases is discussed. We also discuss the behavior of curvature function, its exponents and the correlation factors at non-HS criticalities. We discuss the results and its experimental observabilities in Section.\ref{VI} and finally conclude.

\section{Model and Topological Phase Diagram}\label{II}
\begin{figure}[t]
	\includegraphics[width=6cm,height=6cm]{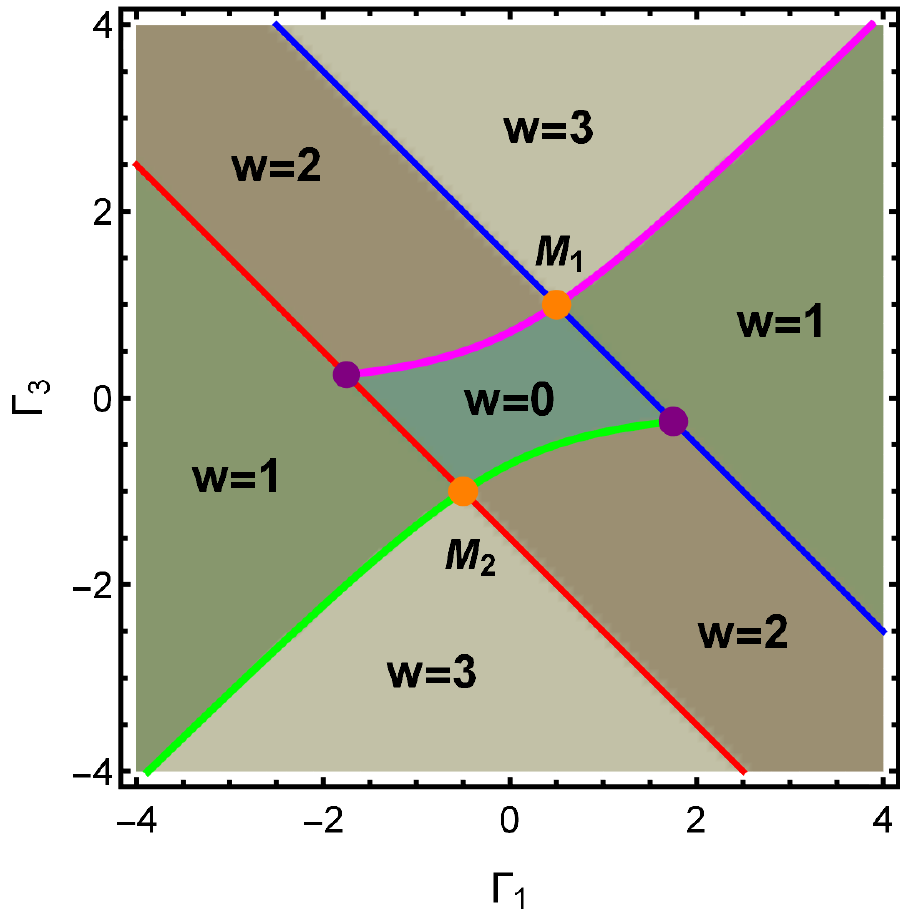}
	\caption{Topological phase diagram. Plotted in the plane $\Gamma_{1}-\Gamma_{3}$ with $\Gamma_{0}=1$, $\Gamma_{2}=0.5$. The gapped phases are identified with integer winding number. These phases are separated by the four critical lines. The HS critical lines are represented in red (for $k_0=0$) and blue (for $k_0=\pm \pi$). The non-HS critical lines are represented in magenta (for $\pm k_0$ in Eq.\ref{k0magenta}) and green (for $\pm k_0$ in Eq.\ref{k0green}). Among the four multicritical points two are identified with quadratic dispersion (purple dots) and other two are identified with linear dispersion (orange  dots) which are labeled $M_{1,2}$.}
	\label{Topo-phase-diag}
\end{figure}

\hspace{0.3cm} We consider one dimensional
lattice chain of spinless fermions in momentum space with onsite potential ($\Gamma_{0}$),
nearest neighbor ($\Gamma_{1}$), next-nearest neighbor ($\Gamma_{2}$), and next-next-nearest neighbor ($\Gamma_{3}$) couplings.
The two-band Bloch
Hamiltonian can be written in the
pseudospin basis as
\begin{equation}
\mathcal{H}(k,\boldsymbol{\Gamma}) = \boldsymbol{\chi(k).\sigma} = \chi_{x} (k) \sigma_x + \chi_{y} (k) \sigma_y +\chi_{z} (k) \sigma_z ,\label{Model-H}
\end{equation}
where $\boldsymbol{\Gamma}=\left\lbrace \Gamma_{0},\Gamma_{1},\Gamma_{2}, \Gamma_{3}\right\rbrace $, $ \chi_{x} (k) = \Gamma_0 + \Gamma_1 \cos k + \Gamma_2 \cos 2k + \Gamma_3 \cos 3k,$ $ \chi_{y} (k) = \Gamma_1 \sin k + \Gamma_2 \sin 2k + \Gamma_3 \sin 3k $, $\chi_{z} (k)=0$ and $\boldsymbol{\sigma}=(\sigma_x,\sigma_y,\sigma_z)$ are the Pauli matrices. The model represents the 1D topological insulator and superconductor with extended nearest neighbor couplings\cite{PhysRevLett.42.1698,kitaev2001unpaired,hsu2020topological,niu2012majorana,kumar2021topological}. Topological distinct gapped phases of the model can be identified with the winding number 
\begin{equation}
w=\frac{1}{2\pi} \oint\limits_{BZ} \frac{\chi_x\partial_k \chi_y - \chi_y \partial_k \chi_x}{\chi_x^2+\chi_y^2} dk, \label{winding}
\end{equation}
where $w\in\mathbb{Z}$ (integer), as shown in Fig.\ref{Topo-phase-diag}. Topological phase transitions between distinct gapped phases are associated with the gap closing, $E_k=\pm\sqrt{\chi_x^2+\chi_y^2}=0$. This dictates the critical lines across which the winding number changes.

\begin{figure}[t]
	\includegraphics[width=4cm,height=2.8cm]{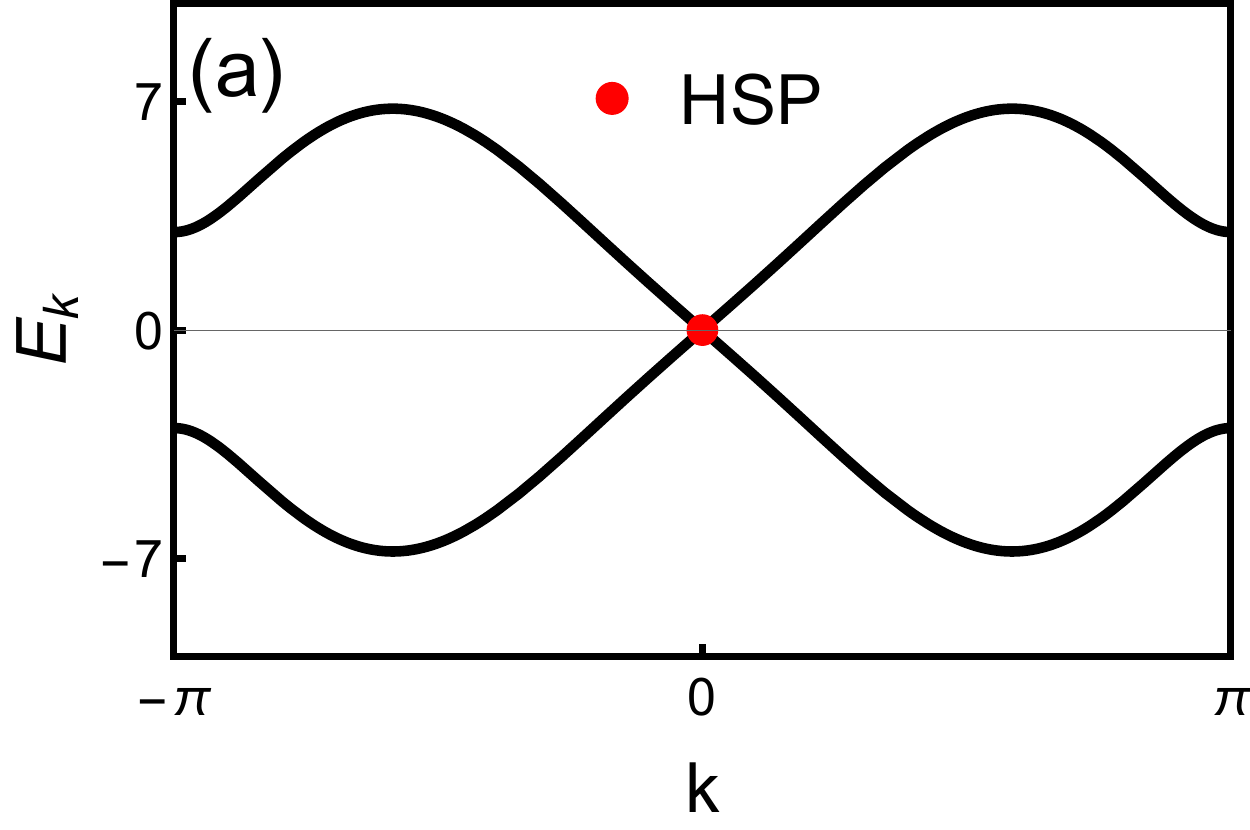}
	\hspace{0.2cm}
	\includegraphics[width=4cm,height=2.8cm]{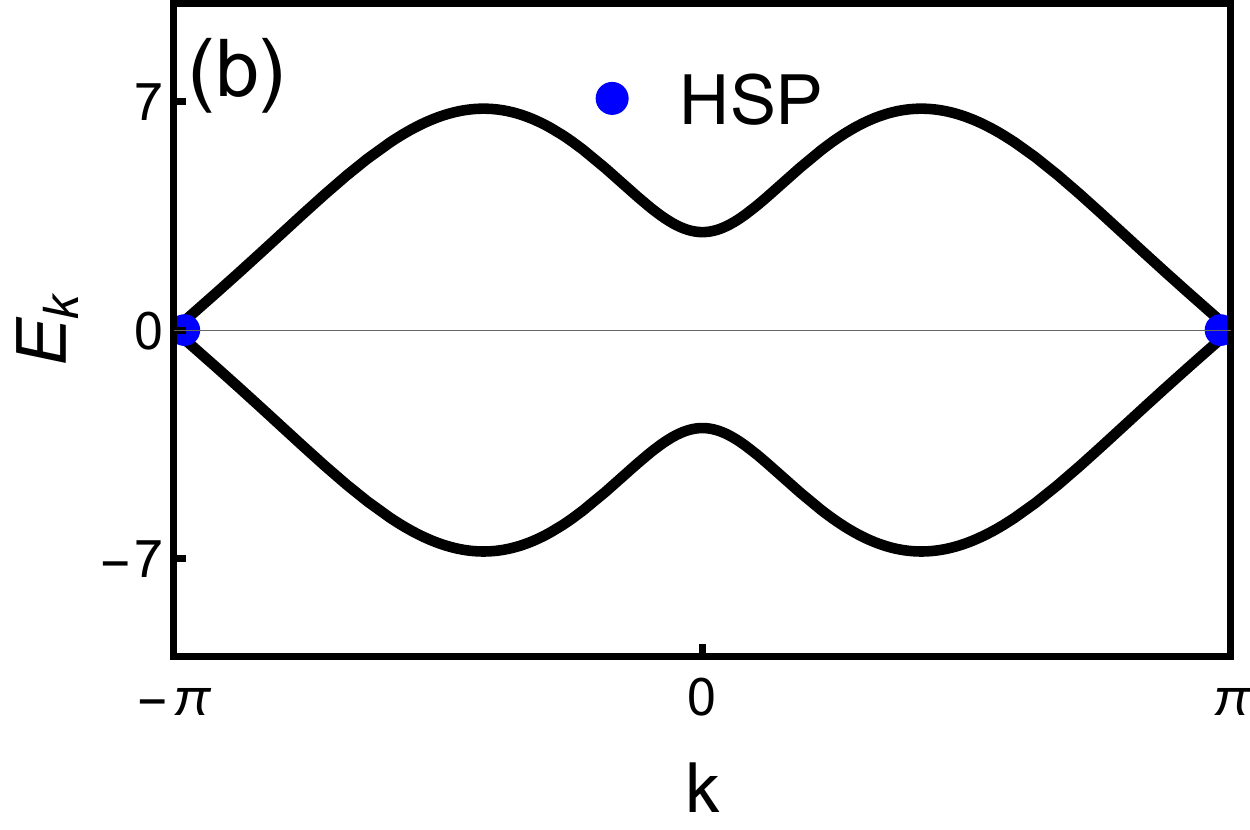}
	\includegraphics[width=4cm,height=2.8cm]{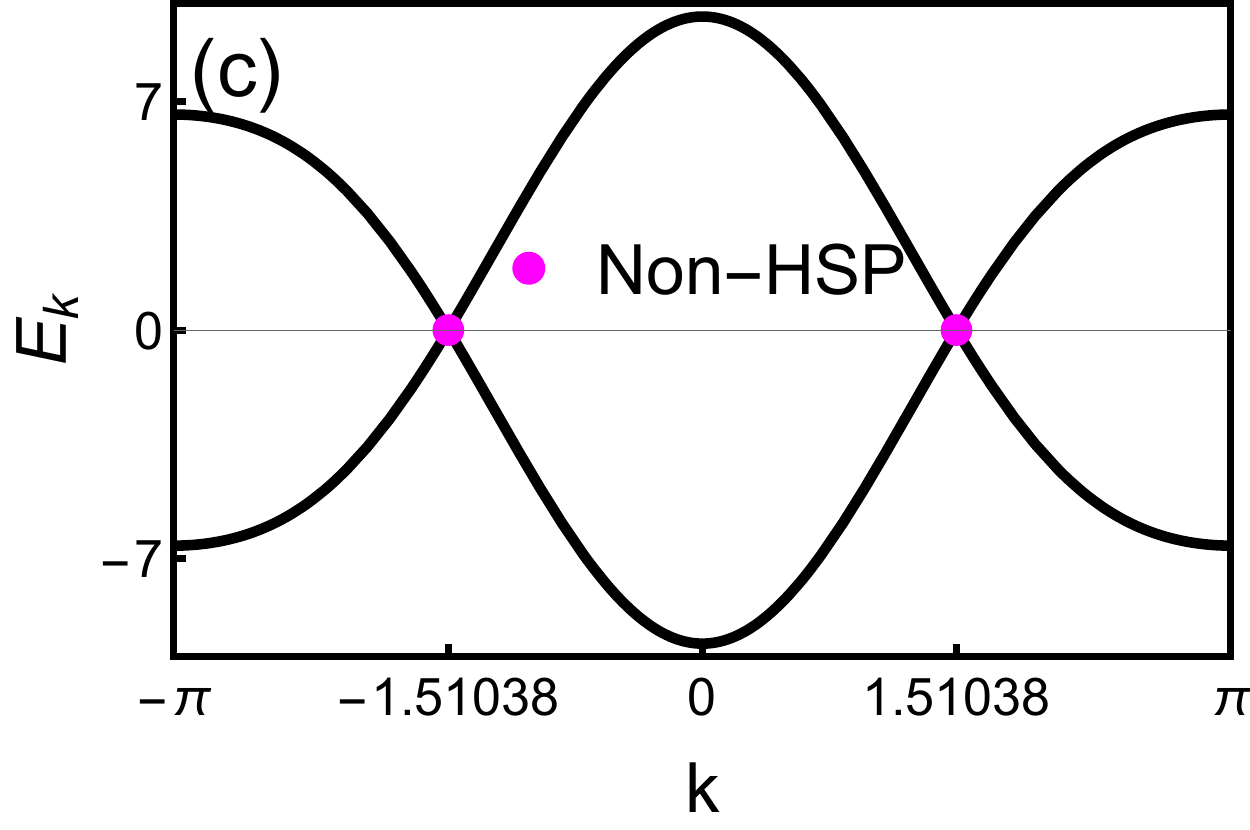}
	\hspace{0.2cm}
	\includegraphics[width=4cm,height=2.8cm]{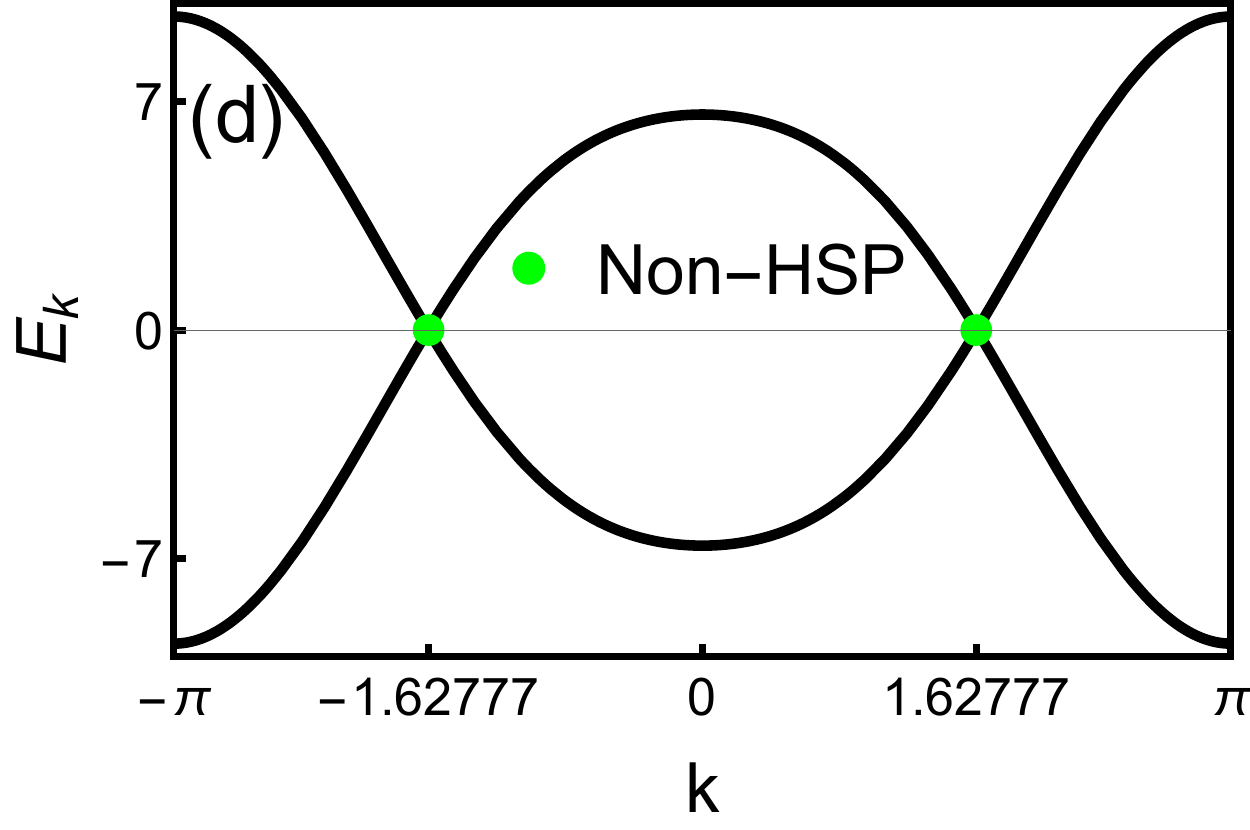}
	\caption{\label{HSnonHS} Energy dispersion for HS and non-HS critical points. (a) HS point at $k_0=0$ which can be obtained for the values of the parameters satisfying the relation $\Gamma_{3}=-(\Gamma_{0}+\Gamma_{1}+\Gamma_{2})$ (red critical line in Fig.\ref{Topo-phase-diag}). (b) HS point at $k_0=\pm \pi$ which can be obtained for the values of the parameters satisfying the relation $\Gamma_{3}=(\Gamma_{0}-\Gamma_{1}+\Gamma_{2})$ (blue critical line in Fig.\ref{Topo-phase-diag}). (c) Non-HS point at $k_0$ in Eq.\ref{k0magenta} where the values of the parameters satisfy $\Gamma_{3}=(\Gamma_{1}+ \sqrt{\Gamma_{1}^2+4\Gamma_{0}(\Gamma_{0}-\Gamma_{2})})/2$ (magenta critical line in Fig.\ref{Topo-phase-diag}). (d) Non-HS point at $k_0$ in Eq.\ref{k0green} where the values of the parameters satisfy $\Gamma_{3}=(\Gamma_{1}- \sqrt{\Gamma_{1}^2+4\Gamma_{0}(\Gamma_{0}-\Gamma_{2})})/2$ (green critical line in Fig.\ref{Topo-phase-diag}).}
\end{figure}

\hspace{0.3cm} The gap closing momenta $k_0$ (critical momenta) in the Brillouin zone can be either HS or non-HS in nature. The momenta with $k_0=-k_0$, (up to a reciprocal lattice vector) are referred to as the HS points and are associated with space-group symmetries~\cite{murakami2011gap,kourtis2017weyl,chen2019universality}. In our model, there are two HS points at $k_0=0,\pm\pi$, as shown in Fig.\ref{HSnonHS}(a) and (b). The distinct gapped phases (i.e. $w=0\leftrightarrow1$, $w=1\leftrightarrow2$, and $w=2\leftrightarrow3$) are separated by HS critical points at which the gap closes at HS momenta in the Brillouin zone. The critical points in the parameter space can be traced with a line on which every point closes the gap at HS momenta and is referred to as a critical line.
In our model, the critical momenta $k_0=0$ yields the critical line $\Gamma_{3}=-(\Gamma_{0}+\Gamma_{1}+\Gamma_{2})$ (red line in Fig.\ref{Topo-phase-diag}), and $k_0=\pm \pi$ yields the critical line $\Gamma_{3}=(\Gamma_{0}-\Gamma_{1}+\Gamma_{2})$ (blue line in Fig.\ref{Topo-phase-diag}).
%can be identified at the the gap closing points across which the winding number changes. This dictates the identification of distinct gapped phases. The distinct gapped phases are separated by a gap closing points (critical points or lines) at which the dispersion
%\begin{equation}
%E_k=\pm\sqrt{\chi_x^2+\chi_y^2}=0.
%\end{equation} 

%\hspace{0.3cm}The gap closing occurs for the {\color{blue}critical} momenta $k_0=0$ which yield the critical line $\Gamma_{3}=-(\Gamma_{0}+\Gamma_{1}+\Gamma_{2})$ (red line in Fig.\ref{Topo-phase-diag}), and $k_0=\pm \pi$ which yield the critical line $\Gamma_{3}=(\Gamma_{0}-\Gamma_{1}+\Gamma_{2})$ (blue line in Fig.\ref{Topo-phase-diag}). As these gap closing points satisfy $k_0=-k_0$ in the Brillouin zone {\color{blue}under periodic boundary condition}, they are regarded as HS points as shown in Fig.\ref{HSnonHS}. 

\hspace{0.3cm} As the nearest-neighbor couplings are increased the gap closing can also occur at arbitrary points in the Brillouin zone, which are referred to as non-HS points~\cite{murakami2011gap,kourtis2017weyl,molignini2018universal}. The distinct gapped phases (i.e. $w=0\leftrightarrow2$, and $w=1\leftrightarrow3$) are also separated by non-HS critical points/lines at which the gap closes at non-HS momenta in the Brillouin zone. Moreover, at each point on the non-HS critical lines, the gap closing occurs at a pair of non-HS momenta, as shown in Fig.\ref{HSnonHS}(c) and (d). 
In our model, these points can be obtained for
\begin{equation}
k_0=\pm\arccos((-2\Gamma_{2}+ \sqrt{4\Gamma_{2}^2-16\Gamma_{3}(\Gamma_{1}-\Gamma_{3})})/8\Gamma_{3})\label{k0magenta}
\end{equation} 
which yield the critical line $\Gamma_{3}=(\Gamma_{1}+ \sqrt{\Gamma_{1}^2+4\Gamma_{0}(\Gamma_{0}-\Gamma_{2})})/2$ (magenta line in Fig.\ref{Topo-phase-diag}), and 
\begin{equation} 
k_0=\pm\arccos((-2\Gamma_{2}- \sqrt{4\Gamma_{2}^2-16\Gamma_{3}(\Gamma_{1}-\Gamma_{3})})/8\Gamma_{3})\label{k0green}
\end{equation} %$k_0={\color{red}\pm}\arccos((-2\Gamma_{2}\pm \sqrt{4\Gamma_{2}^2-16\Gamma_{3}(\Gamma_{1}-\Gamma_{3})})/8\Gamma_{3})$ 
which yield the critical line $\Gamma_{3}=(\Gamma_{1}- \sqrt{\Gamma_{1}^2+4\Gamma_{0}(\Gamma_{0}-\Gamma_{2})})/2$ (green line in Fig.\ref{Topo-phase-diag}). 
Therefore, HS and non-HS critical lines together distinguish the gapped phases with $w=0,1,2,3$ as shown in Fig.\ref{Topo-phase-diag}. 
Note that, the pair of non-HS gap closing points have identical critical properties. Therefore, we address only one point throughout the discussion.
%(unless it is explicitly mentioned).

\hspace{0.3cm} Moreover, the model possess four multicritical points at the intersections of HS and non-HS critical lines. Two of them are identified with quadratic dispersion (purple dots in Fig.\ref{Topo-phase-diag}) whilst the other two are with linear dispersion (orange dots, named $M_{1}$ and $M_{2}$ in Fig.\ref{Topo-phase-diag}). The quadratic multicriticalities are obtained for $\Gamma_{1}=\pm(3\Gamma_{0}+\Gamma_{2})/2$ and linear multicriticalities are obtained for $\Gamma_{1}=\pm\Gamma_{2}$ (here the sign $'\pm'$ represents $M_1$ and $M_2$ respectively). In our model, each non-HS critical line is separated by the multicritical points into two segments. These two segments can be identified with distinct topologies (discussed later). Moreover, the critical lines manifest as critical regions or critical surfaces in the three-dimensional parameter space. Every point on this critical surface is a gap-closing critical point. Therefore, we refer to the different segments of critical lines as critical phases.

\hspace{0.3cm} Localized edge modes living at certain criticalities, lead to a unique topological transition along the critical lines between distinct critical phases \cite{kumar2021topological,kumar2021multi}. In this work, we aim to identify topological distinct nature among the non-HS critical phases and the topological transition between them. We achieve this in three steps. At first, we develop CRG method to address the non-HS criticality and show that it works in identifying the conventional topological transition between gapped phases (Section.\ref{CRG-non}). Later, we construct the model at non-HS criticality and show topological distinct non-HS critical phases and its edge mode solutions both analytically and numerically (Section.\ref{IV}). Finally, we reframe the CRG method to work at non-HS criticality in order to capture the unique topological transition between non-HS critical phases (Section.\ref{V}).
%These critical phases are HS in nature and the transition is mediated by the multicritical points with both quadratic and linear dispersion. 
%In this work, our study involves only non-HSP criticalities since we aim to identify the topological non-HS criticalities and expolre the topological transition between distinct non-HS critical phases.

%Note that, the pair of non-HS gap closing points i.e. $\pm k_0$, have identical critical properties. Therefore, we addres only one point throughout the discussion (unless it is explicitly mention).

%we address only non-HS criticalities throughout the discussion. Among the two non-HS gap closing points we consider only one during the discussion (unless it is explicitly mention), as the critical properties of the two points are identical.

\section{CRG for topological transition between gapped phases through non-HS quantum criticality}\label{CRG-non}
\hspace{0.3cm} The critical behavior of the system can be captured by a scaling scheme based on the divergence of curvature function at a critical point \cite{chen2016scaling}. The curvature function can be defined as 
\begin{equation}
F(k,\mathbf{\Gamma})=\frac{\chi_x\partial_k \chi_y - \chi_y \partial_k \chi_x}{\chi_x^2+\chi_y^2},
\end{equation}
whose integral over the Brillouin zone gives the winding number in Eq.\ref{winding}.
The scaling involves finding a $\mathbf{\Gamma}^{\prime}$ for every $\mathbf{\Gamma}$ such that $F(k_0,\mathbf{\Gamma}^{\prime})=F(k_0+\delta k, \mathbf{\Gamma})$, where $\delta k$ is small deviation from a HS point $k_0$. This procedure gradually reduces the deviation in the curvature function from its fixed point configuration while preserving the topological property\cite{chen2016scaling}. Eventually, the scaling scheme yields RG flow in parameter space which enable the identification of critical points, at which the topological transition occurs, along with fixed points.

\hspace{0.3cm} However, the same scaling scheme does not capture the non-HS critical points, where the peak in the curvature function occur away from HS points and the corresponding $k_0$ changes with every $\mathbf{\Gamma}^{\prime}$. 
Nevertheless, in some cases, the scaling at HS points can reveal the non-HS critical behavior in terms of RG flow lines \cite{malard2020scaling,abdulla2020curvature}. Although, this advantage is not universal and is lost when certain conditions to the parameters are imposed in the model \cite{kumar2021multi}. Therefore, in general, the CRG for HS points fails to capture the non-HS criticalities.  

\hspace{0.3cm} Here we reframe the scaling procedure to obtain an effective scheme which can directly capture the topological transition at non-HS criticality between gapped phases. Similar to the case of HS criticalities, the curvature function shows diverging property by tuning the parameter towards non-HS criticalities as well, i.e. $\mathbf{\Gamma}\rightarrow\mathbf{\Gamma}_c$. The momenta at which the diverging peak occurs ($k_0$) is a set of non-HS points $k_0=\left\lbrace \pm k_0^c,\pm k_0^1,\pm k_0^2,...\right\rbrace$ where $k_0^c$ is the momentum for $\mathbf{\Gamma}=\mathbf{\Gamma}_c$ and the other points are for the parameter values away from the criticality (i.e. $\mathbf{\Gamma}\neq\mathbf{\Gamma}_c$). The diverging peak of the curvature function increases as $\mathbf{\Gamma}\rightarrow\mathbf{\Gamma}_c$ leading to complete divergence at $\mathbf{\Gamma}=\mathbf{\Gamma}_c$ and $k_0=\pm k_0^c$. 
\begin{figure}[t]
	\includegraphics[width=5.4cm,height=3.8cm]{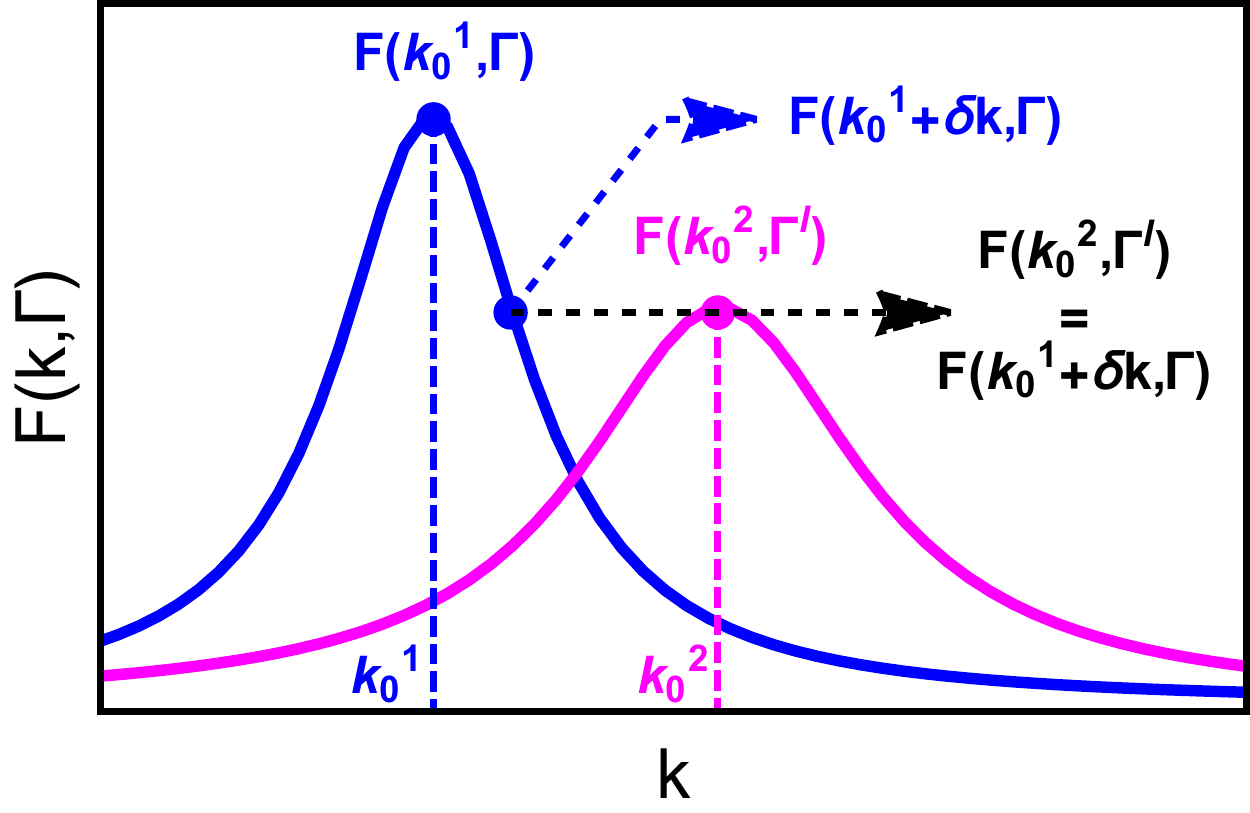}
	\caption{\label{CHSnonHS} Curvature function in the vicinity of non-HS critical point. As one tune the parameter $\mathbf{\Gamma}$ to $\mathbf{\Gamma^{\prime}}$ away from the critical point, peak of the curvature function $F(k,\mathbf{\Gamma})$ decreases and shits from $k_0^1$ to $k_0^2$. This enable one to realize the scaling of the form $F(k_{0}^2,\mathbf{\Gamma}^{\prime})=F(k_{0}^1+\delta k, \mathbf{\Gamma})$.}
\end{figure}

\hspace{0.3cm}Moreover, the curvature function flips the sign as the parameters passes through the non-HS critical point
\begin{equation}
\lim_{\mathbf{\Gamma}\rightarrow \mathbf{\Gamma}_{c}^+}F(k_0,\mathbf{\Gamma})= -\lim_{\mathbf{\Gamma}\rightarrow \mathbf{\Gamma}_{c}^-}F(k_0,\mathbf{\Gamma})=\pm \infty.
\end{equation}
%This flip in the sign of the curvature function involves a change in the topological invariant number. 
The curvature function is symmetric around a non-HS point, $F(k_0+\delta k,\mathbf{\Gamma})=F(k_0-\delta k,\mathbf{\Gamma})$, where $\delta k$ is small deviation from non-HS point $k_0$, and by choosing a proper gauge it can be written in terms Ornstein-Zernike form
\begin{equation}
F(k_0+\delta k,\mathbf{\Gamma})=\frac{F(k_0,\mathbf{\Gamma})}{1+\xi^2\delta k^2}, \label{Lorenzian}
\end{equation}
where $\xi$ is characteristic length scale (inverse of the width of curvature function). This length scale also shows the diverging property on approaching the non-HS critical point. Therefore, one can define the critical exponents for the divergence of curvature function as
\begin{equation}
F(k_0,\mathbf{\Gamma}) \propto |\mathbf{\Gamma}-\mathbf{\Gamma}_c|^{-\gamma},\;\;\;\;\;\; \xi \propto |\mathbf{\Gamma}-\mathbf{\Gamma}_c|^{-\nu}, \label{critical-exponents}
\end{equation}
where the exponents $\gamma$ and $\nu$ dictates the universality class of non-HS criticalities. For one dimensional systems, these exponents obeys the scaling law $\gamma=\nu$, which is imposed by the conservation of topological invariant~\cite{chen2017correlation}.

%\begin{figure}[h]
%	\centering
%	\includegraphics[scale=0.45]{Gen_Cu}
%	\caption{\label{gen-cur}Curvature function in the vicinity of non-HS critical point. As one tune the parameter $\mathbf{\Gamma}$ to $\mathbf{\Gamma^{\prime}}$ away from the critical point, peak of the curvature function $F(k,\mathbf{\Gamma})$ decreses and shits from $k_0^1$ to $k_0^2$. This enable one to realize the scaling of the form $F(k_{0}^2,\mathbf{\Gamma}^{\prime})=F(k_{0}^1+\delta k, \mathbf{\Gamma})$.}
%\end{figure}

\hspace{0.3cm} The striking similarities in the behavior of curvature function between HS and non-HS criticalities, allows one to reframe the scaling theory purely in terms of non-HS points. Let us consider that as $\mathbf{\Gamma}$ is tuned to $\mathbf{\Gamma}^{\prime}$ the peak develops at $k_0^1$ and then shifts to $k_0^2$ respectively, as shown in Fig.\ref{CHSnonHS}.
For this property the scaling can be written as 
\begin{equation}
F(k_{0}^2,\mathbf{\Gamma}^{\prime})=F(k_{0}^1+\delta k, \mathbf{\Gamma}). \label{scaling}
\end{equation}
Expansion of this equation to the leading orders in $\mathbf{\Gamma^{\prime}}$ and $k_0^1$ gives
\begin{multline}
F(k_{0}^1,\mathbf{\Gamma})-F(k_{0}^2,\mathbf{\Gamma}) + \delta k \partial_k F(k,\mathbf{\Gamma})|_{k=k_{0}^1} \\= (\mathbf{\Gamma}^{\prime}-\mathbf{\Gamma}) \partial_{\mathbf{\Gamma}} F(k_{0}^2,\mathbf{\Gamma})
\end{multline}
To obtain the RG equation, without loss of generality, one can choose the parameter values ($\mathbf{\Gamma}$ and $\mathbf{\Gamma}^{\prime}$) in such a way that the non-HS points $k_0^1$ and $k_0^2$ have the closest values i.e. $k_0^1\approxeq k_0^2$, for which the curvature functions are negligibly different i.e. $F(k_{0}^1,\mathbf{\Gamma})\approxeq F(k_{0}^2,\mathbf{\Gamma})$. This approximation yields the generic CRG equation
%\begin{equation}
%\frac{d\mathbf{\Gamma}}{dl} \approx \frac{\partial_k F(k,\mathbf{\Gamma})|_{k=k_{0}^1}}{\partial_{\mathbf{\Gamma}} F(k_{0}^2,\mathbf{\Gamma})}, \label{CRG-General}
%\end{equation}
\begin{equation}
\frac{d\mathbf{\Gamma}}{dl} \approx \frac{\partial_k F(k,\mathbf{\Gamma})|_{k=k_{0}}}{\partial_{\mathbf{\Gamma}} F(k_{0},\mathbf{\Gamma})}, \label{CRG-General}
\end{equation}
where $d\mathbf{\Gamma}=\mathbf{\Gamma}^{\prime}-\mathbf{\Gamma}$ and $dl=\delta k$. The RG flow direction together with the flow rate determines the critical and fixed points in the parameters space~\cite{chen2018weakly} as
\begin{align}
\text{Critical point:}& \hspace{0.2cm} \left| \frac{d\mathbf{\Gamma}}{dl}\right| \rightarrow \infty, \text{flow directs away},\nonumber\\
\text{Stable fixed point:}& \left| \frac{d\mathbf{\Gamma}}{dl}\right| \rightarrow 0, \text{flow directs into},\nonumber\\
\text{Unstable fixed point:}& \left| \frac{d\mathbf{\Gamma}}{dl}\right| \rightarrow 0, \text{flow directs away}.
\label{crit-fixed}
\end{align}
%Tuning the parameters towards a fixed point drives the numerator to zero, while the flow direction will be either in or out distinguishing stable and unstable fixed points respectively. Tuning the parameters towards a critical point drives the denominator to zero with the outward RG flow. 

\hspace{0.3cm} Interestingly, the Wannier-state correlation function, obtained from the charge-polarization correlation between Wannier states at different positions\cite{chen2017correlation}, can be extended to identify the topological transition through non-HS criticality. 
%can be obtained from the Wannier-state representation of the charge-polarization correlation between Wannier states at different positions\cite{PhysRevB.95.075116}. 
It can be obtained from the Fourier transform of the curvature function and the substitution of Ornstein-Zernike form. The correlation function can be written as
\begin{equation}
\lambda_R=\left\langle   R|\hat{r}|0\right\rangle=e^{ik_0R} \frac{F(k_0,\mathbf{\Gamma})}{2\xi}e^{-\frac{R}{\xi}}, \label{correlation}
\end{equation}
where $\hat{r}$ is the position operator for the Wannier states at a distance $R$ from the origin $\ket{0}$, defined as $\ket{R} = \int dk e^{ik(\hat{r}-R)}\ket{u_k}$ with $\ket{u_k}$ being a Bloch state. In Eq.\ref{correlation}, $k_0$ is the non-HS point and the $\xi$ cab be regarded as correlation length, which coincides with the decay length of the edge modes in topological non-trivial  phase\cite{chen2017correlation}. The correlation function $\lambda_R$ decays exponentially near the non-HS critical point and the decay gets slower as we tune towards the criticality with the diverging correlation length $\xi$. This clearly indicate the topological transition through non-HS criticality between gapped phases.
%\subsection{Topological transition between gapped phases/Non-HSP quantum criticality between gapped phases}
\begin{figure}[t]
		\includegraphics[width=4cm,height=2.5cm]{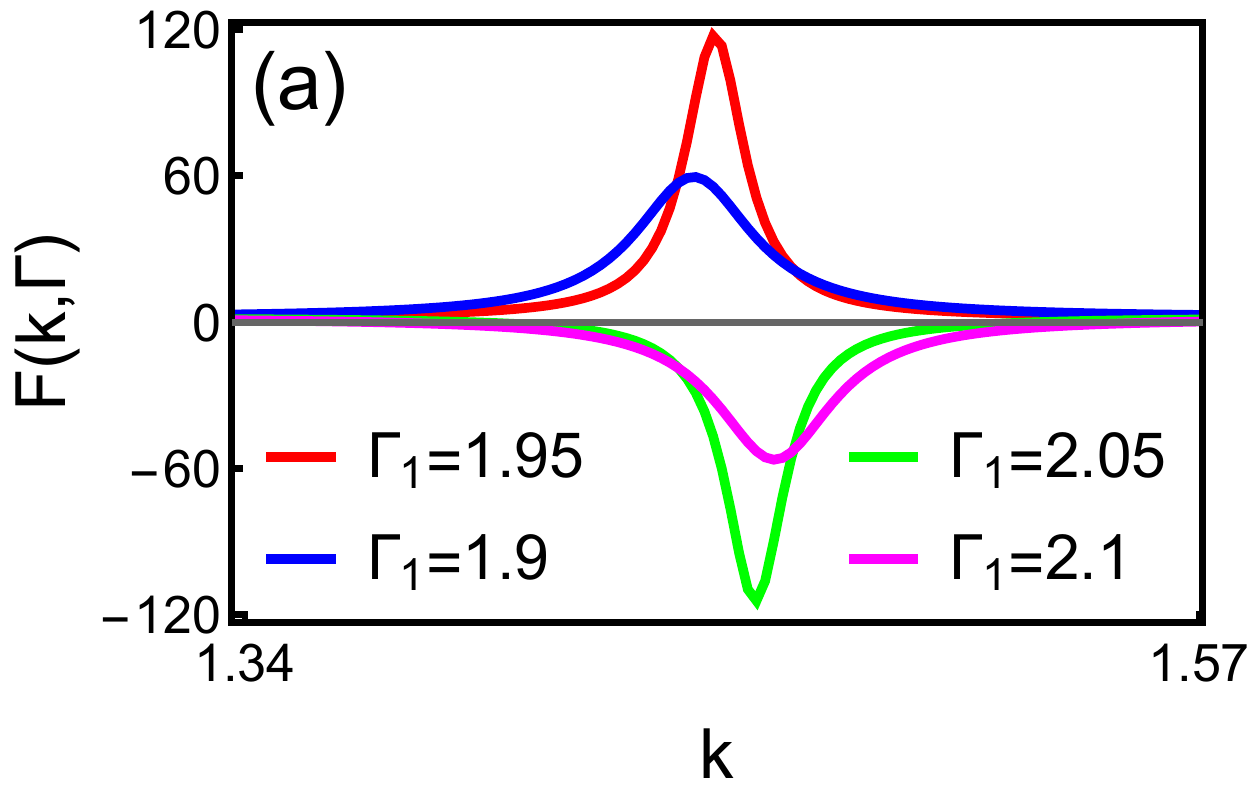} 
		\hspace{0.2cm} 
		\includegraphics[width=4cm,height=2.5cm]{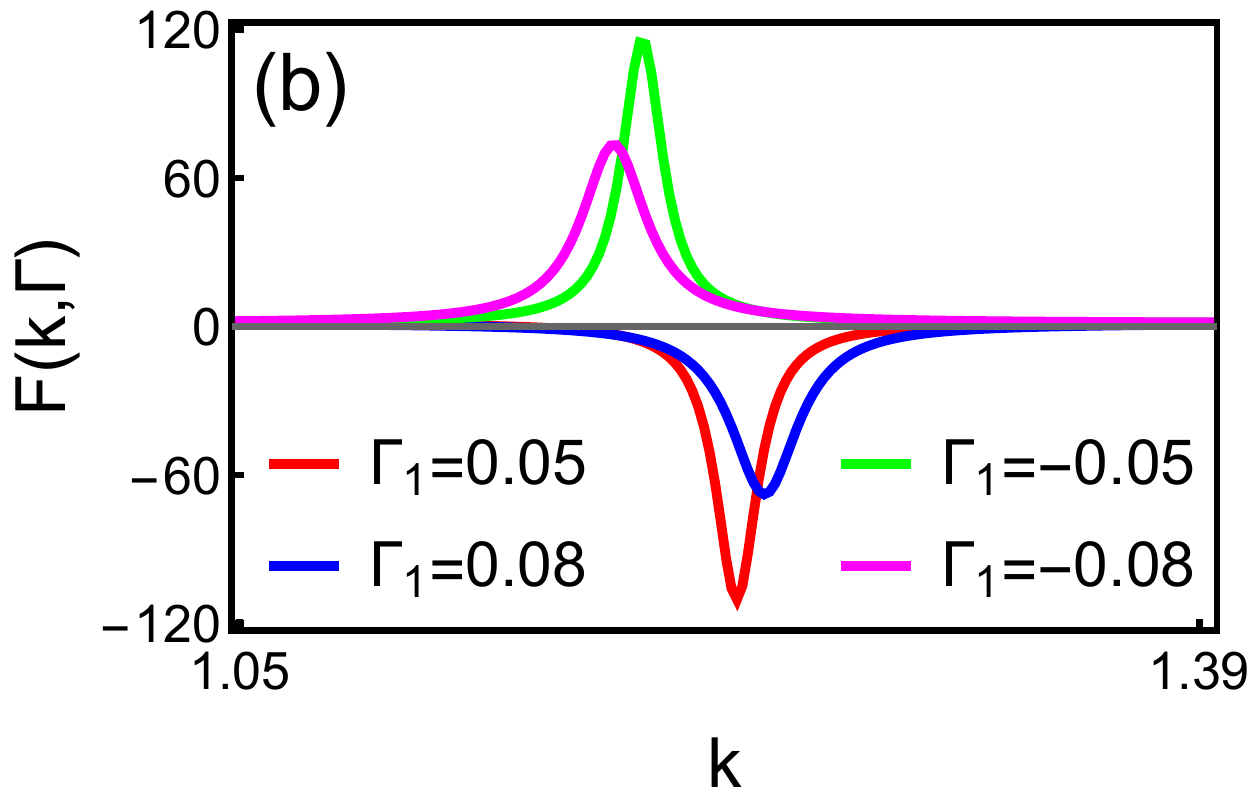}  
		\includegraphics[width=4cm,height=2.5cm]{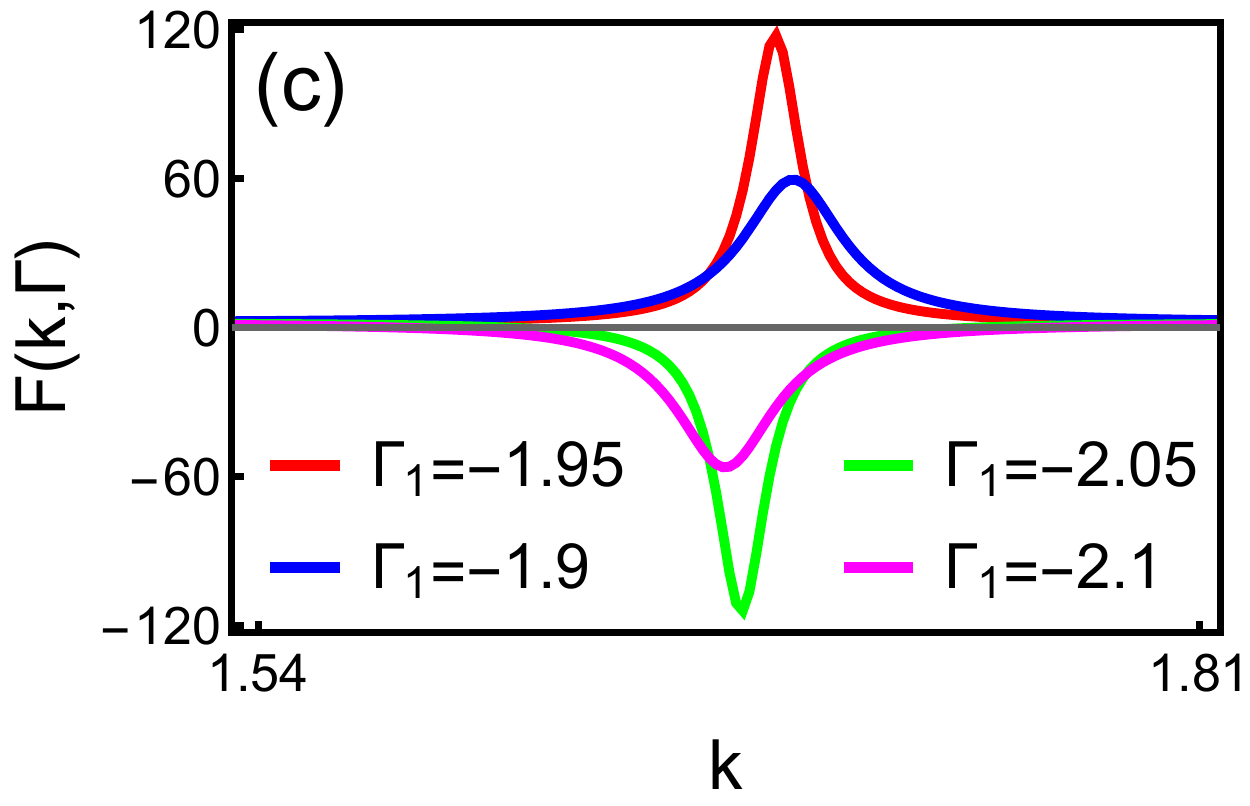}  
		\hspace{0.2cm} 
		\includegraphics[width=4cm,height=2.5cm]{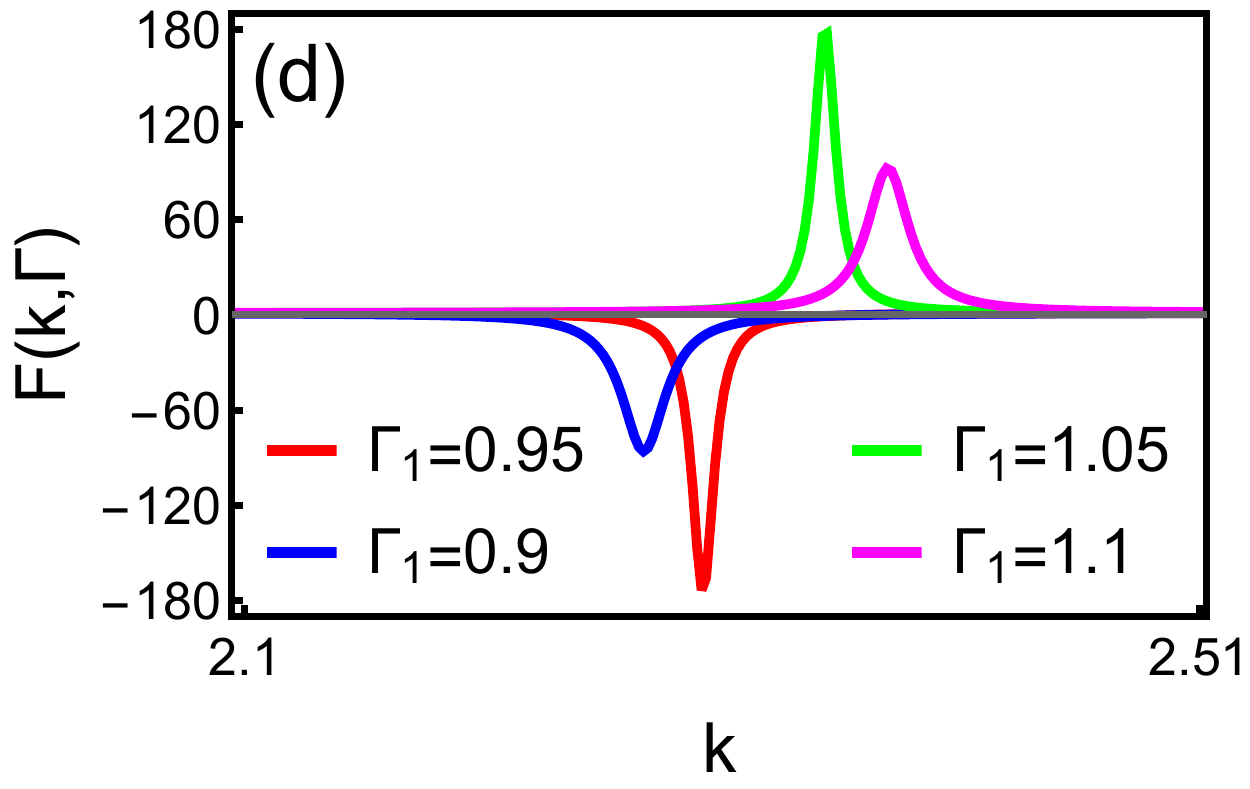}  
	\caption{\label{CF} Nature of curvature function in the vicinity of non-HS criticalities i.e. magenta and green critical lines in Fig.\ref{Topo-phase-diag}. Fixing the values $\Gamma_{0}$, $\Gamma_{2}$ and $\Gamma_{3}$, the parameter $\Gamma_{1}$ is varied around the critical value $\Gamma_{1}^c$. For all the plots $\Gamma_{0}=1$ and $\Gamma_{2}=0.5$. (a) For $\Gamma_{3}=2.2$ and $\Gamma_{1}^c=2$. It is a critical point on the magenta line between the phases $w=3$ and $w=1$. (b) For  $\Gamma_{3}=0.7$ and $\Gamma_{1}^c=0$. It is a critical point on the magenta line between the phases $w=2$ and $w=0$. (c) For $\Gamma_{3}=-2.2$ and $\Gamma_{1}^c=-2$. It is a critical point on the green line between the phases $w=3$ and $w=1$. (d) For $\Gamma_{3}=-0.36$ and $\Gamma_{1}^c=1$. It is a critical point on the green line between the phases $w=2$ and $w=0$.}
\end{figure}
\subsection{Curvature function and critical exponents}\label{expo}
The curvature function of the model in Eq.\ref{Model-H} can be obtained using the pseudo-spin vectors as
\begin{align}
F(k,\mathbf{\Gamma}) &= \frac{\chi_y\partial_k \chi_x - \chi_x \partial_k \chi_y}{\chi_x^2+\chi_y^2} \label{CF-A}  \\\nonumber
&= \frac{A+B\cos(k)+C\cos(2k)+D\cos(3k)}{A^{\prime}+B^{\prime}\cos(k)+C^{\prime}\cos(2k)+D^{\prime}\cos(3k)},
\end{align}
where $A=\Gamma_{1}^2+2\Gamma_{2}^2+3\Gamma_{3}^2$, $A^{\prime}=\Gamma_{0}^2+\Gamma_{1}^2+\Gamma_{2}^2+\Gamma_{3}^2$, 
$B=\Gamma_{0}\Gamma_{1}+3\Gamma_{1}\Gamma_{2}+5\Gamma_{2}\Gamma_{3}$, 
$B^{\prime}=2(\Gamma_{0}\Gamma_{1}+\Gamma_{1}\Gamma_{2}+\Gamma_{2}\Gamma_{3})$, 
$C=2\Gamma_{0}\Gamma_{2}+4\Gamma_{1}\Gamma_{3}$, 
$C^{\prime}=2(\Gamma_{0}\Gamma_{2}+\Gamma_{1}\Gamma_{3})$, 
$D=3\Gamma_{0}\Gamma_{3}$ and 
$D^{\prime}=2\Gamma_{0}\Gamma_{3}$. \hspace{0.3cm}In Fig.\ref{CF}, we show the nature of curvature function in the vicinity of the non-HS criticalities, $\Gamma_{3}=(\Gamma_{1}\pm\sqrt{\Gamma_{1}^2+4\Gamma_{0}(\Gamma_{0}-\Gamma_{2})})/2$ (i.e. magenta and green lines in Fig.\ref{Topo-phase-diag}).
%with corresponding $k_0=\arccos((-2\Gamma_{2}\pm \sqrt{4\Gamma_{2}^2-16\Gamma_{3}(\Gamma_{1}-\Gamma_{3})})/8\Gamma_{3})$ respectively. 
Setting $\Gamma_{0}=1$ and $\Gamma_{2}=0.5$ we tune $\Gamma_{1}$ towards its critical values (say $\Gamma_{1}^c$) for a fixed value of $\Gamma_{3}$. For the non-HS critical point between the gapped phases $w=1$ and $w=3$, $\Gamma_{1}^c=\pm2$ and $\Gamma_{3}=\pm2.2$ respectively for magenta and green criticalities. For $w=0$ and $w=2$, $\Gamma_{1}^c=0$ and $\Gamma_{3}=0.7$ for magenta and $\Gamma_{1}=1$ and $\Gamma_{3}=-0.36$ for green criticalities. Fixing $\Gamma_{3}$, we vary $\Gamma_{1}$ around the critical point, as shown in Fig.\ref{CF}.

\hspace{0.3cm} As one tune the parameter $\Gamma_1$ towards its critical value $\Gamma_1^c$, diverging peak occurs at non-HS points $k_0$, which shifts each time the $\Gamma_1$ is varied. The peak becomes prominent as we approach $\Gamma_1=\Gamma_1^c$ and flips the sign as we tune further across the critical point. 
%and exhibit the negative peak with similar property. 
%As $\Gamma_1$ is varied one can see the shift in the $k_0$ value. 
%either towards or away from $k_0^c$. 
These properties of curvature function can be observed for both non-HS criticalities.
%exhibits the similar property around both the non-HS criticalities of the model. 
Note that, the same nature of curvature function can also be observed at the negative pair of non-HS point. Therefore, the divergence and flipping of curvature function
%, as it involves the change in topological invariant number, 
can be considered as an efficient qualitative observation to identify the topological transition through non-HS criticalities.

\begin{figure}[t]
		\includegraphics[width=4.0cm,height=2.8cm]{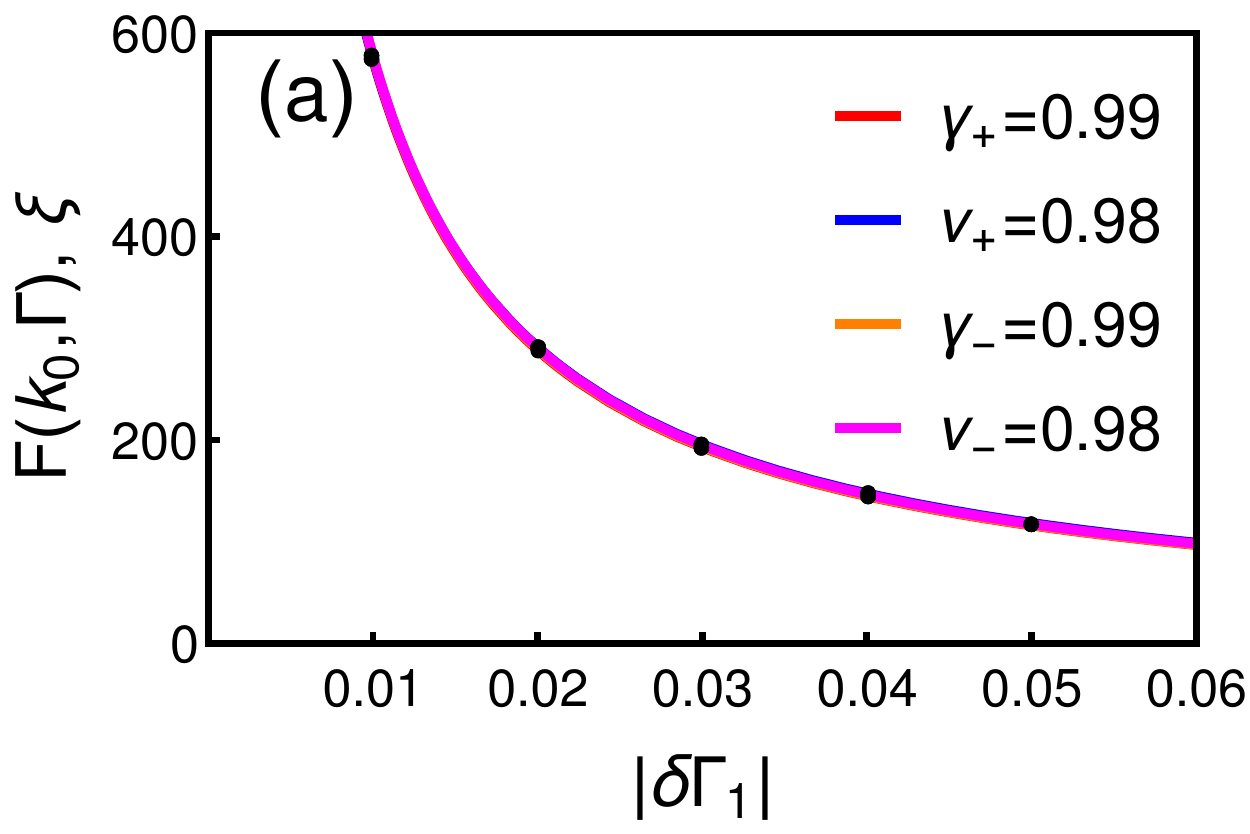}  
		\hspace{0.2cm} 
		\includegraphics[width=4.0cm,height=2.8cm]{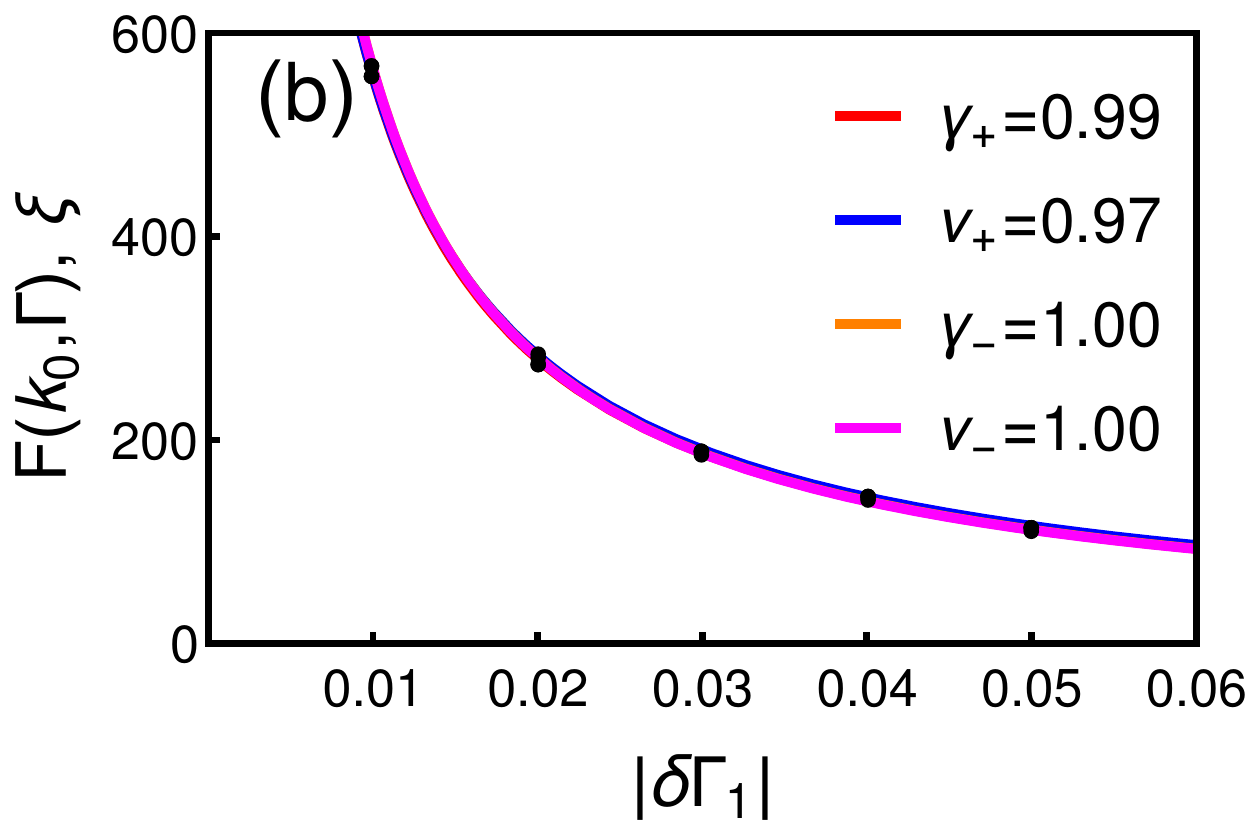}  
		\includegraphics[width=4.0cm,height=2.8cm]{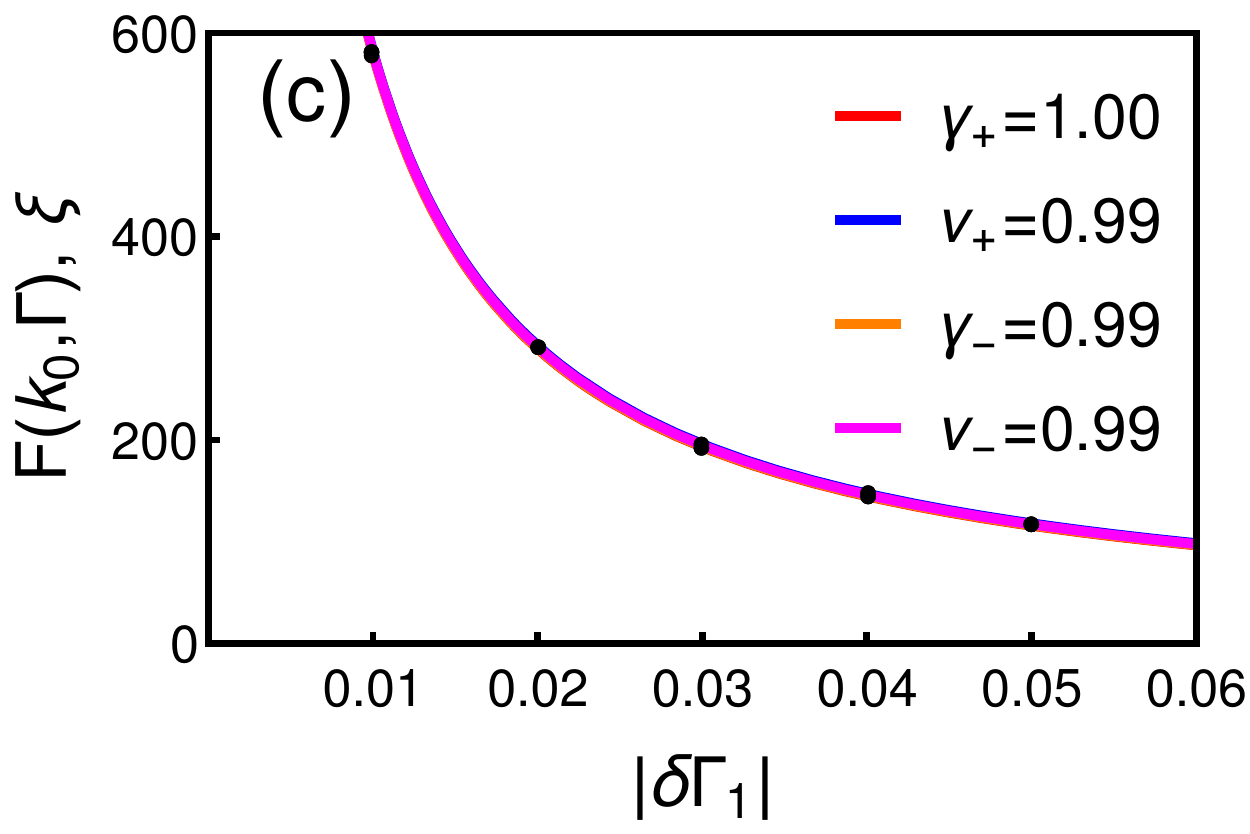}  
		\hspace{0.2cm} 
		\includegraphics[width=4.0cm,height=2.8cm]{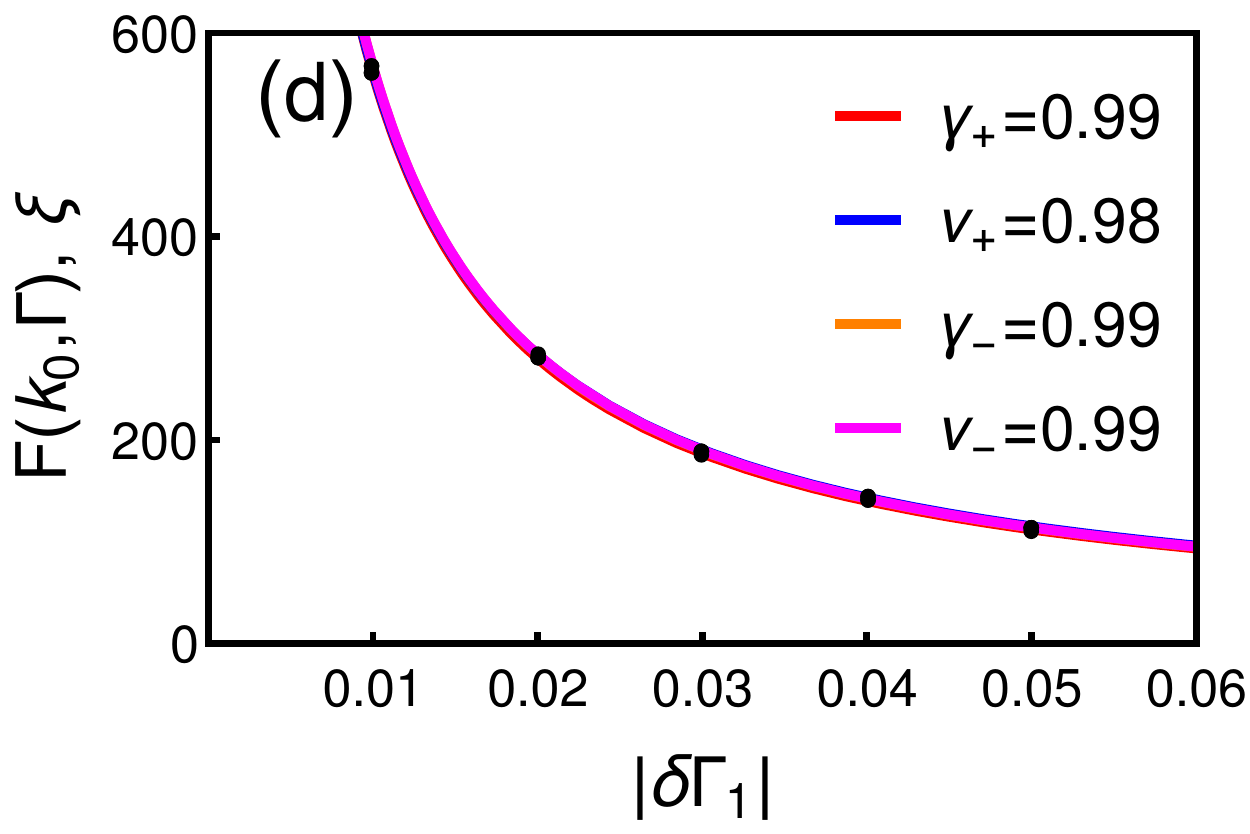}  
		\includegraphics[width=4.0cm,height=2.8cm]{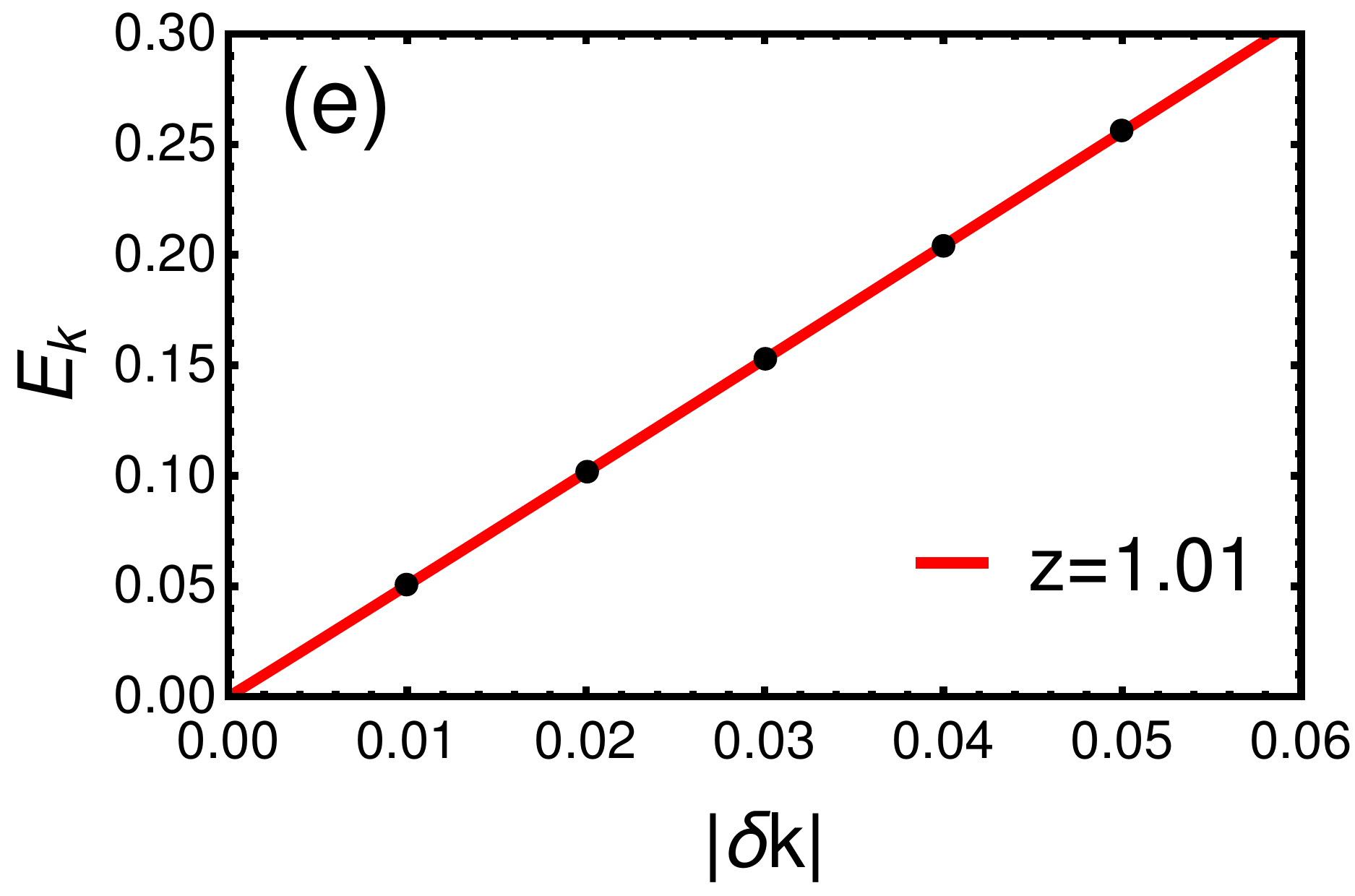} 
		\hspace{0.2cm} 
		\includegraphics[width=4.0cm,height=2.8cm]{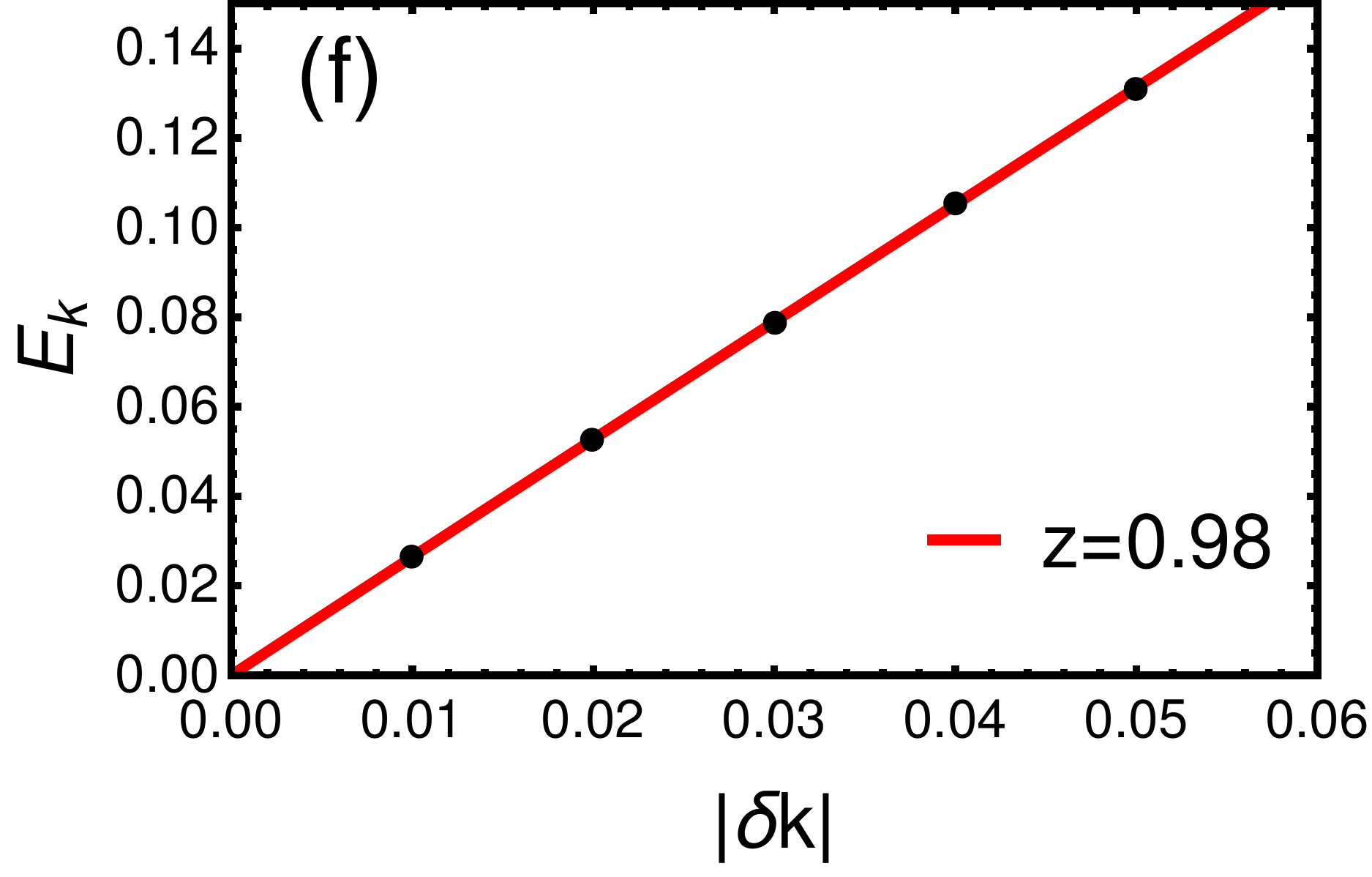} 
	\caption{\label{Expo} Critical exponents for non-HS quantum criticalities. For all the plots $\Gamma_{0}=1$ and $\Gamma_{2}=0.5$. (a) For $\Gamma_{3}=2.2$ and $\Gamma_{1}^c=2$. (b) For $\Gamma_{3}=0.7$ and $\Gamma_{1}^c=0$. (c) For $\Gamma_{3}=-2.2$ and $\Gamma_{1}^c=-2$. (d) For $\Gamma_{3}=-0.7$ and $\Gamma_{1}^c=0$. For all the cases exponents are found to be $\gamma=\gamma_{+/-}\approx 1$ and $\nu=\nu_{+/-}\approx 1$. (e) Dynamical critical exponent for a critical point between the gapped phases $w=1$ and $w=3$. (f) Dynamical critical exponent for a critical point between the gapped phases $w=0$ and $w=2$.}
\end{figure}

%\subsection{Critical exponents}\label{expo}
\hspace{0.3cm} The behavior of the curvature function 
%near non-HS quantum criticalities 
%discussed in the previous section 
can be quantified in terms of critical exponents %According to Eq.\ref{Lorenzian} one can write the curvature function in the form of Ornstein-Zernike equation which enable one to capture the exponents associated with the divergence of the curvature function. 
$\gamma$ and $\nu$ as defined in Eq.\ref{critical-exponents}, which captures the divergences in the height $F(k_0,\mathbf{\Gamma})$ and inverse of the width $\xi$ of curvature function. The values of these exponents can be obtained by numerical fitting of the curvature function with 
\begin{equation}
F_{fitting}=c+\frac{F(k_0,\mathbf{\Gamma})}{1+\xi^2(k-k_0)^2}
\end{equation} 
where $c$ is an arbitrary constant.
The fitting is done by varying $\mathbf{\Gamma}$ in the vicinity of the non-HS critical points with corresponding $k_0$ values. The data points collected for $|F(k_0,\mathbf{\Gamma})|$ and $\xi$ can then be fitted again with Eq.\ref{critical-exponents} to extract the exponents. Fig.\ref{Expo} demonstrates this process and shows the exponent values on approaching the non-HS critical points from either sides (represented as $\gamma_{+/-}$ and $\nu_{+/-}$). The data points are collected by fixing the parameters $\Gamma_{0}$, $\Gamma_{2}$ and $\Gamma_{3}$ and varying $\Gamma_{1}$ by $\delta\Gamma_{1}$ ($\delta\Gamma_{1}=|\Gamma_{1}-\Gamma_{1}^c|$) on either sides of the critical points.
%with the corresponding $k_0$ value. 

\hspace{0.4cm} The critical exponents can also be calculated analytically by expanding the pseudo-spin vector $\boldsymbol{\chi}(k)$ around non-HS point $k_0$ and recasting the curvature function in the Ornstein-Zernike form. The expansion upto first order, $\boldsymbol{\chi}(k)|_{k=k_0} \approx \boldsymbol{\chi}(k_0) + \partial_k\boldsymbol{\chi}(k_0) \delta k$, for the non-HS points $k_0=\arccos((-2\Gamma_{2}\pm \sqrt{4\Gamma_{2}^2-16\Gamma_{3}(\Gamma_{1}-\Gamma_{3})})/8\Gamma_{3})$ yields (the sign `$\pm$' represents the magenta and green criticalities respectively)
\begin{equation}
\chi_{x} \approx \delta\mathbf{\Gamma}+A \delta k \;\;\; \quad \text{and} \quad\;\;\; \chi_{y} \approx B \delta k,
\end{equation}
where 
\begin{gather}
\delta \mathbf{\Gamma} = \Gamma_{0}-\Gamma_{2}/2\mp(1/2)\alpha \nonumber\\
A = \frac{(4\Gamma_{1}\Gamma_{3}-\Gamma_{2}^2\pm\Gamma_{2}\alpha)\sqrt{2\Gamma_{3}(\Gamma_{1}+3\Gamma_{3}-\Gamma_{2}^2\pm\Gamma_{2}\alpha)}}{4\sqrt{2}\Gamma_{3}^2} \nonumber\\
B =\frac{-\Gamma_{2}^3+4\Gamma_{2}\Gamma_{3}(\Gamma_{1}-\Gamma_{3})\pm \Gamma_{2}^2\alpha \mp 2\Gamma_{3}(\Gamma_{1}+3\Gamma_{3}\alpha)}{4\Gamma_{3}^2}
\end{gather}
with $\alpha=\sqrt{\Gamma_{2}^2+4\Gamma_{3}(\Gamma_{3}-\Gamma_{1})}$. Therefore the curvature function in Eq.\ref{CF-A} can be recasted as 
\begin{align}
F(k,\delta\mathbf{\Gamma}) &= \frac{-B/\delta\mathbf{\Gamma}}{1+\left(\frac{2A}{\delta\mathbf{\Gamma}} \right)\delta k +\left(\frac{A^2+B^2}{\delta \mathbf{\Gamma}^2} \right)\delta k^2 } \nonumber\\
&=\frac{F(k_0,\delta\mathbf{\Gamma})}{1+\xi\delta k+\xi^2\delta k^2}
\end{align}
Note that, the term $\xi^2$ is dominant as it diverges more quickly than $\xi$, therefore, one can obtain the Ornstein-Zernike form using only the leading term in the denominator. The critical exponents $\gamma$ and $\nu$ are
\begin{eqnarray}
F(k_0,\delta\mathbf{\Gamma}) =-B\delta\mathbf{\Gamma}^{-1} \implies \gamma=1\\
\xi = \sqrt{(A^2+B^2)}\delta \mathbf{\Gamma}^{-1} \implies \nu=1.
\end{eqnarray}
This clearly demonstrate that, analytical and numerical values of critical exponents agree each other. Therefore, the exponents $\gamma=\nu=1$ for both the non-HS criticalities and they obey the scaling law $\gamma=\nu$ for 1D systems.

\hspace{0.4cm} Moreover, the vanishing 
energy scale of the gap function $\Delta$, defines a gap exponent
\begin{equation}
\Delta \propto |\mathbf{\Gamma}-\mathbf{\Gamma}_{c}|^{y}, \label{dyna-crit}
\end{equation} 
where $y=z\nu$, which is dynamical scaling law \cite{jalal2016topological} 
with $z$ the dynamical critical exponent \cite{malard2020scaling,rufo2019multicritical}. The $z$ dictates the nature of the spectra near the gap closing momenta $k_0$, i.e. $E_k\propto k^z$.
%{\color{red}Write about the analytical way $E_k=\sqrt{A^2+B^2}k \implies z=1$. Can write this in theoretical section while explaining about dispersion.} 
It can be calculated numerically using curve fitting method similar to the previous case. 
%We first collect the data points of $E_k$ very close to the gap closing point and fit the data points collected to the scaling equation. 
This procedure results 
in the Fig.\ref{Expo}(e) and (f), which yields $z=1$ for both the non-HS criticalities. The gap exponent can be obtained as $y=z\nu =1$. 
The critical exponents defines the universality class of the topological transition through both non-HS quantum criticalities between gapped phases. Therefore, both the non-HS criticalities share the same universality class $(\gamma,\nu,z)=(1,1,1)$.

\subsection{Curvature function renormalization group and correlation function}
\hspace{0.3cm}In this section, we perform CRG
%described in Section.\ref{CRG-non} 
to the model in Eq.\ref{Model-H} and obtain RG equations which essentially captures the topological transition between gapped phases through non-HS criticalities. %This is achieved by realizing the scaling in Eq.\ref{scaling} with the curvature function in Eq.\ref{CF-A} of the model under consideration. 
The RG equations in terms of the parameters for the non-HS points can be constructed from Eq.\ref{CRG-General}. For the non-HS point $k_0=\arccos((-2\Gamma_{2}+ \sqrt{4\Gamma_{2}^2-16\Gamma_{3}(\Gamma_{1}-\Gamma_{3})})/8\Gamma_{3})$ (corresponds to magenta line in Fig.\ref{Topo-phase-diag}), the RG equations for $\Gamma_{1}$ and $\Gamma_{3}$ ($\Gamma_{0}=1$ and $\Gamma_{2}=0.5$) can be obtained as
\begin{equation}
\dfrac{d\Gamma_{1}}{dl}=\frac{\alpha\sqrt{\alpha_1}\Lambda_1(\Gamma_{0},\Gamma_{1},\Gamma_{2},\Gamma_{3})}{2\sqrt{2}\Lambda_2(\Gamma_{0},\Gamma_{1},\Gamma_{2},\Gamma_{3})}
\end{equation}
\begin{equation}
\dfrac{d\Gamma_{3}}{dl}=\frac{\Gamma_{3}^2\alpha\sqrt{\alpha_1}\Lambda_1(\Gamma_{0},\Gamma_{1},\Gamma_{2},\Gamma_{3})}{2\sqrt{2}\Lambda_2^{\prime}(\Gamma_{0},\Gamma_{1},\Gamma_{2},\Gamma_{3})}
\end{equation}
Similarly, for the non-HS point $k_0=\arccos((-2\Gamma_{2}- \sqrt{4\Gamma_{2}^2-16\Gamma_{3}(\Gamma_{1}-\Gamma_{3})})/8\Gamma_{3})$ (corresponds to green line in Fig.\ref{Topo-phase-diag}) we obtain
\begin{equation}
\dfrac{d\Gamma_{1}}{dl}=-\frac{\alpha\sqrt{\alpha_1}\Lambda_3(\Gamma_{0},\Gamma_{1},\Gamma_{2},\Gamma_{3})}{\sqrt{2}\Gamma_{3}^2\Lambda_4(\Gamma_{0},\Gamma_{1},\Gamma_{2},\Gamma_{3})}
\end{equation}
\begin{equation}
\dfrac{d\Gamma_{3}}{dl}=-\frac{\alpha\sqrt{\alpha_1}\Lambda_3(\Gamma_{0},\Gamma_{1},\Gamma_{2},\Gamma_{3})}{\sqrt{2}\Lambda_4^{\prime}(\Gamma_{0},\Gamma_{1},\Gamma_{2},\Gamma_{3})}
\end{equation}
\begin{figure}[t]
		\includegraphics[width=4cm,height=4cm]{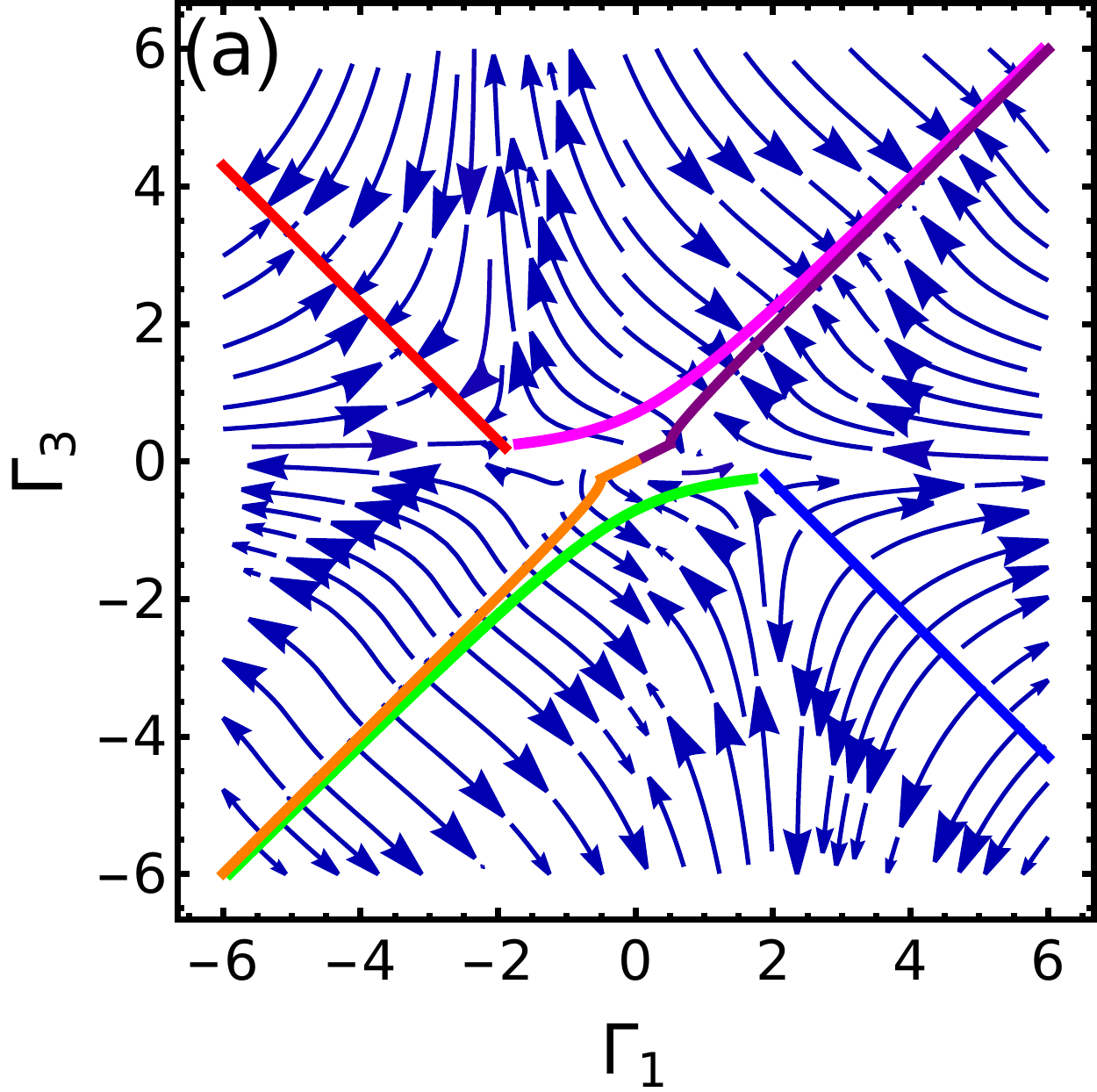}  
		\includegraphics[width=4cm,height=4cm]{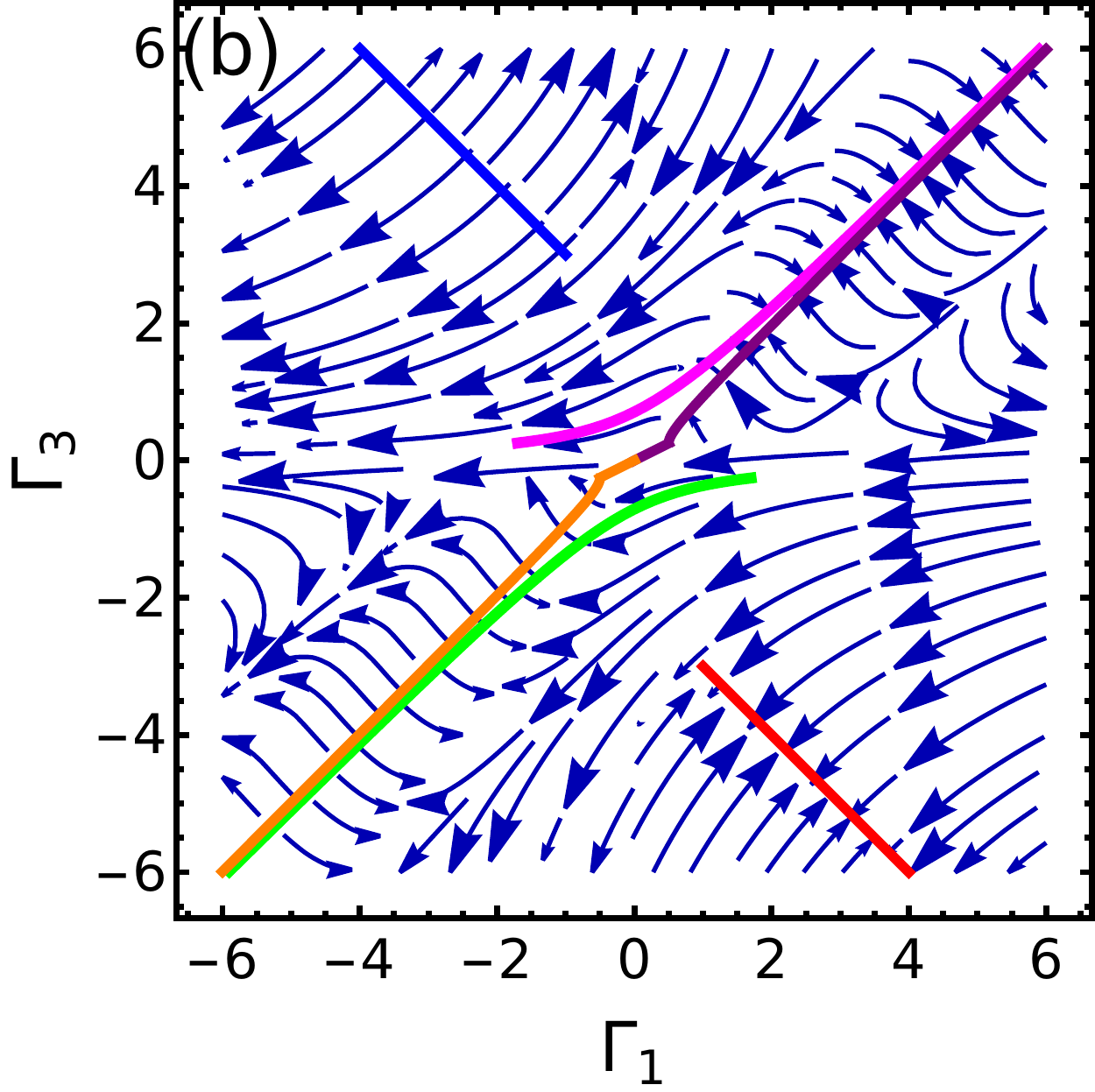}  
	\caption{\label{CRG} CRG flow diagram. Plotted for $\Gamma_{0}=1$ and $\Gamma_{2}=0.5$. (a) RG flow for magenta criticality. (b) RG flow for  green criticality. The magenta and green criticalities manifests as stable and unstable fixed points respectively. These lines coincides with the fixed lines represented in orange and purple. The HS lines, represented as red and blue lines, are partially recognised by the RG flow as stable and unstable fixed lines respectively. %(c) and (d): RG flow for $-k_0$. The flow directions are opposite to the previous case. Here the magenta and green lines are unstable and stable fixed lines and red and blue lines are unstable and stable fixed lines. The orange and purple lines represents the fixed lines.}
	}
\end{figure}
where $\alpha=\sqrt{\Gamma_{2}^2+4\Gamma_{3}(-\Gamma_{1}+\Gamma_{3})}$ and $\alpha_1=\sqrt{(2\Gamma_{3}(\Gamma_{1}+3\Gamma_{3})-\Gamma_{2}^2+\Gamma_{2}\alpha)}/\Gamma_{3}$ (see supplementary material for detailed form of $\Lambda$s). The non-HS criticalities can be identified from the RG flow in the $\Gamma_{1}-\Gamma_{3}$ plane. In general, the RG flow rate and the direction enables the identification the critical and fixed points (stable and unstable) in the parameter space \cite{chen2018weakly}, as explained in Eq.\ref{crit-fixed}. 
%\begin{align}
%\text{Critical point:}& \hspace{0.2cm} \left| \frac{d\mathbf{\Gamma}}{dl}\right| \rightarrow \infty, \text{flow directs away}.\nonumber\\
%\text{Stable fixed point:}& \left| \frac{d\mathbf{\Gamma}}{dl}\right| \rightarrow 0, \text{flow directs into}.\nonumber\\
%\text{Unstable fixed point:}& \left| \frac{d\mathbf{\Gamma}}{dl}\right| \rightarrow 0, \text{flow directs away}.
%\label{crit-fixed}
%\end{align}
However, in this case, the RG flow exhibits an anomalous behavior that both the non-HS criticalities simultaneously satisfy the fixed and critical line conditions. In other words, the non-HS criticalities drives both numerator and denominator of the RG equations to zero individually. This is due to the overlap of critical and fixed lines in the flow diagram.
%of the RG equations for both the paramaters $\Gamma_{1}$ and $\Gamma_{3}$. 

\hspace{0.3cm} The fixed lines can be obtained as $\Gamma_{3}=1/2(\Gamma_{1}+\sqrt{\Gamma_{1}^2-\Gamma_{2}^2})$ which defines stable fixed points and  $\Gamma_{3}=1/2(\Gamma_{1}-\sqrt{\Gamma_{1}^2-\Gamma_{2}^2})$ which defines unstable fixed points (represented as purple and orange lines respectively in Fig.\ref{CRG}). These lines coincide with the non-HS critical lines for higher values of parameters i.e. $|\Gamma_1|$ and $|\Gamma_{2}|$. This results in the manifestation of the magenta critical lines as stable fixed line and green critical line as unstable fixed line, as shown in Fig.\ref{CRG}(a,b). 
%In contrast to the positive pair, for the negative pair of the non-HS point the RG flow can be found in opposite direction. In this case, the magenta line manifests as unstable fixed line and green line as stable fixed line, as shown in Fig.\ref{CRG}(c,d). 
The manifestation of the critical lines as fixed lines can be found in agreement with the observations made in Ref.\cite{malard2020scaling,abdulla2020curvature}.
\begin{figure}[t]
	\centering
		\includegraphics[width=4.2cm,height=3cm]{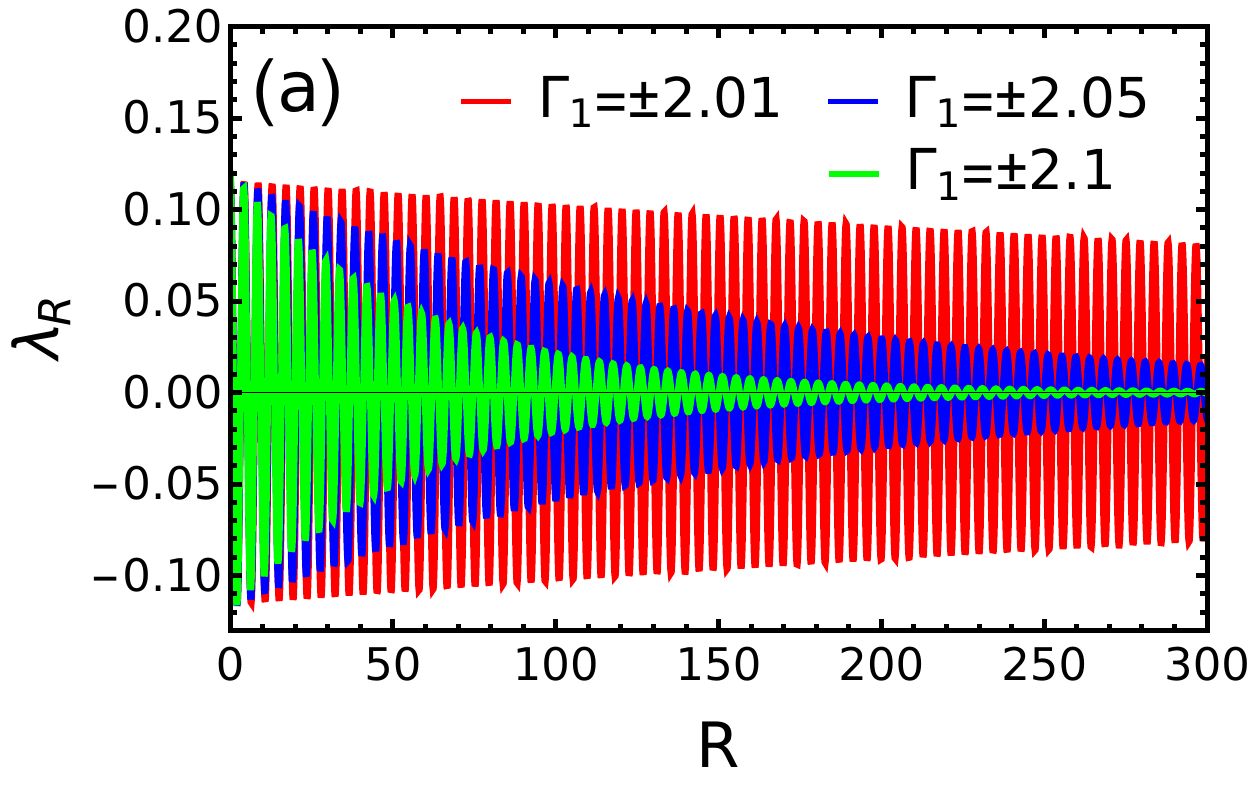}  
		\includegraphics[width=4.2cm,height=3cm]{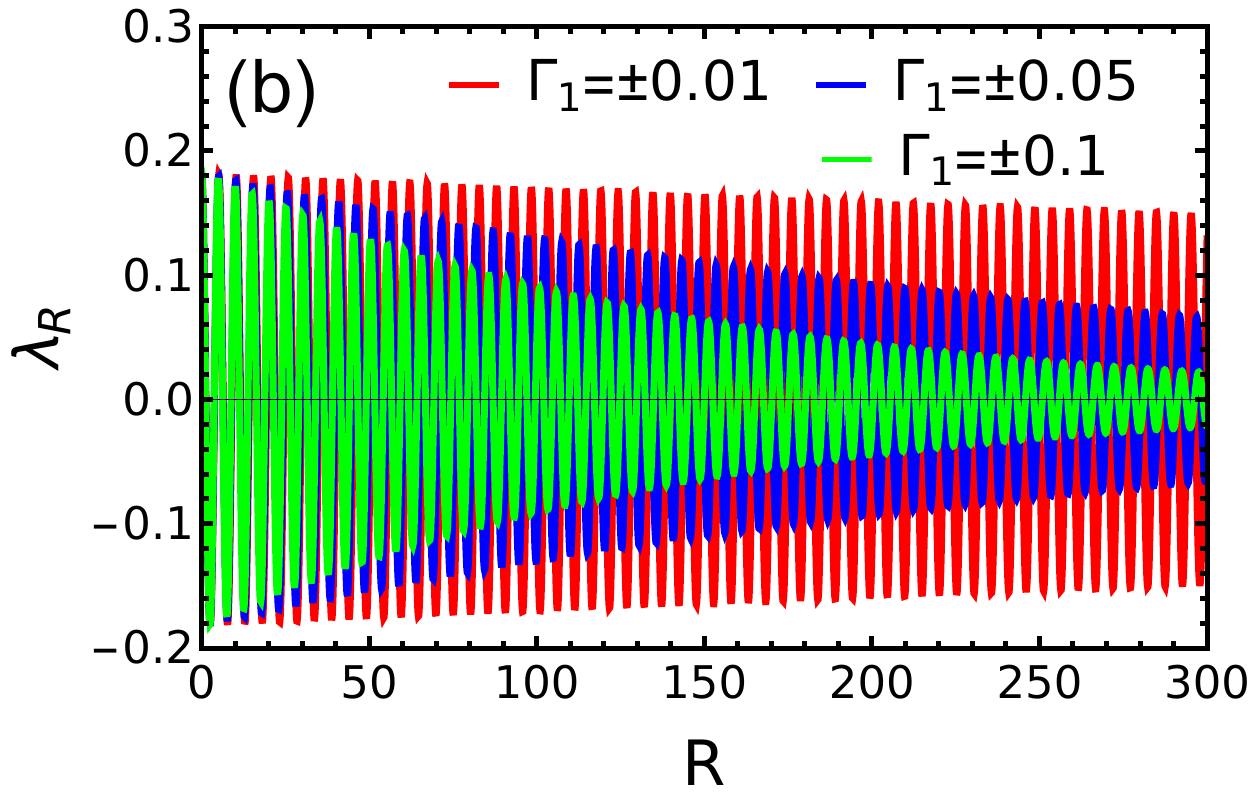}  
	\caption{\label{corr} Wannier state correlation function. The parameter values $\Gamma_{0}=1$ and $\Gamma_{2}=0.5$ are fixed. (a) Represents $\lambda_R$ in the vicinity of a critical point between the gapped phases $w=1$ and $w=3$. For the magenta line we choose $\Gamma_{1}^c=2$ and $\Gamma_{3}=2.2$ and for the green line $\Gamma_{1}^c=-2$ and $\Gamma_{3}=-2.2$. (b) Represents $\lambda_R$ in the vicinity of a critical point between $w=0$ and $w=2$. For the magenta line $\Gamma_{1}^c=0$ and $\Gamma_{3}=0.7$ and green line $\Gamma_{1}^c=0$ and $\Gamma_{3}=-0.7$.}
\end{figure}
Apart from this, interestingly the CRG constructed for non-HS criticalities partially captures the HS criticalities which are also manifested as fixed lines in the flow diagram. As shown in Fig.\ref{CRG} the HS criticalities $\Gamma_{3}=\mp(\Gamma_{0}\pm\Gamma_{1}+\Gamma_{2})$ (the red and blue lines) appear as the stable and unstable fixed points respectively. Therefore, the CRG developed for non-HS criticalities is efficient in detecting the corresponding topological transition between gapped phases and also partially captures the HS topological transitions.

\hspace{0.3cm}The Wannier state correlation function $\lambda_R$ defined in Eq.\ref{correlation}, clearly identify the topological transitions at the non-HS criticalities of the model. Fig.\ref{corr} shows the profile of the correlation function in the vicinity of both the non-HS criticalities. For $k_0=\arccos((-2\Gamma_{2}\pm \sqrt{4\Gamma_{2}^2-16\Gamma_{3}(\Gamma_{1}-\Gamma_{3})})/8\Gamma_{3})$, Eq.\ref{correlation} yields highly oscillatory decay in $\lambda_R$ in the vicinity of the critical points $\Gamma_{1}^c=\pm 2$ and $\Gamma_{1}^c=0$. As the parameter $\Gamma_{1}$ is tuned towards its critical value the decay in $\lambda_R$ slow down leading to the divergence in the length scale $\xi$. This is the typical behavior of the Wannier state correlation function for the topological transitions.

\section{Topological phase transition between non-HS critical phases through multicriticality}\label{IV}
\hspace{0.3cm} In this section, we investigate the existence of edge modes at non-HS critical phases and explore the possible topological transition between non-HS critical phases through multicritical points. To achieve this, at first, we construct the model at criticality using the near-critical approach~\cite{kumar2021topological} in which the Hamiltonian can be considered critical only in parameter space i.e. $\mathcal{H}(\mathbf{\Gamma}_c,k)$ 
%( $\mathbf{\Gamma}\rightarrow \mathbf{\Gamma}_c$ and $k\neq k_0$) 
with $k=k_0+\Delta k$ where $\Delta k<<2\pi$, to avoid the singularity at exact critical point. This method has been efficiently used to study the HS criticalities~\cite{kumar2021topological} and here we show that it is also effective to address the non-HS criticalities.

\begin{figure}[t]
	\includegraphics[width=6.4cm,height=2.5cm]{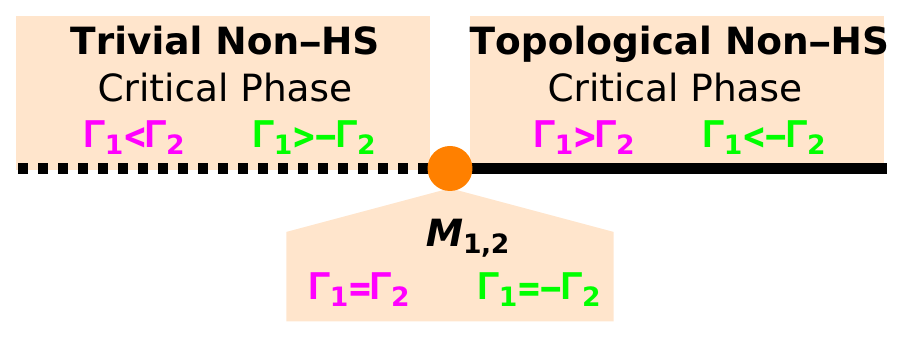}
	\caption{\label{mode-crit}Schematic representation of non-HS criticalities i.e. magenta and green lines in Fig.\ref{Topo-phase-diag}. The trivial and non-trivial phases are represented as dashed and solid lines respectively and are separated by multicritical points $M_{1,2}$. For the magenta line  $\Gamma_{1}<\Gamma_{2}$ ($\Gamma_{1}>\Gamma_{2}$) represents trivial (non-trivial) critical phase. For the green line $\Gamma_{1}>-\Gamma_{2}$ ($\Gamma_{1}<-\Gamma_{2}$) represents trivial (non-trivial) critical phase.  The transition between the trivial and non-trivial phases occur at $\Gamma_{1}=\pm \Gamma_{2}$ on the magenta and green lines respectively.}
\end{figure}

\hspace{0.3cm} To obtain $\mathcal{H}(\mathbf{\Gamma}_c,k)$,  we plug the non-HS critical line expression for $\Gamma_{3}$ into the pseudospin vectors in Eq.\ref{Model-H}.
This yields $ \chi_{x} (k) = \Gamma_0 + \Gamma_1 \cos k + \Gamma_2 \cos 2k + (\Gamma_{1}\pm \sqrt{\Gamma_{1}^2+4\Gamma_{0}(\Gamma_{0}-\Gamma_{2})})/2 \cos 3k,$ and $ \chi_{y} (k) = \Gamma_1 \sin k + \Gamma_2 \sin 2k + (\Gamma_{1}\pm \sqrt{\Gamma_{1}^2+4\Gamma_{0}(\Gamma_{0}-\Gamma_{2})})/2 \sin 3k $. The corresponding dispersion vanishes at multicritical points. Among the four multicritical points only the points with linear dispersion ($M_{1,2}$ in Fig.\ref{Topo-phase-diag}) separates the distinct non-HS critical phases, as schematically shown in Fig.\ref{mode-crit}. Therefore, we study only $M_{1,2}$, which can be obtained for the momenta
%summerized in Table.\ref{mome-swap}.
%\begin{table}[h]
%\begin{tabular}{|c|c|c|c|}
%	\hline \hline
%	 & $k_0^{mc}=0$ & $k_0^{mc}=\pi$ & $k_0^{mc}=\arccos[x]$ \\
%	 \hline
%	 Magenta & -- & $M_1$ & -- \\
%	 \hline
%	 Green & $M_2$ & -- & -- \\
%	 \hline
%	 Red & -- & -- & $M_2$ \\
%	 \hline
%	 Blue & -- & -- & $M_1$ \\
%	 \hline \hline
%\end{tabular}
%\caption{\label{mome-swap} Summary of gap closing momenta $k_0^{mc}$ for multicritical points $M_{1,2}$ ($\Gamma_{1}=\pm \Gamma_{2}$). We have $x=(-2\Gamma_{2}+\sqrt{4\Gamma_{2}^2}-16(\pm\Gamma_{0}-\Gamma_{1}\pm\Gamma_{2})(\Gamma_{1}-(\pm\Gamma_{0}-\Gamma_{1}\pm\Gamma_{2})))/8(\pm\Gamma_{0}-\Gamma_{1}\pm\Gamma_{2})$ with signs $\pm$ for blue and red criticalities repsctively. The solution of multicritical points is obtained for both HS and non-HS points as the $M_{1,2}$ are the intersection points of HS and non-HS criticalities.}
%\end{table}
\begin{multline}
k_0^{mc}=\pi,\\ \pm\arccos[\frac{-2\Gamma_{2}+\sqrt{-8(\Gamma_{1}+\frac{1}{2}(-\Gamma_{1}- \alpha))(\Gamma_{1}+\alpha+4\Gamma_{2}^2)}}{4(\Gamma_{1}+\alpha)}] \label{mome-crit1}
\end{multline}
for $M_1$ (multicritical point on the magenta line) and
\begin{multline}
k_0^{mc}=0,\\ \pm\arccos[\frac{-2\Gamma_{2}-\sqrt{-8(\Gamma_{1}+\frac{1}{2}(-\Gamma_{1}+ \alpha))(\Gamma_{1}-\alpha+4\Gamma_{2}^2)}}{4(\Gamma_{1}+ \alpha)}] \label{mome-crit2}
\end{multline}
for $M_2$ (multicritical point on the green line). Here $\alpha=\sqrt{\Gamma_{1}^2+4\Gamma_{0}(\Gamma_{0}-\Gamma_{2})}$.
%where %$x=\frac{-2\Gamma_{2}+\sqrt{4\Gamma_{2}^2}-16(\pm\Gamma_{0}-\Gamma_{1}\pm\Gamma_{2})(\Gamma_{1}-(\pm\Gamma_{0}-\Gamma_{1}\pm\Gamma_{2}))}{8(\pm\Gamma_{0}-\Gamma_{1}\pm\Gamma_{2})}$ and 
%the signs $\pm$ are for magenta and green criticalities respectively. 
At $M_{1,2}$ the gap closing occurs at three points in the Brillouin zone. One of them is HS point and the other two are non-HS points. This is due to the fact that $M_{1,2}$ are the intersection points of non-HS and HS critical lines (see Fig.\ref{Topo-phase-diag}). 
%As a consequence of this intersection, we observe an interesting \textit{swaping} property of $k_0^{mc}$. It is evident from Table.\ref{mome-swap} that the solution for $M_{1,2}$ ($\Gamma_{1}=\pm\Gamma_{2}$) on the non-HS criticalities (magenta and green lines) is obtained for HS points i.e. $k_0^{mc}=0$ and  $k_0^{mc}=\pi$. Similarly, the solution for $M_{1,2}$ on the HS criticalities (blue and red lines) is obtained for non-HS points i.e. $k_0^{mc}=\arccos[(-2\Gamma_{2}+\sqrt{4\Gamma_{2}^2}-16(\pm\Gamma_{0}-\Gamma_{1}\pm\Gamma_{2})(\Gamma_{1}-(\pm\Gamma_{0}-\Gamma_{1}\pm\Gamma_{2})))/8(\pm\Gamma_{0}-\Gamma_{1}\pm\Gamma_{2})]$. This is the generaization of the swaping property discussed in Ref.~\cite{kumar2021topological}. We discuss the further consequence of this property in the curvature function section.
Now driving the parameters towards the multicritical point involves both $\mathbf{\Gamma}_c \rightarrow \mathbf{\Gamma}_{mc}$ and $k \rightarrow k_0^{mc}$.
%Note that, there are HS values for $k_0^{mc}$ in Eq.\ref{mome-crit} along with non-HS values.This is due to the fact that the multicritical points are the intersection points of non-HS and HS critical lines. Among the four multicritical points only the points with linear dispersion ($M_{1,2}$ in Fig.\ref{Topo-phase-diag}) separates the distinct non-HS critical phases, as schematically shown in Fig.\ref{mode-crit}. 
In the following subsections we show topological trivial and non-trivial characters of the non-HS critical phases. %We obtain the edge mode solutions analytically and numerically at these critical phases. 
\begin{figure}[t]
	\includegraphics[width=4.0cm,height=2.6cm]{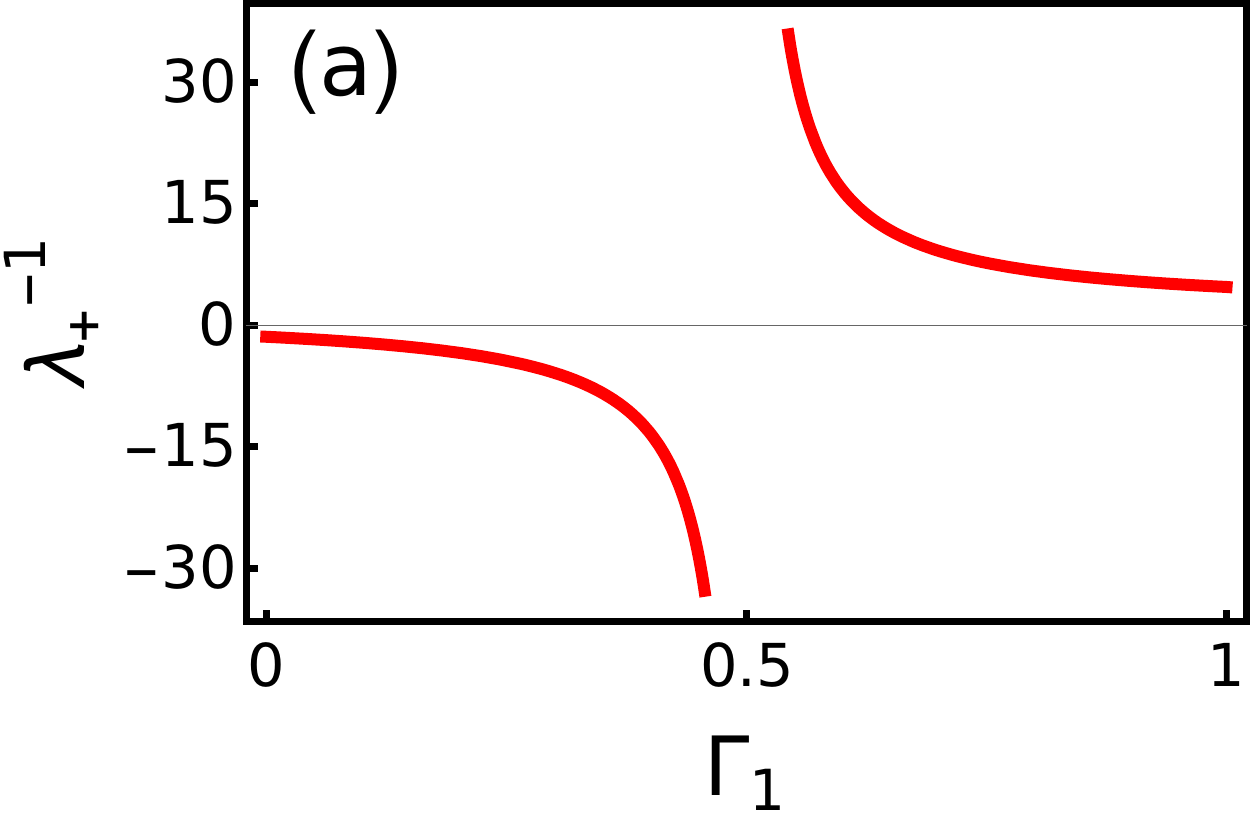}  
	\includegraphics[width=4.0cm,height=2.6cm]{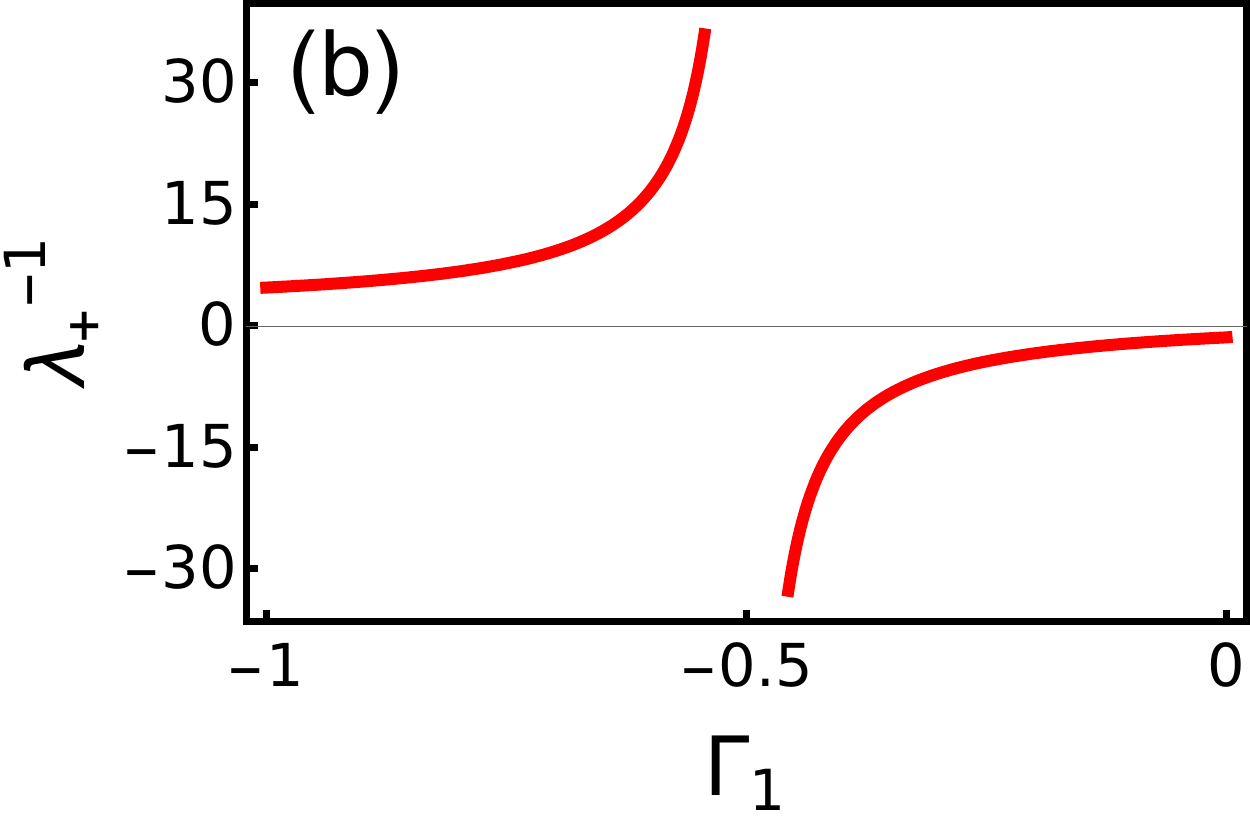}  
	\caption{\label{decay-L}Decay length of edge modes at non-HS criticalities. The parameters $\Gamma_{0}=1$ and $\Gamma_{2}=0.5$ are fixed. (a) On the magenta line. The decay length $\lambda^{-1}_+$ can be found to be positive (negative) at non-trivial (trivial) critical phase $\Gamma_{1}>0.5$ ($\Gamma_{1}<0.5$). (b) On the green line. The decay length $\lambda^{-1}_+$ can be found positive (negative) at non-trivial (trivial) critical phase $\Gamma_{1}<-0.5$ ($\Gamma_{1}>-0.5$).}
\end{figure} 

\subsection{Decay length of edge modes at non-HS criticalities}
\hspace{0.3cm} To enable the identification of the trivial and topological non-HS critical phases, we calculate the edge mode decay length using the Dirac equation~\cite{shen2011topological,lu2011non,jackiw1976solitons,verresen2020topology,kumar2021topological} at non-HS criticalities. The multicritical points are the phase boundaries, between distinct non-HS critical phases, at which the gap closes for $k_0^{mc}$. We expand the Hamiltonian defined at criticality around $k_0^{mc}=\pi,0$ (for magenta and green criticalities respectively) to obtain 
\begin{equation}
\mathcal{H}(k)=(m+\epsilon_1 k^2)\sigma_x+(\epsilon_2 k)\sigma_y
\end{equation}
where $m=\Gamma_{0}\mp 3\Gamma_{1}/2-(1/2)\alpha+\Gamma_{2}$, $\epsilon_1=\pm\Gamma_{1}+9/2(\Gamma_{1}+\alpha)-4\Gamma_{2}$ and $\epsilon_2=\mp 5\Gamma_{1}/2-(3/2)\alpha+2\Gamma_{2}$ with $\alpha=\sqrt{\Gamma_{1}^2+4\Gamma_{0}(\Gamma_{0}-\Gamma_{2})}$ (the sign `$\pm$' are for magenta and green lines respectively). The zero energy solution in real space $\mathcal{H}(-i\partial_x)\psi(x)=0$ (with $\hbar=1$) can be obtained by multiplying $\sigma_y$ from right hand side. This implies the wavefunction $\psi(x)=\rho_{\eta}\phi(x)$ is an eigenstate of $\sigma_z\rho_{\eta}=\eta\rho_{\eta}$. Using the trial wavefunction $\phi(x)\propto e^{-x \lambda}$ we get
\begin{equation}
-\eta\epsilon_1\lambda^{2}+\epsilon_2\lambda+\eta m=0
\end{equation}
where $\lambda$ is the inverse of the decay length which can be obtained as
\begin{equation}
\lambda_+=\frac{m}{\eta \epsilon_2}
\end{equation}
with $\eta=sign(\epsilon_2)$. The decay length remain positive for $\Gamma_{1}>\Gamma_{2}$ and negative for $\Gamma_{1}<\Gamma_{2}$ on the magenta line. This means that the critical phase $\Gamma_{1}>\Gamma_{2}$ is the topological phase with the edge modes and $\Gamma_{1}<\Gamma_{2}$ is the trivial critical phase with no edge modes. Similarly, on the green line, $\Gamma_{1}<-\Gamma_{2}$ has positive decay length and is non-trivial critical phase with edge mode while  $\Gamma_{1}>-\Gamma_{2}$ is trivial critical phase with no edge modes. The term $m$ plays the role of mass and is zero at the multicritical points $\Gamma_{1}=\pm\Gamma_{2}$. Therefore, as $m\rightarrow 0$ the decay length diverges, as shown in Fig.\ref{decay-L}, implying the delocalization of the edge modes into the bulk. This clearly indicates that the multicritical points are the topological phase transition points between the distinct non-HS critical phases.
\begin{figure}[t]
	\includegraphics[width=4.0cm,height=2.8cm]{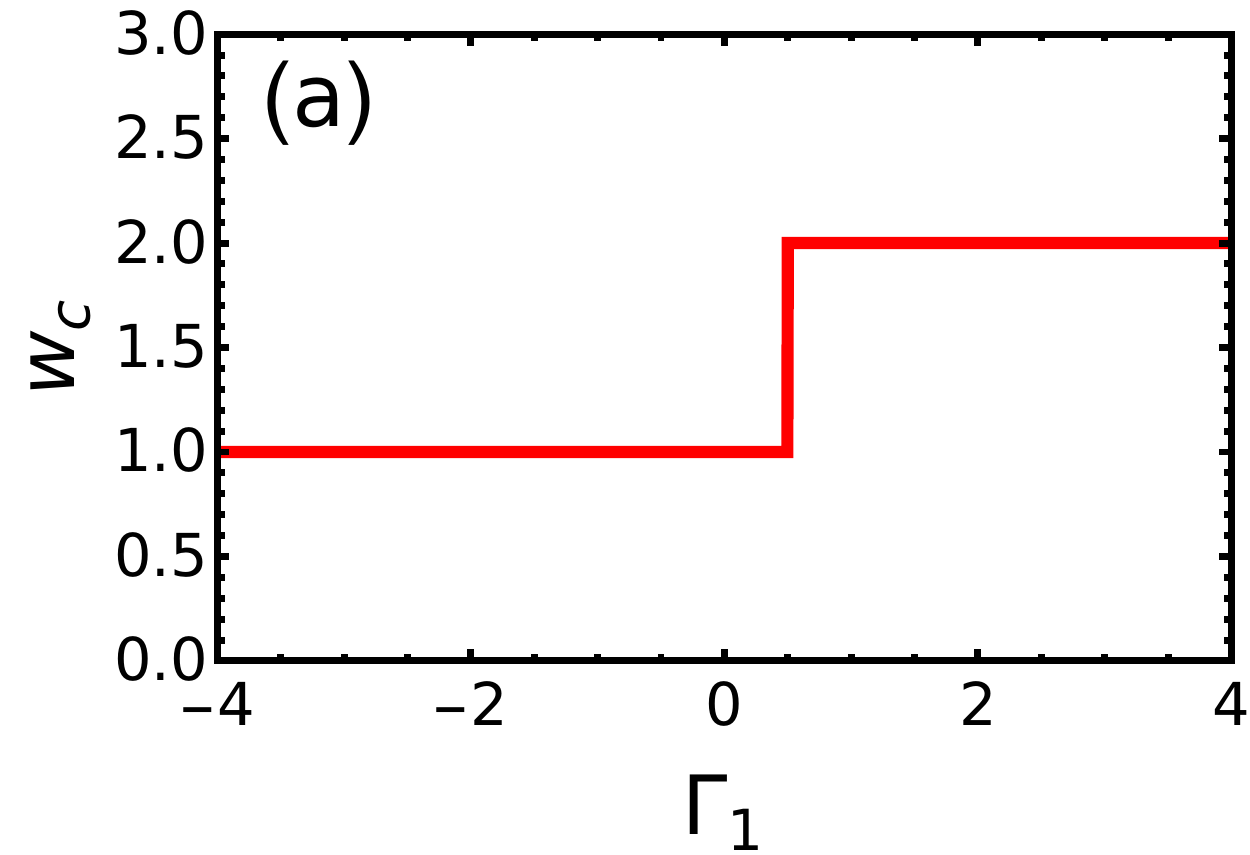}  
	\includegraphics[width=4.0cm,height=2.8cm]{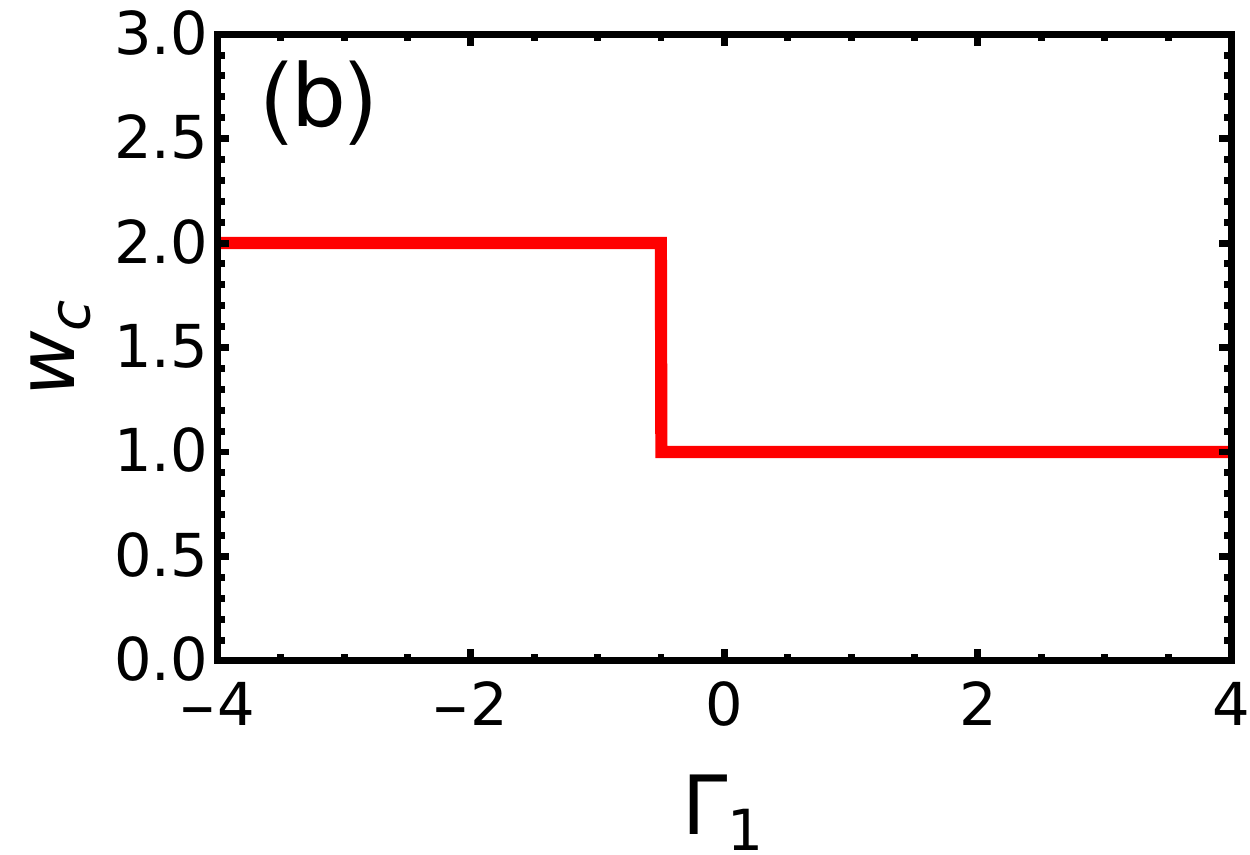}  
	\caption{\label{wind}Winding number at non-HS criticalities. The parameters $\Gamma_{0}=1$ and $\Gamma_{2}=0.5$ are fixed. (a) On the magenta line. For trivial phase, $\Gamma_1<0.5$, winding number $w_c=1$ and for non-trivial phase $\Gamma_1>0.5$ the winding number $w_c=2$. Transition occurs at the multicritical point $M_1$ at $\Gamma_1=0.5$. (b) On the green line. For trivial phase, $\Gamma_1>-0.5$, winding number $w_c=1$ and for non-trivial phase $\Gamma_1<-0.5$ the winding number $w_c=2$. Transition occurs at the multicritical point $M_2$ at $\Gamma_1=-0.5$.}
\end{figure} 

\subsection{Winding number at non-HS criticalities}
\hspace{0.3cm}Topological trivial and non-trivial characters of non-HS critical phase can also be identified using winding numbers. The winding number is a topological invariant number which quantify the edge excitations with gapped bulk, i.e. the bulk-boundary correspondence~\cite{kitaev2001unpaired,kane2005quantum}. As shown in Eq.\ref{winding}, it is defined as the integral of the curvature function over the Brillouin zone  which yields
%of Bloch wavefunction $\psi_k(r)=u_k(r) e^{ikr}$,
%\begin{equation}
%w=\frac{1}{2\pi}\oint\limits_{BZ}  F(k,\mathbf{\Gamma})  dk, \label{winding}
%\end{equation}
%where $ F(k,\mathbf{\Gamma}) = i \left\langle u_k\left| \partial_k\right| u_k\right\rangle$. 
integer values $\mathbb{Z}$, and it features a quantized jump at the topological phase transition point \cite{hasan2010colloquium}. However, the definition in Eq.\ref{winding} fails at the transition point due to the divergence of the integrand (curvature function). Therefore, in order to quantify the edge modes at criticality one has to exclude the singular point and can write~\cite{verresen2020topology,kumar2021topological}
\begin{equation}
w_c=\frac{1}{2\pi}\lim_{\delta \rightarrow 0}\oint\limits_{|k-k_0|>\delta} F(k,\mathbf{\Gamma}_c) dk. \label{wind-crit}
\end{equation} 
This defines the winding number at criticality and dictates the fractional values $(\mathbb{Z}/2)$ for HS critical phases~\cite{verresen2020topology}. The quantized jump of the fractional winding number at the transition points indicate the topological transition between HS critical phases. The winding number at non-HS criticalities can also be obtained from Eq.\ref{wind-crit}. In this case, one has to avoid two singular point in the set $k_0$, which yields $w_c$ integer values, as shown in Fig.\ref{wind}. As each gap closing point can contribute a factor of $(1/2)$, the winding number at the trivial non-HS critical phase is $w_c=1$. The non-trivial critical phase with one edge mode is assigned with winding number $w_c=2$. Fig.\ref{wind} shows the topological transition between non-HS critical phases through multicritical point at magenta and green lines. 
%{\color{red}To obtain the winding number at criticality which correctly count the edge modes at non-HS critical phases one has to consider the difference in their values. }

\hspace{0.3cm}Based on these observations one can argue that the bulk-boundary correspondence can be realised by identifying the difference between the winding numbers of topological non-trivial and trivial phases (either gapped or critical). 
\begin{equation}
\delta w=w^{non-trivial}-w^{trivial},\label{wind-diff}
\end{equation}
where $\delta w$ will be non-zero integer and counts the number of edge modes in the corresponding non-trivial phase. 
In the case of gapped phases, Eq.\ref{wind-diff} looks trivial as the winding number for a trivial phase $w^{trivial}=0$. However, for the critical phases it provides proper physical picture as it correctly counts the edge modes at the non-trivial critical phase. 
%gives more physical picture as the difference is alsways an interger number. This integer number counts the edge modes at the non-trivial critical phase. 
In case of HS criticality the trivial winding number is $w_c^{trivial}=1/2$ and non-trivial winding number is $w_c^{non-trivial}=\mathbb{Z}/2$ (where $\mathbb{Z}$ is non-zero integer)~\cite{verresen2020topology,kumar2021topological}. Therefore, the number of edge modes at the non-trivial HS critical phase is $\delta w_c=w_c^{non-trivial}-w_c^{trivial}=\mathbb{Z}$, this can be found in agreement with Ref \cite{kumar2021topological}.

\hspace{0.3cm}In the case of non-HS criticalities, there exists two gap closing points in the momentum space. Therefore, the trivial winding number itself turns out to be an integer (as each gap closing point contributes a factor of $1/2$). The trivial non-HS critical phase is now identified with $w_c=1$ and a non-trivial non-HS critical phase is with $w_c=\mathbb{Z}$ (where $\mathbb{Z}\geq 2$). In order to obtain correct number of edge modes at the non-trivial critical phase, one has to identify the difference $\delta w_c=w_c^{non-trivial}-w_c^{trivial}=\mathbb{Z}-1$. In our model the non-trivial non-HS critical phases are identified with $w_c^{non-trivial}=2$ (see Fig.\ref{wind}) for both the criticalities. Therefore, the proper number of edge modes at these phases can be obtained to be $\delta w_c=1$. 
%This can be found to be in agreement with analytical and numerical results obtained in the previous subsections.

\hspace{0.3cm} The analytical results of winding number and decay length of edge modes at criticality are found to be in agreement with the edge mode solutions obtained numerically under open boundary condition (we refer to supplementary material for the detailed discussion).

\section{CRG for topological transition between non-HS critical phases}\label{V}
\hspace{0.3cm}
%In this section, we discuss the CRG to identify the unique topological transition between the distinct non-HS critical phases. 
Scaling theory for the topological transition at non-HS critical point between gapped phases is developed in Section.\ref{CRG-non}. Here we reframe this scaling scheme in order to capture the topological transition between non-HS critical phases. This is possible based on the fact that the curvature function defined at criticality using near-critical approach inherits the diverging property. Divergence occurs at the momentum $k_0^{mc}$ as one tunes the parameter $\mathbf{\Gamma}_{c}\rightarrow \mathbf{\Gamma}_{mc}$. The diverging peak flips sign as the parameters tuned across the multicritical points
\begin{equation}
\lim_{\mathbf{\Gamma}_c\rightarrow \mathbf{\Gamma}_{mc}^+}F(k_0^{mc},\mathbf{\Gamma}_c)= -\lim_{\mathbf{\Gamma}_c\rightarrow \mathbf{\Gamma}_{mc}^-}F(k_0^{mc},\mathbf{\Gamma}_c)=\pm \infty.
\end{equation}
The curvature function at criticality is also symmetric and acquires the Ornstein-Zernike form
\begin{equation}
F(k_0^{mc}+\delta k,\mathbf{\Gamma}_c)=\frac{F(k_0^{mc},\mathbf{\Gamma}_c)}{1+\xi_c^2\delta k^2}, \label{Lorenzian-crit}
\end{equation}
where $\xi_c$ is the characteristic length scale at criticality. Corresponding critical exponents can be obtained as
%The curvature function $F(k_0^{mc}, \mathbf{\Gamma}_{c})$ along with $\xi_c$ defines the critical exponents
\begin{equation}
F(k_0^{mc},\mathbf{\Gamma}_c) \propto |\mathbf{\Gamma}_c-\mathbf{\Gamma}_{mc}|^{-\gamma},\;\;\;\;\;\; \xi_c \propto |\mathbf{\Gamma}_c-\mathbf{\Gamma}_{mc}|^{-\nu}.
\label{crit-exp}
\end{equation}
In order to construct a scaling scheme similar to the gapped case, we consider the non-HS points of $k_0^{mc}$, 
%since the first order derivative of the curvature function at criticality is zero for HS points. Moreover, the non-HS points 
which effectively captures the scaling at multicritical points (a reverse technique is used in Ref ~\cite{malard2020scaling}, where scaling at HS points identify the non-HS criticalities).
The scaling now can be recasted as
\begin{equation}
F(k^{mc}_{02},\mathbf{\Gamma^{\prime}}_c)=F(k^{mc}_{01}+\delta k, \mathbf{\Gamma}_c).
\end{equation}
Considering the same approximation employed in the case of gapped phases (Section.\ref{CRG-non}), the generic RG equation at non-HS criticality can be obtained as
%\begin{equation}
%\frac{d\mathbf{\Gamma}_c}{dl} \approx \frac{\partial_k F(k,\mathbf{\Gamma}_c)|_{k=k^{mc}_{01}}}{\partial_{\mathbf{\Gamma}_c} F(k^{mc}_{02},\mathbf{\Gamma}_c)},
%\label{RG-crit}
%\end{equation}
%Following the same procedure as in the case of gapped phases, the generic RG equation at criticality is obtained to be
\begin{equation}
\frac{d\mathbf{\Gamma}_c}{dl} \approx \frac{\partial_k F(k,\mathbf{\Gamma}_c)|_{k=k^{mc}_{0}}}{\partial_{\mathbf{\Gamma}_c} F(k^{mc}_{0},\mathbf{\Gamma}_c)},
\label{RG-crit}
\end{equation}
where $d\mathbf{\Gamma}_c=\mathbf{\Gamma^{\prime}}_c-\mathbf{\Gamma}_c$ and $dl=\delta k$ (with $\delta k$ being small deviation away from $k^{mc}_0$). The RG flow lines identify the topological transition between non-HS critical phases.

\hspace{0.3cm}The correlation function in terms of Wannier state representation can also be written at non-HS criticalities to characterize the topological transition. At criticality one can write
\begin{equation}
\lambda^c_{R}=\frac{e^{i k_0^{mc} R}}{2\xi_c} F(k_0^{mc},\mathbf{\Gamma}_c) e^{R/\xi_c}, \label{corr-crit}
\end{equation}
where $\xi_c$ is the correlation length. The correlation function $\lambda^c_R$ decays as the parameters tune towards the multicritical point. The decay rate decreases near the point and gets sharper as we tune away from the point. This typical behavior of $\lambda^c_{R}$ confirms the topological transition at multicritical points between non-HS critical phases.

\subsection{Curvature function and critical exponents}
\begin{figure}[t]
	\includegraphics[width=4.2cm,height=2.8cm]{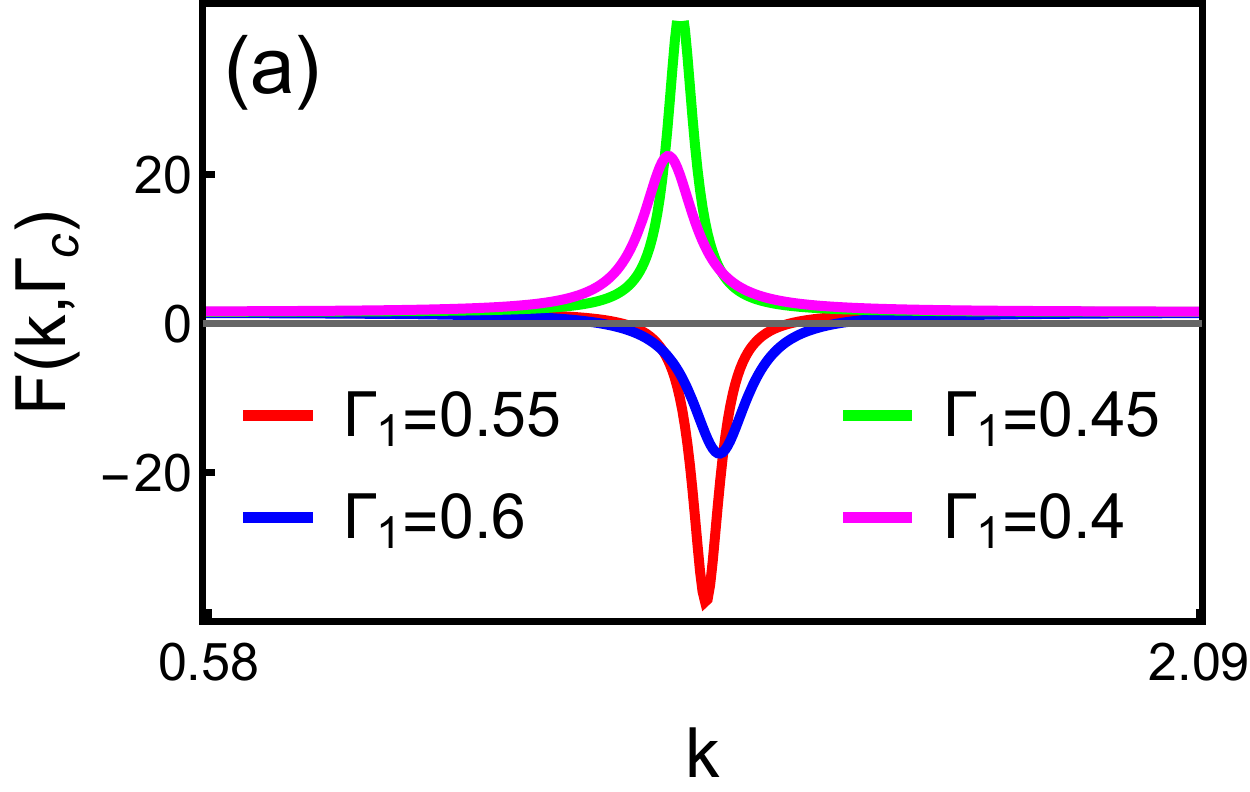}  
	\includegraphics[width=4.2cm,height=2.8cm]{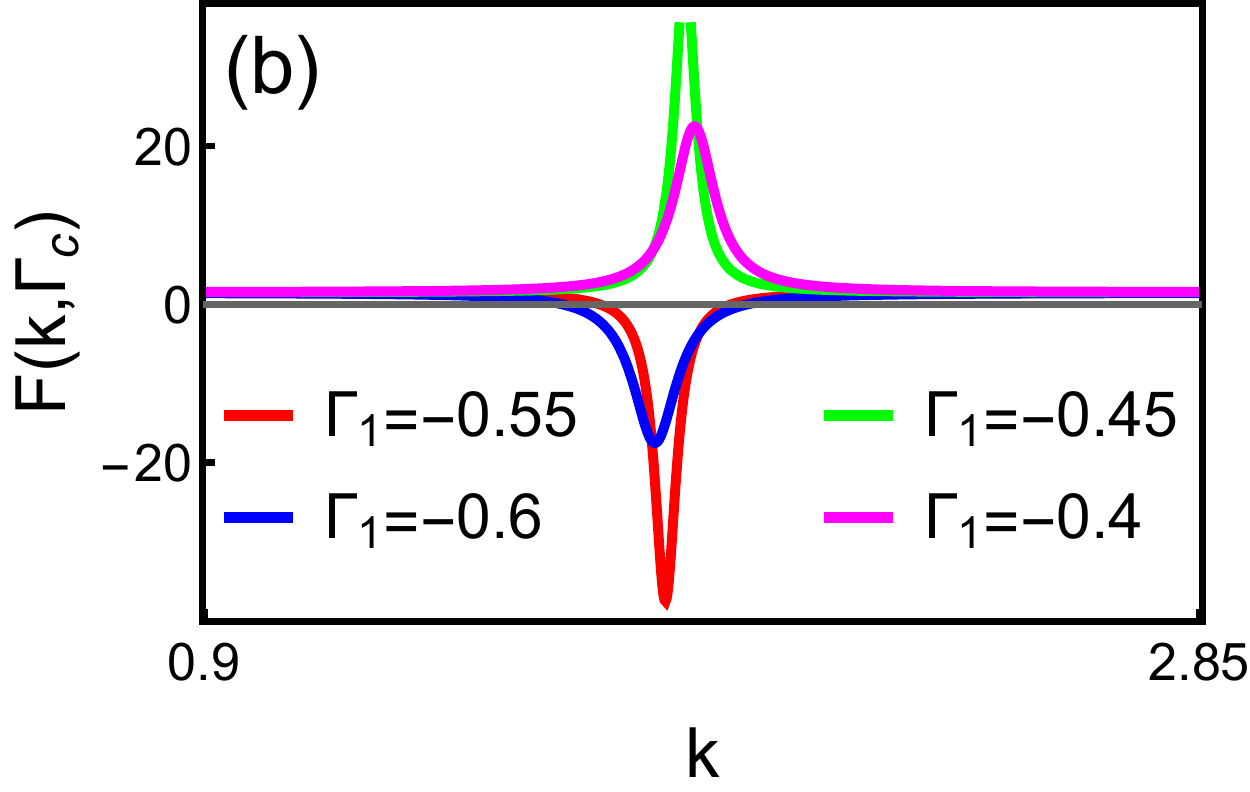}  
	\caption{\label{crit-curve} Curvature function at non-HS criticalities. Plotted for $\Gamma_{0}=1$ and $\Gamma_{2}=0.5$. (a) Curvature function at magenta line in the vicinity of the multicritical point $\Gamma_{1}=0.5$ ($M_1$) for 
		%(a) HS $k_0^{mc}$, (b) 
		one of the non-HS $k_0^{mc}$. %of Eq.\ref{mome-crit1}.  %The divergence occurs at $k_0^{mc}=\pi$. 
		(b) Curvature function at green line in the vicinity of the multicritical point $\Gamma_{1}=-0.5$ ($M_2$) for 
		%(c) HS $k_0^{mc}$, (d) 
		one of the non-HS $k_0^{mc}$. %of Eq.\ref{mome-crit2}.  %The divergence occurs at $k_0^{mc}=0$.
	}
\end{figure}

Curvature function at non-HS criticalities can be written using the components $ \chi_{x} (k) = \Gamma_0 + \Gamma_1 \cos k + \Gamma_2 \cos 2k + (\Gamma_{1}\pm \sqrt{\Gamma_{1}^2+4\Gamma_{0}(\Gamma_{0}-\Gamma_{2})})/2 \cos 3k,$ and $ \chi_{y} (k) = \Gamma_1 \sin k + \Gamma_2 \sin 2k + (\Gamma_{1}\pm \sqrt{\Gamma_{1}^2+4\Gamma_{0}(\Gamma_{0}-\Gamma_{2})})/2 \sin 3k $.
\begin{equation}
F(k,\mathbf{\Gamma}_{c})= \frac{A+B \cos(k)+C \cos(2k)}{2 D^2+E^2},
\end{equation}
where $A=6\Gamma_{0}^2+\Gamma_{1}(3\Gamma_{1}\pm\alpha)-8\Gamma_{0}\Gamma_{2}+4\Gamma_{2}^2$, $B=-2\Gamma_{0}\Gamma_{1}+\Gamma_{2}(\pm7\Gamma_{1}+\alpha)$, $C=6\Gamma_{0}(\Gamma_{2}-\Gamma_{0})$, $D=(\Gamma_{1}+\Gamma_{2} \cos(k))^2$ and $E=(\Gamma_{2}-2\Gamma_{0})^2 \sin(k)^2$ with $\alpha=\sqrt{\Gamma_{1}^2+4\Gamma_{0}(\Gamma_{0}-\Gamma_{2})}$.
The diverging peak of the curvature function can be observed at $k_0^{mc}$, as one tune the parameters $\mathbf{\Gamma}_{c}$ towards $\mathbf{\Gamma}_{mc}$.  The peaks at non-HS $k_0^{mc}$ points
%as a consequence of the swaping property discussed earlier, 
are shown in Fig.\ref{crit-curve}. 
%The peak at non-HS points can be observed by approaching the multicritical points from HS criticalities, as shown in Fig.\ref{crit-curve}(c,d). Therefore, even though the $k_0^{mc}$ has both HS and non-HS points, the divergence occurs at HS points as we are on the non-HS criticalities. 
The diverging peak flips the sign (similar to the case of gapped phases) as we tune across the multicritical points ($M_{1,2}$) signaling the topological transition between non-HS critical phases.
%as shown in Fig.\ref{crit-curve}. 
In Fig.\ref{crit-curve}(a), curvature function in the vicinity of the multicritical point $M_1$, i.e. $\Gamma_{1}=0.5$ 
%(with $\Gamma_{0}=1$ and $\Gamma_{2}=0.5$) 
on the magenta line is shown. In Fig.\ref{crit-curve}(b), multicritical point $M_2$, i.e. $\Gamma_{1}=-0.5$ on the green line is shown. Therefore, both the non-HS criticalities shows the similar behavior of curvature function at criticality. The multicritical points $M_{1,2}$
%, i.e. $\Gamma_{1}=\pm 0.5$ respectively at magenta and green lines 
are the topological phase transition points between non-HS critical phases.
\begin{figure}[t]
	\includegraphics[width=4.0cm,height=2.8cm]{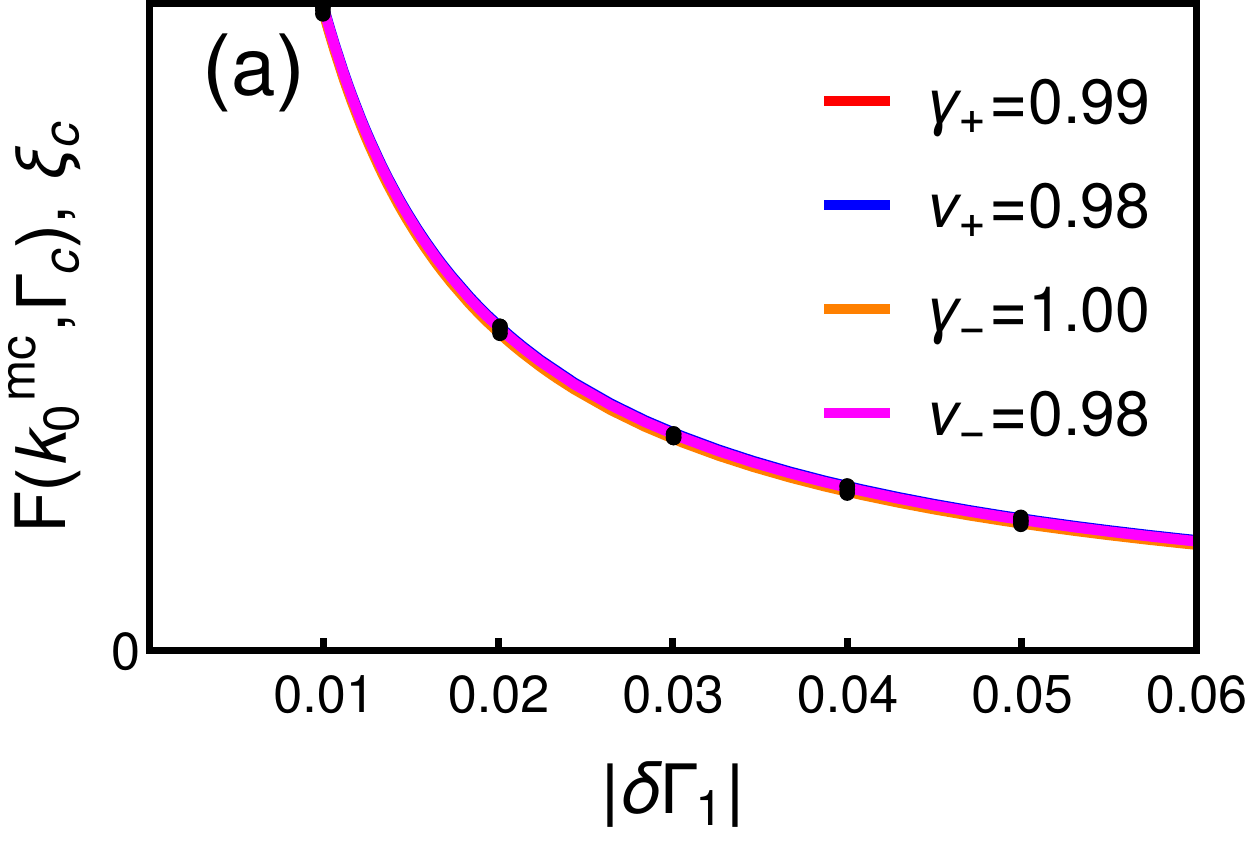}  
	\includegraphics[width=4.0cm,height=2.8cm]{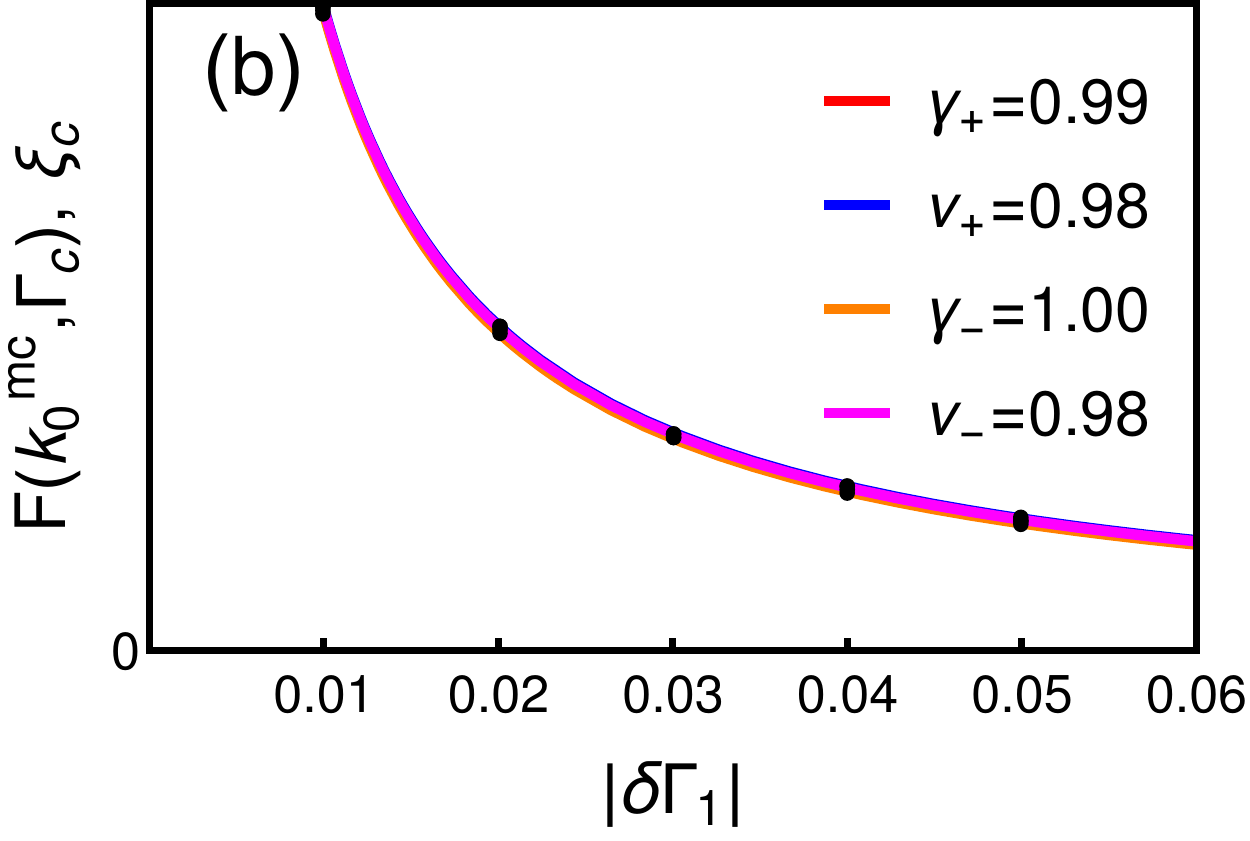}  
	\caption{\label{crit-expo} Critical exponents for multicritical points. Plotted for $\Gamma_{0}=1$ and $\Gamma_{2}=0.5$. (a) For $\Gamma_{1}=0.5$ ($M_1$). (b) For $\Gamma_{1}=-0.5$ ($M_2$). The exponents are found to be $\gamma\approx\nu\approx1$ for both the multicritical points.}
\end{figure}

\hspace{0.3cm}The critical exponents of the curvature function at criticality near the multicritical points can be obtained from Eq.\ref{crit-exp}. The Ornstein-Zernike form of the curvature function in the vicinity of the multicritical points allows one to extract the exponent values numerically using the fitting equation
\begin{equation}
F_{fitting}=c+\frac{F(k_0^{mc},\mathbf{\Gamma}_c)}{1+\xi_c^2(k-k_0^{mc})^2}
\end{equation} 
Fig.\ref{crit-expo} shows 
%the obtained values of critical exponents for both $M_{1,2}$
%the multicritical points at both the non-HS criticalities. 
%It can be observed 
that, one can extract the exponents values as $\gamma\approx\nu\approx1$ for both the multicritical points $M_{1,2}$.

\hspace{0.3cm}The exponents can also be evaluated analytically similar to the case of gapped phases. Expansion of the components $\chi_{x,y}$ around $k_0^{mc}$ yields
\begin{equation}
\chi_{x} \approx \delta\mathbf{\Gamma}_c+A \delta k^2 \;\;\; \quad and \quad\;\;\; \chi_{y} \approx B \delta k,
\end{equation}
where 
\begin{align}
\delta \mathbf{\Gamma}_c &= \Gamma_{0}\mp 3\Gamma_{1}/2-(1/2)\alpha+\Gamma_{2}\nonumber\\
A &= \pm\Gamma_{1}+9/2(\Gamma_{1}+\alpha)-4\Gamma_{2}\\
B &=\mp 5\Gamma_{1}/2-(3/2)\alpha+2\Gamma_{2} \nonumber
\end{align}
with $\alpha=\sqrt{\Gamma_{1}^2+4\Gamma_{0}(\Gamma_{0}-\Gamma_{2})}$. This allows one to write the curvature function at criticality in Ornstein-Zernike form as
\begin{align}
F(k,\delta\mathbf{\Gamma}_c) =\frac{F(k_0^{mc},\delta\mathbf{\Gamma}_c)}{1+\xi_c^2\delta k^2}
\end{align}
where $F(k_0^{mc},\delta\mathbf{\Gamma}_c)=B\delta\mathbf{\Gamma}_c^{-1} \implies \gamma=1$ and $\xi_c=B\delta\mathbf{\Gamma}_c^{-1}\implies \nu=1$. Therefore, both numerical and analytical values of the exponents are found to be same and they obey the scaling law $\gamma=\nu$.
\begin{figure}[t]
	\includegraphics[width=4.0cm,height=4cm]{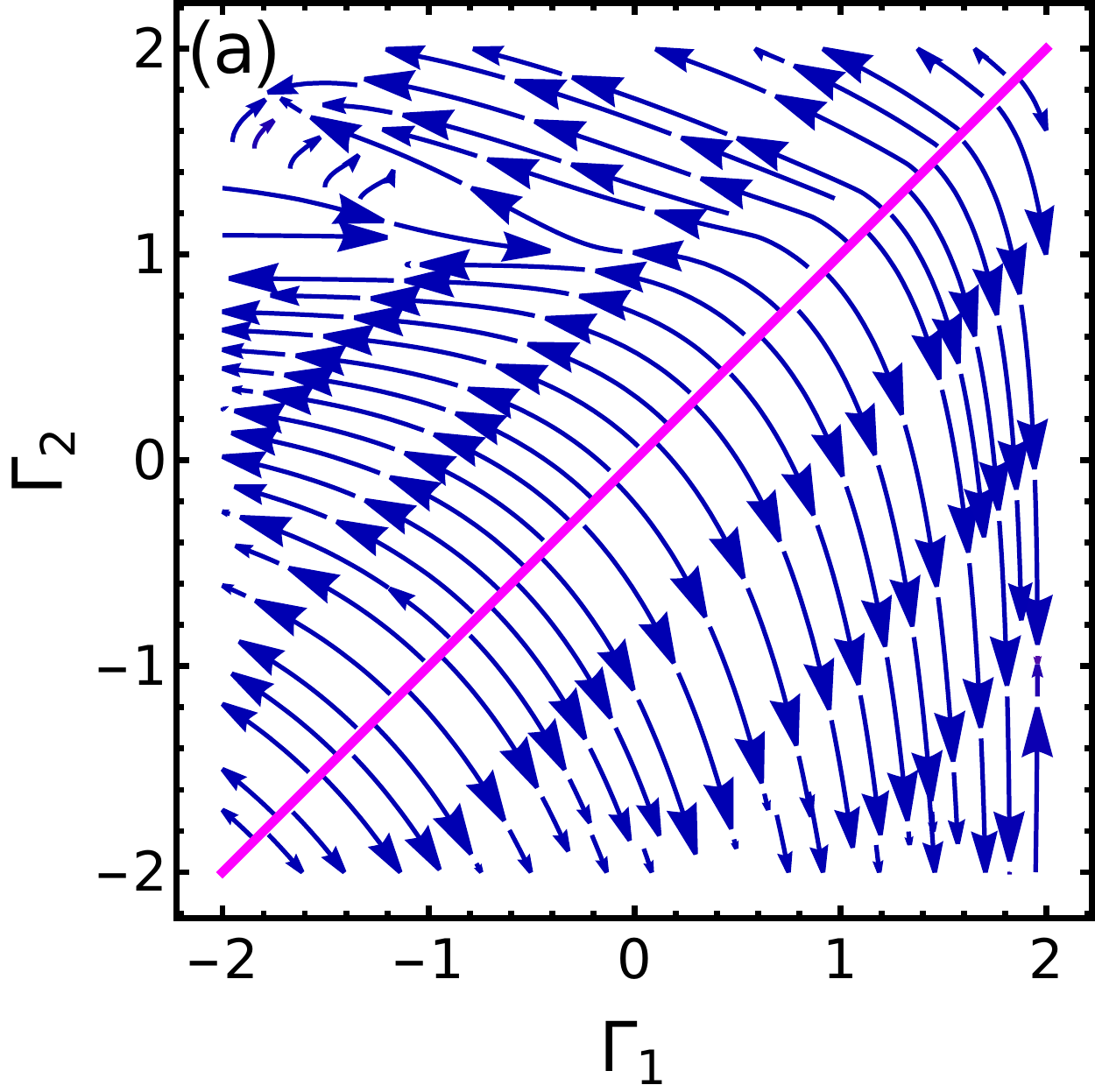}  
	\includegraphics[width=4.0cm,height=4cm]{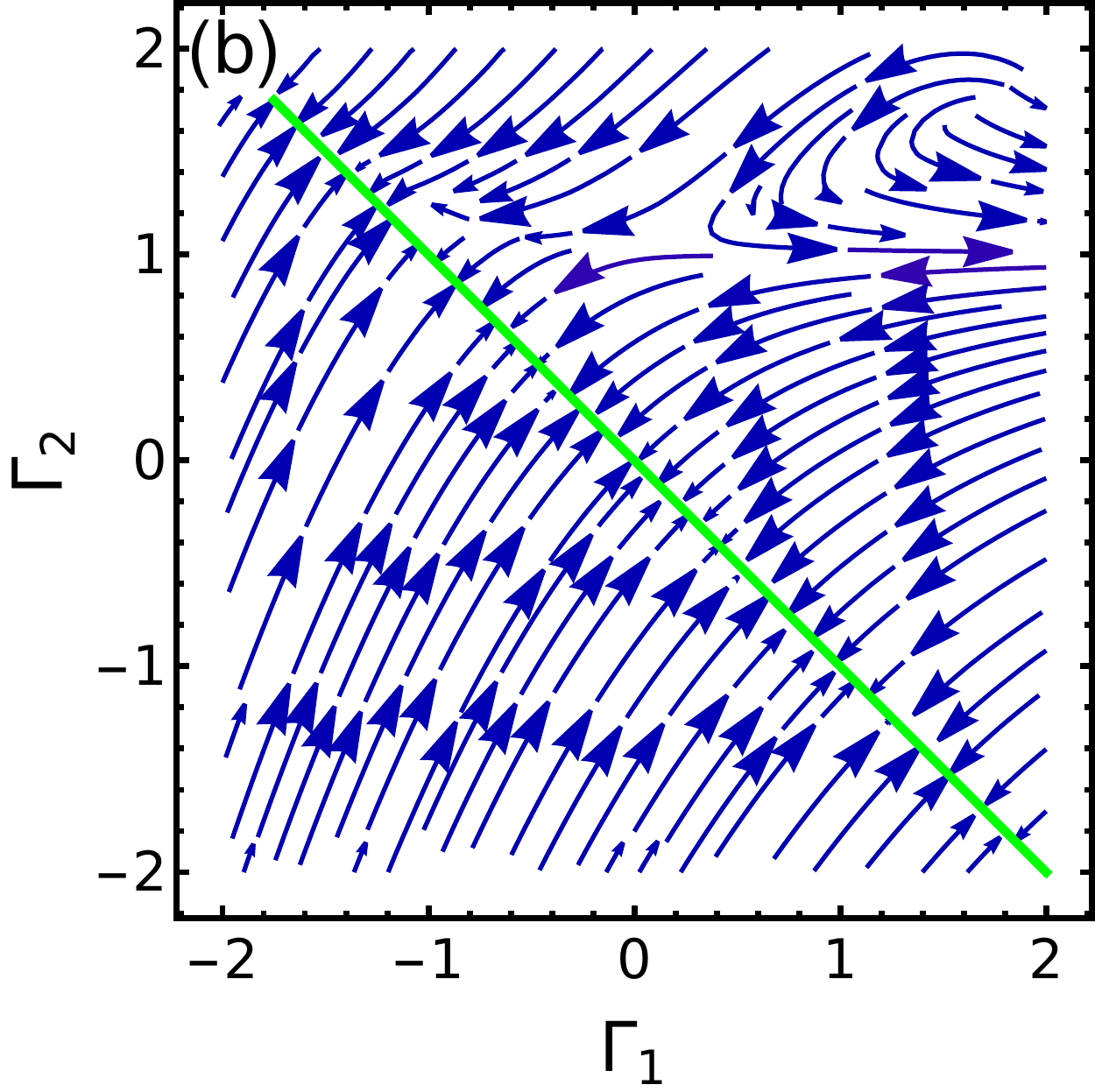}  
	\caption{\label{crit-CRG} CRG at non-HS criticalities. (a) On the magenta line which capture the transition at $M_1$ ($\Gamma_{1}=\Gamma_{2}$). The transition line is characterized with the outward RG flow. (b) On the green line to capture the transition at $M_2$ ($\Gamma_{1}=-\Gamma_{2}$). The transition line is characterized with the inward RG flow.}
\end{figure}
\begin{figure}[t]
	\includegraphics[width=4.0cm,height=2.8cm]{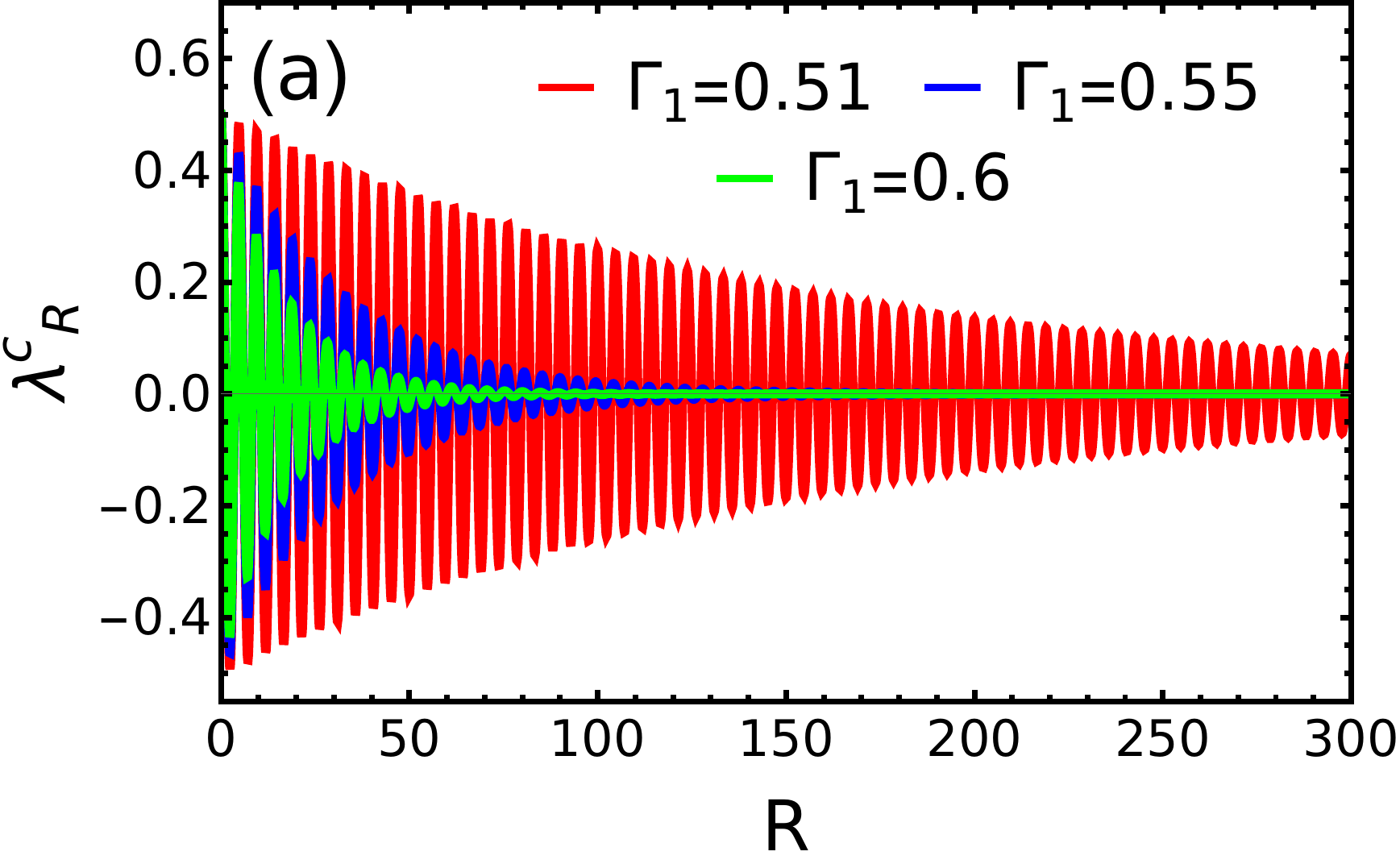}  
	\includegraphics[width=4.0cm,height=2.8cm]{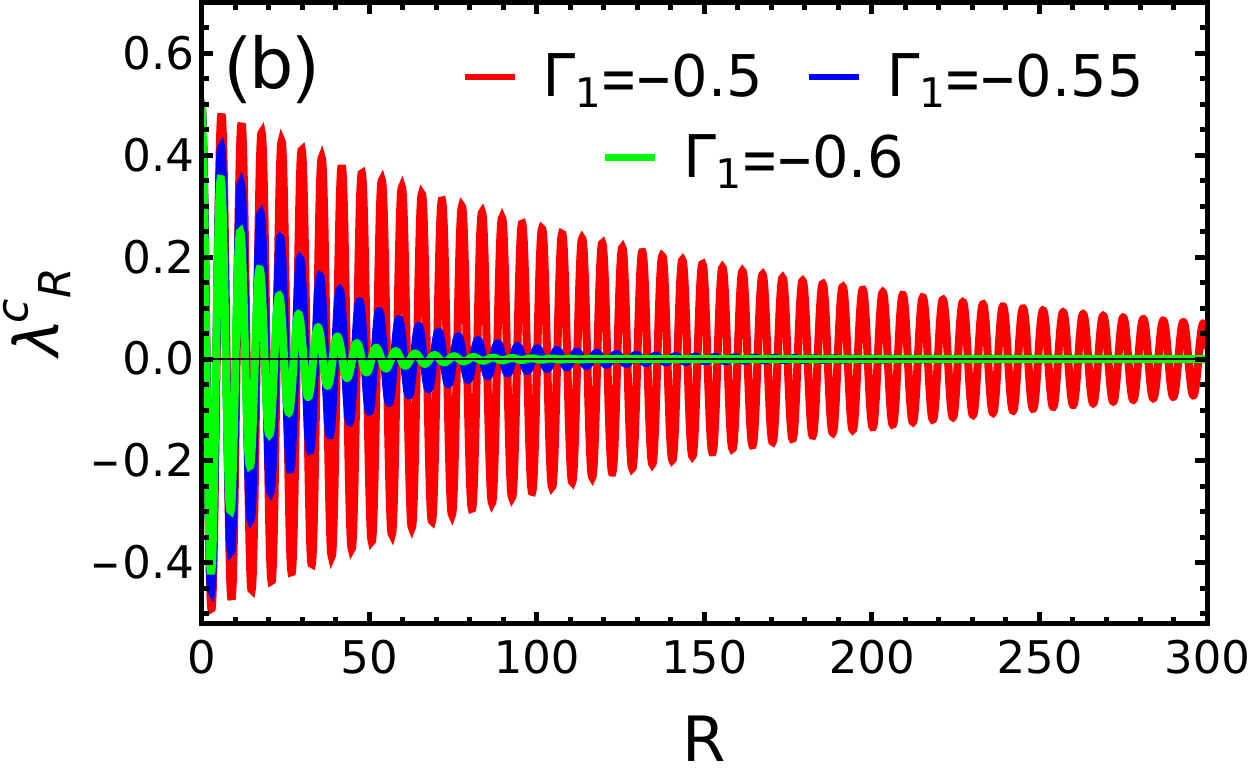}
	\caption{\label{corre-crit} Wannier state correlation function at criticality. The parameter values $\Gamma_{0}=1$ and $\Gamma_{2}=0.5$ are fixed. (a) Represents $\lambda_R^c$ in the vicinity of $M_1$ i.e. $\Gamma_{1}=0.5$. (b) Represents $\lambda_R^c$ in the vicinity of $M_2$ i.e. $\Gamma_{1}=-0.5$.}
\end{figure}

\subsection{Curvature function renormalization group and correlation function}
\hspace{0.3cm}We perform the scaling scheme at non-HS criticality and obtain RG equations of the model which essentially captures the topological transition between non-HS critical phases. From Eq.\ref{RG-crit}, the RG equations for the magenta line can be obtained for the parameters $\Gamma_{1}$ and $\Gamma_{2}$ with $\Gamma_{0}=1$ as
\begin{equation}
\frac{d\Gamma_{1}}{dl}=\frac{(\Gamma_{1}+\alpha)^4\alpha\Lambda_5(\Gamma_{0},\Gamma_{1},\Gamma_{2})}{\Lambda_6(\Gamma_{0},\Gamma_{1},\Gamma_{2})}
\end{equation}
\begin{equation}
\frac{d\Gamma_{2}}{dl}=-\frac{(\Gamma_{1}+\alpha)^4\alpha\alpha_1^{\prime}\Lambda_5(\Gamma_{0},\Gamma_{1},\Gamma_{2})}{\Lambda_6^{\prime}(\Gamma_{0},\Gamma_{1},\Gamma_{2})}
\end{equation}
Similarly, for the green line we get
\begin{equation}
\frac{d\Gamma_{1}}{dl}=-\frac{(\Gamma_{1}-\alpha)^4\alpha\Lambda_7(\Gamma_{0},\Gamma_{1},\Gamma_{2})}{\Lambda_8(\Gamma_{0},\Gamma_{1},\Gamma_{2})}
\end{equation}
\begin{equation}
\frac{d\Gamma_{2}}{dl}=-\frac{(\Gamma_{1}-\alpha)^4\alpha\alpha_1^{\prime}\Lambda_7(\Gamma_{0},\Gamma_{1},\Gamma_{2})}{\Lambda_8^{\prime}(\Gamma_{0},\Gamma_{1},\Gamma_{2})}
\end{equation}
where $\alpha=\sqrt{\Gamma_{1}^2+4\Gamma_{0}(\Gamma_{0}-\Gamma_{2})}$ and $\alpha_1^{\prime}=(\Gamma_{2}-2\Gamma_{0})$ (see supplementary material  for the detailed form of $\Lambda$s). The multicritical points $\Gamma_{1}=\pm\Gamma_{2}$ ($M_{1,2}$) are identified using the RG flow directions in $\Gamma_{1}-\Gamma_{2}$ plane. As shown in Fig.\ref{crit-CRG}, $M_1$ ($\Gamma_{1}=\Gamma_{2}$) manifest as a critical line with the flow lines flowing away. Similarly, $M_2$ ($\Gamma_{1}=-\Gamma_{2}$) manifest as a fixed line with flow lines flowing into, as explained in Eq.\ref{crit-fixed}. This clearly demonstrates that the multicritical points are indeed the topological phase transition points between non-HS critical phases at both the non-HS criticalities.

\hspace{0.3cm}Apart from CRG, the correlation function defined at criticality in Eq.\ref{corr-crit} can be obtained to identify the topological transition between non-HS critical phases. The profile of the $\lambda_R^c$, for the non-HS $k_0^{mc}$, in the vicinity of multicritical points 
%at both the non-HS criticalities 
are shown in Fig.\ref{corre-crit}. The correlation function decay slowly near $M_{1,2}$, i.e. $\Gamma_{1}=\pm 0.5$. The decay gets sharper as the parameters are tuned away the multicritical points. %The difference between the two multicriticalities is the oscillatory nature of $\lambda_R^c$ emerging from the term $e^{i k_0^{mc} R}$ in Eq.\ref{corr-crit}. For $M_1$, due to $k_0^{mc}=\pi$ we observe oscillatory behavior as shown in Fig.\ref{corre-crit}(a) and for $M_2$, we have $k_0^{mc}=0$ which results in the plane/non-oscillatory exponential decay as shown in Fig.\ref{corre-crit}(b). 
Therefore, this clearly shows that the multicritical points are the transition points between the non-HS critical phases.

\section{Conclusions} \label{VI}
\hspace{0.3cm}In summary, we have identified a unique topological phase transition between non-HS critical phases through multicritical points. A generic model of topological insulators and superconductors have been constructed at criticality using the near-critical approach~\cite{kumar2021topological}, which provides an effective platform to study the edge mode solutions and topological transitions at non-HS criticalities. 

\hspace{0.3cm} The decay length of edge modes and winding number associated to the non-HS critical phases enables the characterization of trivial and topological non-HS critical phases both qualitatively and quantitatively. 
The decay length remains positive for non-trivial critical phases and negative for  trivial critical phases. The multicritical point is associated with the divergence of the decay length, which indicate the delocalization of edge modes into the bulk. On the other hand, the winding number, which count the number of edge modes, acquires non-zero integer values at non-HS criticalities. This gives $w_c=2$ to a topological non-HS critical phase where only one edge mode is localized. Therefore, we have suggested to consider the difference in the winding numbers between trivial and non-trivial critical phases which yields the correct count of the edge modes localized at the non-trivial critical phase. 
The numerical solutions in the open boundary condition are found to be in agreement with these results.

\hspace{0.3cm}We have also generalized the scaling theory to capture the topological transition at non-HS critical points. The scaling theory based on the divergence of curvature function, characterize both the conventional and the unique topological transition in terms of RG flow, critical exponents and correlation functions. Investigating the conventional topological transition between gapped phases, we have found that the CRG method is efficient to capture the non-HS critical points. Reframing the CRG method to work at criticality, we have identified the topological transition between non-HS critical phases through multicritical points. The critical and fixed line behaviors of the CRG equations in the parameter space is identified with RG flow rate and directions. In addition, the exponential decay of the correlation function near the 
%critical and 
multicritical points clearly evidence the unique topological transition between non-HS critical phases.
Moreover, the divergence in curvature function along with the flipping of its sign across the transition points, locate the non-HS critical and multicritical points. %The dispersion at the multicritical points (which favor the unique transition between non-HS critical phases) was found to be linear in $k$ with the dynamical critical exponent $z=1$. Moreover, 
The critical exponents of curvature function, calculated both analytically and numerically yields $\gamma=\nu=1$, which establish the universality class of non-HS critical points and multicritical points. 

\hspace{0.3cm} The model discussed in this work can be efficiently simulated using the superconducting circuit of a single
qubit driven by the microwave pulses~\cite{PhysRevB.101.035109,niu2021simulation} and the ultracold atoms in optical lattices~\cite{goldman2016topological,xie2019topological,an2018engineering,meier2016observation,kraus2012preparing,jiang2011majorana,an2018engineering}. Therefore, using the good control over the nearest neighbors provided by these platforms one can study the results discussed in this work. 
As the non-HS criticality becomes prominent with increasing nearest-neighbor couplings~\cite{hsu2020topological,niu2012majorana,kartik2021topological}, an interesting question is whether the unique topological transition survive in truly long-range models. Moreover, the study of this interesting phenomena in non-Hermitian systems~\cite{rahul2022topological}, spin systems~\cite{niu2012majorana} and driven systems~\cite{PhysRevLett.121.076802,molignini2018universal} sets the future direction of the work. In addition, the fate of the edge modes and topological transition at non-HS criticality in the presence of interactions is an intriguing open problem. Therefore, we hope that our work will provide a step forward towards the understanding of the interesting interplay between topology and criticality. 
Moreover, the model considered in this study is not a true long-range model with decaying coupling strengths~\cite{PhysRevLett.113.156402,PhysRevB.95.195160}. Nevertheless, the results discussed in this work will remain effective even with power-law decaying nearest-neighbor coupling strengths~\cite{kartik2021topological}. However, a detailed study, specifically, the topological transition between non-HS critical phases in truly long-range models remains a future scope of our work.
\section{ACKNOWLEDGMENTS}
RRK and SS would like to acknowledge DST (Department of Science and Technology, Government of India-CRG/2021/00996) for the the funding and support. YRK would like to thank AMEF (Admar Mutt Education Foundation) for the funding and support. Authors would like to thank Nilanjan Roy and Rahul S for the useful discussions.

%\bibliography{edgemode}

\begin{thebibliography}{10}
	
	\bibitem{haldane1988model}
	F~Duncan~M Haldane.
%	\newblock Model for a quantum hall effect without landau levels: Condensed-matter realization of the" parity anomaly".
	\href{https://journals.aps.org/prl/abstract/10.1103/PhysRevLett.61.2015}{\newblock {\em Phys. Rev. Lett.} \textbf{61}, 2015 (1988).}
	
	\bibitem{hasan2010colloquium}
	M~Zahid Hasan and Charles~L Kane.
%	\newblock Colloquium: topological insulators.
	\href{https://journals.aps.org/rmp/abstract/10.1103/RevModPhys.82.3045}{\newblock {\em Rev. Mod. Phys.} \textbf{82}, 3045 (2010).}
	
	\bibitem{wang2017topological}
	Jing Wang and Shou-Cheng Zhang.
%	\newblock Topological states of condensed matter.
	\href{https://www.nature.com/articles/nmat5012}{\newblock {\em Nat. Mater.} \textbf{16}, 1062--1067 (2017).}
	
	\bibitem{goldman2016topological}
	Nathan Goldman, Jan~C Budich, and Peter Zoller.
%	\newblock Topological quantum matter with ultracold gases in optical lattices.
	\href{https://www.nature.com/articles/nphys3803}{\newblock {\em Nat. Phys.} \textbf{12}, 639--645 (2016).}
	
	\bibitem{narang2021topology}
	Prineha Narang, Christina~AC Garcia, and Claudia Felser.
%	\newblock The topology of electronic band structures.
	\href{https://www.nature.com/articles/s41563-020-00820-4}{\newblock {\em Nat. Mater.} \textbf{20}, 293--300 (2021).}
	
	\bibitem{kitaev2001unpaired}
	A~Yu Kitaev.
%	\newblock Unpaired majorana fermions in quantum wires.
	\href{https://iopscience.iop.org/article/10.1070/1063-7869/44/10S/S29}{\newblock {\em Phys. Usp.} \textbf{44}, 131 (2001).}
	
	\bibitem{kane2005quantum}
	Charles~L Kane and Eugene~J Mele.
%	\newblock Quantum spin hall effect in graphene.
	\href{https://journals.aps.org/prl/abstract/10.1103/PhysRevLett.95.226801}{\newblock {\em Phys. Rev. Lett.} \textbf{95}, 226801 (2005).}
	
	\bibitem{thouless1982quantized}
	D. J. Thouless, M. Kohmoto, M. P. Nightingale, and M. den Nijs.
%	\newblock Quantized Hall conductance in a two-dimensional periodic potential.
	\href{https://journals.aps.org/prl/abstract/10.1103/PhysRevLett.49.405}{\newblock {\em Phys. Rev. Lett.} \textbf{49}, 405 (1982).}
	
	\bibitem{berry1984quantal}
	Michael V Berry.
%	\newblock Quantal phase factors accompanying adiabatic changes.
	\href{https://royalsocietypublishing.org/doi/10.1098/rspa.1984.0023}{\newblock {\em Proc. R. Soc. Lond. A. Mathematical and Physical Sciences}, \textbf{392}, 45--57 (1984).}
	
	\bibitem{zak1989berry}
	J.~Zak.
%	\newblock Berry's phase for energy bands in solids.
	\href{https://journals.aps.org/prl/abstract/10.1103/PhysRevLett.62.2747}{\newblock {\em Phys. Rev. Lett.}, \textbf{62}, 2747--2750 (1989).}
	
	
	\bibitem{altland1997symmetry}
	Alexander Altland and Martin~R. Zirnbauer.
%	\newblock Nonstandard symmetry classes in mesoscopic normal-superconducting hybrid structures.
	\href{https://journals.aps.org/prb/abstract/10.1103/PhysRevB.55.1142}{\newblock {\em Phys. Rev. B}, \textbf{55}, 1142--1161 (1997).}
	
	
	\bibitem{sarkar2018quantization}
	Sujit Sarkar.
%	\newblock Quantization of geometric phase with integer and fractional topological characterization in a quantum ising chain with long-range interaction.
	\href{https://www.nature.com/articles/s41598-018-24136-1}{\newblock {\em Sci. Rep.}, \textbf{8}, 1--20 (2018).}
	
	\bibitem{rahul2019interplay}
	Rahul S, Ranjith Kumar R, Y R Kartik, Amitava Banerjee, Sujit Sarkar.
%	\newblock An interplay of topology and quantized geometric phase for two different symmetry-class hamiltonians.
	\href{https://iopscience.iop.org/article/10.1088/1402-4896/ab1d7b}{\newblock {\em Phys. Scr.}, \textbf{94}, 115803 (2019).}
	
	
	\bibitem{kartik2021topological}
	Y R~Kartik, Ranjith~R Kumar, S~Rahul, Nilanjan Roy, and Sujit Sarkar.
%	\newblock Topological quantum phase transitions and criticality in a longer-range kitaev chain.
	\href{https://journals.aps.org/prb/abstract/10.1103/PhysRevB.104.075113}{\newblock {\em Phys. Rev. B.} \textbf{104}, 075113 (2021).}
	
	\bibitem{chen2016scaling}
	Wei Chen.
%	\newblock Scaling theory of topological phase transitions.
	\href{https://iopscience.iop.org/article/10.1088/0953-8984/28/5/055601}{\newblock {\em J. Condens. Matter Phys.} \textbf{28}, 055601 (2016).}
	
	\bibitem{chen2016scalinginvariant}
	Wei Chen, Manfred Sigrist, and Andreas~P Schnyder.
%	\newblock Scaling theory of topological invariants.
	\href{https://iopscience.iop.org/article/10.1088/0953-8984/28/36/365501/meta}{\newblock {\em J. Condens. Matter Phys.} \textbf{28}, 365501 (2016).}
	
	\bibitem{chen2017correlation}
	Wei Chen, Markus Legner, Andreas R\"uegg, and Manfred Sigrist.
%	\newblock Correlation length, universality classes, and scaling laws associated with topological phase transitions.
	\href{https://journals.aps.org/prb/abstract/10.1103/PhysRevB.95.075116}{\newblock {\em Phys. Rev. B.} \textbf{95}, 075116 (2017).}
	
	\bibitem{chen2018weakly}
	Wei Chen.
%	\newblock Weakly interacting topological insulators: Quantum criticality and the renormalization group approach.
	\href{https://journals.aps.org/prb/abstract/10.1103/PhysRevB.97.115130}{\newblock {\em Phys. Rev. B.} \textbf{97}, 115130 (2018).}
	
	\bibitem{chen2019universality}
	Wei Chen and Andreas~P Schnyder.
%	\newblock Universality classes of topological phase transitions with higher-order band crossing.
	\href{https://iopscience.iop.org/article/10.1088/1367-2630/ab2a2d}{\newblock {\em New J. Phys.} \textbf{21}, 073003 (2019).}
	
	\bibitem{molignini2018universal}
	Paolo Molignini, Wei Chen, and Ramasubramanian Chitra.
%	\newblock Universal quantum criticality in static and floquet-majorana chains.
	\href{https://journals.aps.org/prb/abstract/10.1103/PhysRevB.98.125129}{\newblock {\em Phys. Rev. B.} \textbf{98}, 125129 (2018).}
	
	\bibitem{panahiyan2020fidelity}
	S~Panahiyan, W~Chen, and S~Fritzsche.
%	\newblock Fidelity susceptibility near topological phase transitions in quantum walks.
	\href{https://journals.aps.org/prb/abstract/10.1103/PhysRevB.102.134111}{\newblock {\em Phys. Rev. B}, \textbf{102}, 134111 (2020).}
	
	
	\bibitem{molignini2020generating}
	Paolo Molignini, Wei Chen, and R~Chitra.
%	\newblock Generating quantum multicriticality in topological insulators by periodic driving.
	\href{https://journals.aps.org/prb/abstract/10.1103/PhysRevB.101.165106}{\newblock {\em Phys. Rev. B.} \textbf{101}, 165106 (2020).}
	
	\bibitem{abdulla2020curvature}
	Faruk Abdulla, Priyanka Mohan, and Sumathi Rao.
%	\newblock Curvature function renormalization, topological phase transitions, and multicriticality.
	\href{https://journals.aps.org/prb/abstract/10.1103/PhysRevB.102.235129}{\newblock {\em Phys. Rev. B.} \textbf{102}, 235129 (2020).}
	
	\bibitem{malard2020scaling}
	M~Malard, H~Johannesson, and W~Chen.
%	\newblock Scaling behavior in a multicritical one-dimensional topological insulator.
	\href{https://journals.aps.org/prb/abstract/10.1103/PhysRevB.102.205420}{\newblock {\em Phys. Rev. B.} \textbf{102}, 205420 (2020).}
	
	\bibitem{molignini2020unifying}
	Paolo Molignini, R~Chitra, and Wei Chen.
%	\newblock Unifying topological phase transitions in non-interacting, interacting, and periodically driven systems.
	\href{https://iopscience.iop.org/article/10.1209/0295-5075/128/36001}{\newblock {\em Europhys. Lett.} \textbf{128}, 36001 (2020).}
	
	\bibitem{kumar2021multi}
	Ranjith~R Kumar, Y R~Kartik, S~Rahul, and Sujit Sarkar.
%	\newblock Multi-critical topological transition at quantum criticality.
	\href{https://www.nature.com/articles/s41598-020-80337-7}{\newblock {\em Sci. Rep.} \textbf{11}, 1--20 (2021).}
	
	\bibitem{continentino2020finite}
	Mucio~A Continentino, Sabrina Rufo, and Griffith~M Rufo.
	\newblock Finite size effects in topological quantum phase transitions.
	\href{https://link.springer.com/chapter/10.1007/978-3-030-35473-2_12}{\newblock {\em Strongly Coupled Field Theories for Condensed Matter and Quantum Information Theory}}, Springer Proceedings in Physics 239 (2020).
	
	\bibitem{verresen2018topology}
	Ruben Verresen, Nick~G Jones, and Frank Pollmann.
%	\newblock Topology and edge modes in quantum critical chains.
	\href{https://journals.aps.org/prl/abstract/10.1103/PhysRevLett.120.057001}{\newblock {\em Phys. Rev. Lett.} \textbf{120}, 057001 (2018).}
	
	\bibitem{verresen2019gapless}
	Ruben Verresen, Ryan Thorngren, Nick~G Jones, and Frank Pollmann.
%	\newblock Gapless topological phases and symmetry-enriched quantum criticality.
	\href{https://journals.aps.org/prx/abstract/10.1103/PhysRevX.11.041059}{\newblock {\em Phys. Rev. X.} \textbf{11}, 041059 (2021).} 
	
	\bibitem{jones2019asymptotic}
	Nick~G Jones and Ruben Verresen.
%	\newblock Asymptotic correlations in gapped and critical topological phases of 1d quantum systems.
	\href{https://link.springer.com/article/10.1007/s10955-019-02257-9}{\newblock {\em J. Stat. Phys.} \textbf{175}, 1164--1213 (2019).}

\bibitem{verresen2020topology}
	Ruben Verresen.
%	\newblock Topology and edge states survive quantum criticality between topological insulators.
	\href{https://arxiv.org/abs/2003.05453}{\newblock {\em  arXiv:2003.05453v1 [cond-mat.str-el]}} (2020).

	
	\bibitem{rahul2021majorana}
	S~Rahul, Ranjith~R Kumar, Y R~Kartik, and Sujit Sarkar.
%	\newblock Majorana zero modes and bulk-boundary correspondence at quantum criticality.
	\href{https://journals.jps.jp/doi/abs/10.7566/JPSJ.90.094706?journalCode=jpsj}{\newblock {\em J. Phys. Soc. Jpn.} \textbf{90}, 094706 (2021).}

	\bibitem{niu2021emergent}
	Sen Niu, Yucheng Wang, and Xiong-Jun Liu.
%	\newblock Emergent gapless topological luttinger liquid.
	\href{https://arxiv.org/abs/2106.13400}{\newblock {\em arXiv:2106.13400v2 [cond-mat.str-el]}} (2021).


	\bibitem{PhysRevB.104.075132}
	Ryan Thorngren, Ashvin Vishwanath, and Ruben Verresen.
%	\newblock Intrinsically gapless topological phases.
	\href{https://journals.aps.org/prb/abstract/10.1103/PhysRevB.104.075132}{\newblock {\em Phys. Rev. B.} \textbf{104}, 075132 (2021).}

	\bibitem{PhysRevResearch.3.043048}
	Oleksandr Balabanov, Daniel Erkensten, and Henrik Johannesson.
%	\newblock Topology of critical chiral phases: Multiband insulators and superconductors.
	\href{https://journals.aps.org/prresearch/abstract/10.1103/PhysRevResearch.3.043048}{\newblock {\em Phys. Rev. Res.} \textbf{3}, 043048 (2021).}

	
	\bibitem{fraxanet2021topological}
	Joana Fraxanet, Daniel Gonz{\'a}lez-Cuadra, Tilman Pfau, Maciej Lewenstein, Tim
	Langen, and Luca Barbiero.
%	\newblock Topological quantum critical points in the extended bose-hubbard model.
	\href{https://journals.aps.org/prl/abstract/10.1103/PhysRevLett.128.043402}{\newblock {\em Phys. Rev. Lett.} \textbf{128}, 043402 (2022).}

	\bibitem{keselman2015gapless}
	Anna Keselman, Erez Berg.
%	Gapless symmetry-protected topological phase of fermions in one dimension.
	\href{https://journals.aps.org/prb/abstract/10.1103/PhysRevB.91.235309}{\newblock {\em Phys. Rev. B.} \textbf{91}, 235309 (2015).}

	\bibitem{scaffidi2017gapless}
	Thomas Scaffidi, Daniel E. Parker, Romain Vasseur.
%	Gapless symmetry-protected topological order.
	\href{https://journals.aps.org/prx/abstract/10.1103/PhysRevX.7.041048}{\newblock {\em Phys. Rev. X.} \textbf{7}, 041048 (2017).}
	
	\bibitem{duque2021topological}
	Carlos M. Duque, Hong-Ye Hu, Yi-Zhuang You, Vedika Khemani, Ruben Verresen, and Romain Vasseur.
%	Topological and symmetry-enriched random quantum critical points.
	\href{https://journals.aps.org/prb/abstract/10.1103/PhysRevB.103.L100207}{\newblock {\em Phys. Rev. B.} \textbf{103}, L100207 (2021).}
	
	\bibitem{kumar2021topological}
	Ranjith~R Kumar, Nilanjan Roy, Y R~Kartik, S~Rahul, and Sujit Sarkar.
%	\newblock Topological phase transition at quantum criticality.
	\href{https://arxiv.org/pdf/2112.02485.pdf}{\newblock {\em arXiv:2112.02485v2 [cond-mat.str-el]}, (2021).}
	
	
	\bibitem{rufo2019multicritical}
	Rufo, S., Lopes, N., Continentino, M.A. and Griffith, M.A.R.
%	\newblock Multicritical behavior in topological phase transitions.
	\href{https://journals.aps.org/prb/abstract/10.1103/PhysRevB.100.195432}{\newblock {\em Phys. Rev. B.} \textbf{100}, 195432 (2019).}
	
	\bibitem{malard2020multicriticality}
	Mariana Malard, David Brandao, Paulo~Eduardo de~Brito, and Henrik Johannesson.
%	\newblock Multicriticality in a one-dimensional topological band insulator.
	\href{https://journals.aps.org/prresearch/abstract/10.1103/PhysRevResearch.2.033246}{\newblock {\em Phys. Rev. Res.}, \textbf{2}, 033246 (2020).}
	
	\bibitem{sim2022quench}
	Karin Sim, R~Chitra, and Paolo Molignini.
%	\newblock Quench dynamics and scaling laws in topological nodal loopsemimetals.
	\href{https://doi.org/10.48550/arXiv.2207.10676}{\newblock {\em arXiv:2207.10676v2 [cond-mat.stat-mech]}, (2022).}
	
	
	\bibitem{hsu2020topological}
	Hsiu-Chuan Hsu and Tsung-Wei Chen.
%	\newblock Topological anderson insulating phases in the long-range su-schrieffer-heeger model.
	\href{https://journals.aps.org/prb/abstract/10.1103/PhysRevB.102.205425}{\newblock {\em Phys. Rev. B.} \textbf{102}, 205425 (2020).}
	
	\bibitem{niu2012majorana}
	Yuezhen Niu, Suk~Bum Chung, Chen-Hsuan Hsu, Ipsita Mandal, S~Raghu, and Sudip
	Chakravarty.
%	\newblock Majorana zero modes in a quantum ising chain with longer-ranged interactions.
	\href{https://journals.aps.org/prb/abstract/10.1103/PhysRevB.85.035110}{\newblock {\em Phys. Rev. B.} \textbf{85}, 035110 (2012).}
	
	\bibitem{PhysRevLett.42.1698}
	W.~P. Su, J.~R. Schrieffer, and A.~J. Heeger.
%	\newblock Solitons in polyacetylene.
	\href{https://journals.aps.org/prl/abstract/10.1103/PhysRevLett.42.1698}{\newblock {\em Phys. Rev. Lett.} \textbf{42}, 1698--1701 (1979).}
	
	\bibitem{murakami2011gap}
	Shuichi Murakami.
	%	\newblock Gap closing and universal phase diagrams in topological insulators.
	\href{https://www.sciencedirect.com/science/article/abs/pii/S1386947710004339}{\newblock {\em Physica E: Low-dimensional Systems and Nanostructures.} \textbf{43}, 748--754 (2011).}
	
	\bibitem{kourtis2017weyl}
	Stefanos Kourtis, Titus Neupert, Christopher Mudry, Manfred Sigrist, and Wei Chen.
	%	\newblock Weyl-type topological phase transitions in fractional quantum Hall like systems.
	\href{https://journals.aps.org/prb/abstract/10.1103/PhysRevB.96.205117}{\newblock {\em Phys. Rev. B.} \textbf{96}, 205117 (2017).}
	
	\bibitem{jalal2016topological}
	Somenath Jalal, Rishabh Khare, and Siddhartha Lal.
	\href{https://www.researchgate.net/publication/309572927_Topological_transitions_in_Ising_models}{\newblock Topological transitions in ising models (2016).} 

	\bibitem{shen2011topological}
	Shun-Qing Shen, Wen-Yu Shan, and Hai-Zhou Lu.
	\newblock Topological insulator and the dirac equation.
	\href{https://www.worldscientific.com/doi/abs/10.1142/S2010324711000057}{\newblock In {\em Spin}, \textbf{1}, 33--44, World Scientific (2011).}
	
	\bibitem{lu2011non}
	Jie Lu, Wen-Yu Shan, Hai-Zhou Lu, and Shun-Qing Shen.
%	\newblock Non-magnetic impurities and in-gap bound states in topological insulators.
	\href{https://iopscience.iop.org/article/10.1088/1367-2630/13/10/103016}{\newblock {\em New J. Phys.}, \textbf{13}, 103016 (2011).}

	
	\bibitem{jackiw1976solitons}
	Roman Jackiw and Cl{\'a}udio Rebbi.
%	\newblock Solitons with fermion number $1/2$.
	\href{https://journals.aps.org/prd/abstract/10.1103/PhysRevD.13.3398}{\newblock {\em Phys. Rev. D}, \textbf{13}, 3398 (1976).}

	
	\bibitem{PhysRevB.101.035109}
	Ziyu Tao, Tongxing Yan, Weiyang Liu, Jingjing Niu, Yuxuan Zhou, Libo Zhang, Hao
	Jia, Weiqiang Chen, Song Liu, Yuanzhen Chen, and Dapeng Yu.
%	\newblock Simulation of a topological phase transition in a kitaev chain with long-range coupling using a superconducting circuit.
	\href{https://journals.aps.org/prb/abstract/10.1103/PhysRevB.101.035109}{\newblock {\em Phys. Rev. B.} \textbf{101}, 035109, (2020).}
	
	\bibitem{niu2021simulation}
	Jingjing Niu, Tongxing Yan, Yuxuan Zhou, Ziyu Tao, Xiaole Li, Weiyang Liu, Libo
	Zhang, Hao Jia, Song Liu, Zhongbo Yan, et~al.
%\newblock Simulation of higher-order topological phases and related topological phase transitions in a superconducting qubit.
	\href{https://www.sciencedirect.com/science/article/pii/S2095927321001766}{\newblock {\em Sci. Bull.} \textbf{66}, 1168--1175 (2021).}
	
	\bibitem{xie2019topological}
	Dizhou Xie, Wei Gou, Teng Xiao, Bryce Gadway, and Bo~Yan.
%\newblock Topological characterizations of an extended su--schrieffer--heeger model.
	\href{https://www.nature.com/articles/s41534-019-0159-6}{\newblock {\em npj Quantum Inf.} \textbf{5}, 1--5 (2019).}
	
	\bibitem{an2018engineering}
	Fangzhao~Alex An, Eric~J Meier, and Bryce Gadway.
%\newblock Engineering a flux-dependent mobility edge in disordered zigzag chains.
	\href{https://journals.aps.org/prx/abstract/10.1103/PhysRevX.8.031045}{\newblock {\em Phys. Rev. X.} \textbf{8}, 031045 (2018).}
	
	\bibitem{meier2016observation}
	Eric~J Meier, Fangzhao~Alex An, and Bryce Gadway.
%\newblock Observation of the topological soliton state in the su--schrieffer--heeger model.
	\href{https://www.nature.com/articles/ncomms13986}{\newblock {\em Nat. Commun.} \textbf{7}, 1--6 (2016).}
	
	\bibitem{kraus2012preparing}
	Christina~V Kraus, Sebastian Diehl, Peter Zoller, and Mikhail~A Baranov.
%	\newblock Preparing and probing atomic majorana fermions and topological order in optical lattices.
	\href{https://iopscience.iop.org/article/10.1088/1367-2630/14/11/113036}{\newblock {\em New J. Phys.} \textbf{14}, 113036 (2012).}	
	
	\bibitem{jiang2011majorana}
	Liang Jiang, Takuya Kitagawa, Jason Alicea, AR~Akhmerov, David Pekker, Gil
	Refael, J~Ignacio Cirac, Eugene Demler, Mikhail~D Lukin, and Peter Zoller.
%\newblock Majorana fermions in equilibrium and in driven cold-atom quantum wires.
	\href{https://journals.aps.org/prl/abstract/10.1103/PhysRevLett.106.220402}{\newblock {\em Phys. Rev. Lett.} \textbf{106}, 220402 (2011).}
	
	\bibitem{rahul2022topological}
	S~Rahul and Sujit Sarkar.
%	\newblock Topological quantum criticality in non-hermitian extended kitaev chain.
	\href{https://www.nature.com/articles/s41598-022-11126-7}{\newblock {\em Sci. Rep.}, \textbf{12}, 1--12 (2022).}
	
	
	\bibitem{PhysRevLett.121.076802}
	Daniel Yates, Yonah Lemonik, Aditi Mitra.
%	\newblock Central Charge of Periodically Driven Critical Kitaev Chains.
	\href{https://link.aps.org/doi/10.1103/PhysRevLett.121.076802}{\newblock {\em Phys. Rev. Lett.} \textbf{121}, 076802 (2018).}
	
	\bibitem{PhysRevLett.113.156402}
	Davide Vodola, Luca Lepori, Elisa Ercolessi, Alexey V. Gorshkov, and Guido Pupillo.
	%	\newblock Kitaev Chains with Long-Range Pairing.
	\href{https://journals.aps.org/prl/abstract/10.1103/PhysRevLett.113.156402}{\newblock {\em Phys. Rev. Lett.} \textbf{113}, 156402 (2014).}
	
	\bibitem{PhysRevB.95.195160}
	Antonio Alecce and Luca Dell'Anna.
	%	\newblock Extended Kitaev chain with longer-range hopping and pairing.
	\href{https://journals.aps.org/prb/abstract/10.1103/PhysRevB.95.195160}{\newblock {\em Phys. Rev. B.} \textbf{95}, 195160 (2017).}
	
\end{thebibliography}

\end{document}

\title{Topological phase transition between non-high symmetry critical phases and curvature function renormalization group}
	
\author{Ranjith R Kumar}
\email{ranjith.btd6@gmail.com}
\affiliation{Theoretical Sciences Division, Poornaprajna Institute of Scientific Research, Bidalur, Bengaluru-562164, India.}
\affiliation{Graduate Studies, Manipal Academy of Higher Education, Madhava Nagar, Manipal-576104, India.}
\author{Y R Kartik}
\email{yrkartik@gmail.com}
\affiliation{Theoretical Sciences Division, Poornaprajna Institute of Scientific Research, Bidalur, Bengaluru-562164, India.}
\affiliation{Graduate Studies, Manipal Academy of Higher Education, Madhava Nagar, Manipal-576104, India.}
\author{Sujit Sarkar}
\email{Corresponding author: sujit.tifr@gmail.com}
\affiliation{Theoretical Sciences Division, Poornaprajna Institute of Scientific Research, Bidalur, Bengaluru-562164, India.}
	
\maketitle
\section{CRG equations for topological transition between gapped phases at non-HS critical points}
For the non-HS point $k_0=\arccos((-2\Gamma_{2}+ \sqrt{4\Gamma_{2}^2-16\Gamma_{3}(\Gamma_{1}-\Gamma_{3})})/8\Gamma_{3})$, which corresponds to magenta line in Fig.1 of the main text, we have
\begin{equation}
\dfrac{d\Gamma_{1}}{dl}=\frac{\alpha\sqrt{\alpha_1}\Lambda_1(\Gamma_{0},\Gamma_{1},\Gamma_{2},\Gamma_{3})}{2\sqrt{2}\Lambda_2(\Gamma_{0},\Gamma_{1},\Gamma_{2},\Gamma_{3})},
\end{equation}
\begin{equation}
\dfrac{d\Gamma_{3}}{dl}=\frac{\Gamma_{3}^2\alpha\sqrt{\alpha_1}\Lambda_1(\Gamma_{0},\Gamma_{1},\Gamma_{2},\Gamma_{3})}{2\sqrt{2}\Lambda_2^{\prime}(\Gamma_{0},\Gamma_{1},\Gamma_{2},\Gamma_{3})}.
\end{equation}
The explicit form of the $\Lambda$s can be obtained as

\begin{equation}
\begin{split}
\Lambda_1(\Gamma_{0},\Gamma_{1},\Gamma_{2},\Gamma_{3})&=\frac{1}{\Gamma_{3}}[2\Gamma_{3}(\Gamma_{2}\Gamma_{3}+\Gamma_{1}\alpha)+\Gamma_{0}(\Gamma_{2}^2-4\Gamma_{1}\Gamma_{3}
-\Gamma_{2}\alpha)][\Gamma_{3}(\Gamma_{2}^2(\Gamma_{1}+3\Gamma_{3})-4\Gamma_{3}(\Gamma_{1}-\Gamma_{3})(\Gamma_{1}+3\Gamma_{3})\\ 
&-\Gamma_{1}\Gamma_{2}\alpha+5\Gamma_{2}\Gamma_{3}\alpha)+\Gamma_{0}(-\Gamma_{2}^3+4\Gamma_{2}\Gamma_{3}(\Gamma_{1}-\Gamma_{3})+\Gamma_{2}^2\alpha-2\Gamma_{3}(\Gamma_{1}
+3\Gamma_{3})\alpha)]+[2\Gamma_{0}^2+2\Gamma_{3}(\Gamma_{3}-\Gamma_{1})\\ 
&-2\Gamma_{0}(\Gamma_{2}+\alpha)+\Gamma_{2}(\Gamma_{2}+\alpha)][2\Gamma_{3}
(-5\Gamma_{2}\Gamma_{3}+\Gamma_{1}(\Gamma_{2}-4\alpha))
+\Gamma_{0}(16\Gamma_{1}\Gamma_{3}+5\Gamma_{2}(-\Gamma_{2}+\alpha))],
\end{split}
\end{equation}

\begin{equation}
\begin{split}
\Lambda_2(\Gamma_{0},\Gamma_{1},\Gamma_{2},\Gamma_{3})&=\Gamma_{0}[\Gamma_{2}^3+2\Gamma_{2}\Gamma_{3}(\Gamma_{1}-13\Gamma_{3})-\Gamma_{2}^2\alpha
-16\Gamma_{1}\Gamma_{3}\alpha]
+4\Gamma_{0}\Gamma_{3}[\Gamma_{3}(\Gamma_{1}-\Gamma_{3})(3\Gamma_{3}-7\Gamma_{1})
+\Gamma_{2}^2(\Gamma_{1}+3\Gamma_{3})\\ 
&+2\Gamma_{1}\Gamma_{2}\alpha+2\Gamma_{2}\Gamma_{3}\alpha]
-2\Gamma_{3}[\Gamma_{2}^3(\Gamma_{1}+\Gamma_{3})-\Gamma_{2}\Gamma_{3}(5\Gamma_{1}^2-6\Gamma_{1}\Gamma_{3}+\Gamma_{3}^2)
+\Gamma_{1}\Gamma_{2}^2\alpha-2\Gamma_{1}^2\Gamma_{3}\alpha \\ 
&+\Gamma_{2}^2\Gamma_{3}\alpha+4\Gamma_{1}\Gamma_{3}^2\alpha-2\Gamma_{3}^3\alpha]
+4\Gamma_{0}^3[\Gamma_{3}(3\Gamma_{1}+\Gamma_{3})
-\Gamma_{2}^2+\Gamma_{2}\alpha],
\end{split}
\end{equation}

\begin{equation}
\begin{split}
\Lambda_2^{\prime}(\Gamma_{0},\Gamma_{1},\Gamma_{2},\Gamma_{3})&=2\Gamma_{0}^3[-\Gamma_{2}^4+2\Gamma_{2}^2\Gamma_{3}(2\Gamma_{1}-\Gamma_{3})-2\Gamma_{3}^2(\Gamma_{1}^2
-3\Gamma_{1}\Gamma_{3}+6\Gamma_{3^2})+\Gamma_{2}^2\alpha-2\Gamma_{1}\Gamma_{2}\Gamma_{3}\alpha]
+\Gamma_{0}^2\Gamma_{3}[\Gamma_{2}^3(7\Gamma_{1} \\ 
&-8\Gamma_{3})+2\Gamma_{2}\Gamma_{3}
(-11\Gamma_{1}^2+9\Gamma_{1}\Gamma_{3}+14\Gamma_{3}^2)-7\Gamma_{1}\Gamma_{2}^2\alpha+8\Gamma_{1}^2\Gamma_{3}\alpha+8\Gamma_{2}^2\Gamma_{3}\alpha
-16\Gamma_{1}\Gamma_{3}^2\alpha
+24\Gamma_{3}^3\alpha]\\
&-4\Gamma_{0}\Gamma_{3}^2[-5\Gamma_{1}^3\Gamma_{3}+2\Gamma_{1}^2(\Gamma_{2}^2+6\Gamma_{3}^2-\Gamma_{2}\alpha)+\Gamma_{1}\Gamma_{3}(-7\Gamma_{2}^2
-13\Gamma_{3}^2
+2\Gamma_{2}\alpha)+\Gamma_{3}^2(9\Gamma_{2}^2+6\Gamma_{3}^2+4\Gamma_{2}\alpha)]\\
&+2\Gamma_{3}^2[\Gamma_{1}^3(\Gamma_{2}-2\alpha) +4\Gamma_{1}^2\Gamma_{3}(-\Gamma_{2}+\alpha)-\Gamma_{1}(\Gamma_{2}^3-\Gamma_{2}\Gamma_{3}^2
+\Gamma_{2}^2\alpha+2\Gamma_{3}^2\alpha)+\Gamma_{2}\Gamma_{3}(2\Gamma_{3}^2
+3\Gamma_{2}(\Gamma_{2}+\alpha))].
\end{split}
\end{equation}

For the non-HS point $k_0=\arccos((-2\Gamma_{2}- \sqrt{4\Gamma_{2}^2-16\Gamma_{3}(\Gamma_{1}-\Gamma_{3})})/8\Gamma_{3})$, which corresponds to green line in Fig.1 of the main text, we have

\begin{equation}
\dfrac{d\Gamma_{1}}{dl}=-\frac{\alpha\sqrt{\alpha_1}\Lambda_3(\Gamma_{0},\Gamma_{1},\Gamma_{2},\Gamma_{3})}{\sqrt{2}\Gamma_{3}^2\Lambda_4(\Gamma_{0},\Gamma_{1},\Gamma_{2},\Gamma_{3})},
\end{equation}
\begin{equation}
\dfrac{d\Gamma_{3}}{dl}=-\frac{\alpha\sqrt{\alpha_1}\Lambda_3(\Gamma_{0},\Gamma_{1},\Gamma_{2},\Gamma_{3})}{\sqrt{2}\Lambda_4^{\prime}(\Gamma_{0},\Gamma_{1},\Gamma_{2},\Gamma_{3})},
\end{equation}

where the $\Lambda$s can be obtained as

\begin{equation}
\begin{split}
\Lambda_3(\Gamma_{0},\Gamma_{1},\Gamma_{2},\Gamma_{3})&=\Gamma_{0}^2[\Gamma_{2}^5+\Gamma_{2}^3\Gamma_{3}(-7\Gamma_{1}+\Gamma_{3})+6\Gamma_{2}\Gamma_{3}^2
(2\Gamma_{1}^2-\Gamma_{1}\Gamma_{3}+3\Gamma_{3}^2)+\Gamma_{2}^4\alpha+4\Gamma_{1}\Gamma_{3}^2(\Gamma_{1}-3\Gamma_{3})\alpha
-\Gamma_{2}^2\Gamma_{3}(5\Gamma_{1}\\ 
&+\Gamma_{3})\alpha]
+\Gamma_{0}^3\Gamma_{3}^2[-16\Gamma_{1}\Gamma_{3}+5\Gamma_{2}(\Gamma_{2}+\alpha)]
+\Gamma_{3}^2(\Gamma_{1}-\Gamma_{3})
[-2\Gamma_{2}\Gamma_{3}(\Gamma_{2}^2-\Gamma_{3}^2)
-4\Gamma_{1}^2\Gamma_{3}(\Gamma_{2}+\alpha)+\Gamma_{1}\\ 
&(+\Gamma_{2}^3+10+\Gamma_{2}+\Gamma_{3}^2)
+\Gamma_{2}^2\alpha-4\Gamma_{3}^2\alpha]
+\Gamma_{0}\Gamma_{3}[-16\Gamma_{1}^3\Gamma_{3}^2
+4\Gamma_{1}^2\Gamma_{3}(3\Gamma_{2}^2+4\Gamma_{3}^2+2\Gamma_{2}\alpha)
+\Gamma_{2}\Gamma_{3}(2\Gamma_{2}^3\\
&-7\Gamma_{2}\Gamma_{3}^2+2\Gamma_{2}^2\alpha+3\Gamma_{3}^2\alpha)
-\Gamma_{1}\Gamma_{2}(2\Gamma_{2}^3+13\Gamma_{2}\Gamma_{3}^2
+2\Gamma_{2}^2\alpha+3\Gamma_{3}^2\alpha)],
\end{split}
\end{equation}

\begin{equation}
\begin{split}
\Lambda_4(\Gamma_{0},\Gamma_{1},\Gamma_{2},\Gamma_{3})&=-\Gamma_{0}^2[\Gamma_{2}^3+2\Gamma_{2}\Gamma_{3}(\Gamma_{1}-13\Gamma_{3})+\Gamma_{2}^2\alpha
+16\Gamma_{1}\Gamma_{3}\alpha]
+4\Gamma_{0}\Gamma_{3}[\Gamma_{3}(7\Gamma_{1}-3\Gamma_{3})(\Gamma_{1}-\Gamma_{3})
-\Gamma_{2}^2(\Gamma_{1}+3\Gamma_{3})\\ \nonumber
&+2\Gamma_{1}\Gamma_{2}\alpha+2\Gamma_{2}\Gamma_{3}\alpha]
+2\Gamma_{3}[\Gamma_{2}^3(\Gamma_{1}+\Gamma_{3})-\Gamma_{2}\Gamma_{3}(5\Gamma_{1}^2-6\Gamma_{1}\Gamma_{3}
+\Gamma_{3}^2)
-\Gamma_{1}\Gamma_{2}^2\alpha+2\Gamma_{1}^2\Gamma_{3}\alpha-\Gamma_{2}^2\Gamma_{3}\alpha\\
&-4\Gamma_{1}\Gamma_{3}^2\alpha+2\Gamma_{3}^3\alpha]
+4\Gamma_{0}^3
[-\Gamma_{3}(3\Gamma_{1}+\Gamma_{3})+\Gamma_{2}(\Gamma_{2}+\alpha)],
\end{split}
\end{equation}

\begin{equation}
\begin{split}
\Lambda_4^{\prime}(\Gamma_{0},\Gamma_{1},\Gamma_{2},\Gamma_{3})&=2\Gamma_{0}^3[\Gamma_{2}^4+2\Gamma_{2}^2\Gamma_{3}(-2\Gamma_{1}+\Gamma_{3})+2\Gamma_{3}^2(\Gamma_{1}^2
-3\Gamma_{1}\Gamma_{3}+6\Gamma_{3}^2)+\Gamma_{2}^3\alpha-2\Gamma_{1}\Gamma_{2}\Gamma_{3}\alpha]
+\Gamma_{0}^2\Gamma_{3}[\Gamma_{1}^2(22\Gamma_{2}\Gamma_{3}\\
&+8\Gamma_{3}\alpha)+4\Gamma_{3}(2\Gamma_{2}^3-7\Gamma_{2}\Gamma_{3}^2+2\Gamma_{2}^2\alpha+6\Gamma_{3}^2\alpha)-\Gamma_{1}(7\Gamma_{2}^3+18\Gamma_{2}\Gamma_{3}^2
+7\Gamma_{2}^2\alpha+16\Gamma_{3}^2\alpha)]
+2\Gamma_{3}^2[4\Gamma_{1}^2\Gamma_{3}\\
&(\Gamma_{2}+\alpha)-\Gamma_{1}^3(\Gamma_{2}+2\alpha)-\Gamma_{1}(-\Gamma_{2}^3
+\Gamma_{2}\Gamma_{3}^2+\Gamma_{2}^2\alpha+2\Gamma_{3}^2\alpha)
+\Gamma_{2}\Gamma_{3}(-2\Gamma_{3}^2+3\Gamma_{2}(-\Gamma_{2}+\alpha))]\\
&+4\Gamma_{0}\Gamma_{3}^2
[-5\Gamma_{1}^3\Gamma_{3}+\Gamma_{3}^2(9\Gamma_{2}^2+6\Gamma_{3}^2-4\Gamma_{2}\alpha)
-\Gamma_{1}\Gamma_{3}(7\Gamma_{2}^2+13\Gamma_{3}^2+2\Gamma_{2}\alpha)+2\Gamma_{1}^2(6\Gamma_{3}^2+\Gamma_{2}(\Gamma_{2}+\alpha))].
\end{split}
\end{equation}

In all the equations $\alpha=\sqrt{\Gamma_{2}^2+4\Gamma_{3}(-\Gamma_{1}+\Gamma_{3})}$ and $\alpha_1=\sqrt{(2\Gamma_{3}(\Gamma_{1}+3\Gamma_{3})-\Gamma_{2}^2+\Gamma_{2}\alpha)}/\Gamma_{3}$.

\section{CRG equations for topological transition between non-HS critical phases at multicritical points}
The RG equations on the magenta line can be written as

\begin{equation}
\frac{d\Gamma_{1}}{dl}=\frac{(\Gamma_{1}+\alpha)^4\alpha\Lambda_5(\Gamma_{0},\Gamma_{1},\Gamma_{2})}{\Lambda_6(\Gamma_{0},\Gamma_{1},\Gamma_{2})},
\end{equation}
\begin{equation}
\frac{d\Gamma_{2}}{dl}=-\frac{(\Gamma_{1}+\alpha)^4\alpha\alpha_1^{\prime}\Lambda_5(\Gamma_{0},\Gamma_{1},\Gamma_{2})}{\Lambda_6^{\prime}(\Gamma_{0},\Gamma_{1},\Gamma_{2})},
\end{equation}
where the $\Lambda$s are obtained as
\begin{equation}
\begin{split}
\Lambda_5(\Gamma_{0},\Gamma_{1},\Gamma_{2})&=(6\Gamma_{0}^2+4\Gamma_{1}(\Gamma_{1}+\alpha)-6\Gamma_{0}\Gamma_{2}+\Gamma_{2}
(-\Gamma_{2}+(\Gamma_{2}-2\Gamma_{0})))^{1/2}/(\Gamma_{1}+\alpha)(1/2(2\Gamma_{0}\Gamma_{1}-(7\Gamma_{1}+\alpha)\Gamma_{2}\\
&+(12\Gamma_{0}(\Gamma_{0-\Gamma_{2}})\alpha_1)/(\Gamma_{1}+\alpha))(6\Gamma_{0}^2+\Gamma_{1}(\Gamma_{1}+\alpha)-6\Gamma_{0}\Gamma_{2}+\Gamma_{2}\alpha_1)
+2(\Gamma_{1}\Gamma_{2}+(2\Gamma_{0}(\Gamma_{2}-\Gamma_{0})\\
&\alpha_1)/(\Gamma_{1}+\alpha))6\Gamma_{0}^2+\Gamma_{1}(3\Gamma_{1}+\alpha)
-8\Gamma_{0}\Gamma_{2}+4\Gamma_{2}^2+(-2\Gamma_{0}\Gamma_{1}+(7\Gamma_{1}+\alpha)\Gamma_{2})\alpha_1/(2(\Gamma_{1}+\alpha))\\
&+6\Gamma_{0}(\Gamma_{2}-\Gamma_{0})\cos[2\sec^{-1}[2(\Gamma_{1}+\alpha)/\alpha_1]]),
\end{split}
\end{equation}

\begin{equation}
\begin{split}
\Lambda_6(\Gamma_{0},\Gamma_{1},\Gamma_{2})&=8\sqrt{2}\Gamma_{0}(24\Gamma_{0}^7+\Gamma_{1}^5(\Gamma_{1}+\alpha)\alpha_1-2\Gamma_{0}^6
(43\Gamma_{2}+3\alpha_1^{\prime})+\Gamma_{0}\Gamma_{1}(4\Gamma_{1}^5+4\Gamma_{1}^4\alpha+2\Gamma_{1}^2\alpha\Gamma_{2}(3\Gamma_{2}-\alpha_1^{\prime})\\
&-4\Gamma_{1}^3\Gamma_{2}(-2\Gamma_{2}+\alpha_1^{\prime})+\Gamma_{1}\Gamma_{2}^3(\Gamma_{2}+\alpha_1^{\prime})+\alpha\Gamma_{2}^3(\Gamma_{2}+\alpha_1^{\prime}))-\Gamma_{0}^4
(24\Gamma_{2}^2(3\Gamma_{2}+\alpha_1^{\prime})
+\Gamma_{1}^2(133\Gamma_{2}+\alpha_1^{\prime})\\
&+\Gamma_{1}\alpha(59\Gamma_{2}+3\alpha_1^{\prime}))
-\Gamma_{0}^2(2\Gamma_{1}^3\alpha(13\Gamma_{2}-\alpha_1^{\prime})-4\Gamma_{1}^4(-9\Gamma_{2}+\alpha_1^{\prime})+2\Gamma_{2}^4(\Gamma_{2}+\alpha_1^{\prime})
\Gamma_{1}^2\Gamma_{2}^2(23\Gamma_{2}+3\alpha_1^{\prime})\\
&+\Gamma_{1}\alpha\Gamma_{2}^2(13\Gamma_{2}+5\alpha_1^{\prime}))+4\Gamma_{0}^5(14\Gamma_{1}^2+6\Gamma_{1}\alpha
+\Gamma_{2}(29\Gamma_{2}+5\alpha_1^{\prime}))+\Gamma_{0}^3(28\Gamma_{1}^4+20\Gamma_{1}^3\alpha+3\Gamma_{1}^2\Gamma_{2}(33\Gamma_{2}+\alpha_1^{\prime})\\
&+4\Gamma_{2}^3(5\Gamma_{2}+3\alpha_1^{\prime})+\Gamma_{1}\alpha\Gamma_{2}(47\Gamma_{2}+7\alpha_1^{\prime}))),
\end{split}
\end{equation}

\begin{equation}
\begin{split}
\Lambda_6^{\prime}(\Gamma_{0},\Gamma_{1},\Gamma_{2})&=4\sqrt{2}\Gamma_{0}(-8\Gamma_{0}^7(4\Gamma_{1}+3\alpha)+\Gamma_{0}^6
(\Gamma_{1}+\alpha)
-2\Gamma_{0}\Gamma_{1}^4(\Gamma_{1}^3+\Gamma_{1}^2\alpha+2\alpha\Gamma_{2}\alpha_1^{\prime}-3\Gamma_{1}\Gamma_{2}\alpha_1)
+\Gamma_{0}^2(\Gamma_{1}^3\Gamma_{2}^2\\
&(13\Gamma_{2}-5\alpha_1^{\prime})+2\Gamma_{1}\Gamma_{2}^4(\Gamma_{2}+\alpha_1^{\prime})+2\alpha\Gamma_{2}^4(\Gamma_{2}+\alpha_1^{\prime})+3\Gamma_{1}^4\alpha
(3\Gamma_{2}+\alpha_1^{\prime})+\Gamma_{1}^2\alpha\Gamma_{2}^2(7\Gamma_{2}+\alpha_1^{\prime})+\Gamma_{1}^5(15\Gamma_{2}\\
&+\alpha_1^{\prime}))-2\Gamma_{0}^3(3\Gamma_{1}^5
+\Gamma_{1}^4\alpha+5\Gamma_{1}^3\Gamma_{2}(4\Gamma_{2}+3\alpha_1^{\prime})+2\alpha\Gamma_{2}^3(5\Gamma_{2}+4\alpha_1^{\prime})+\Gamma_{1}\Gamma_{2}^3(15\Gamma_{2}
+11\alpha_1^{\prime})
\Gamma_{1}^2\alpha\Gamma_{2}(11\Gamma_{2}\\
&+14\alpha_1^{\prime}))
+4\Gamma_{0}^6(3\alpha(7\Gamma_{2}+2\alpha_1^{\prime})
+\Gamma_{1}(31\Gamma_{2}
+17\alpha_1^{\prime}))+\Gamma_{0}^4(10\alpha\Gamma_{2}^2(7\Gamma_{2}+5\alpha_1^{\prime})+\Gamma_{1}^2\alpha(17\Gamma_{2}+27\alpha_1^{\prime})\\
&+\Gamma_{1}^3(29\Gamma_{2}+35\alpha_1^{\prime})+2\Gamma_{1}\Gamma_{2}(57\Gamma_{2}+53\alpha_1^{\prime}))
-2\Gamma_{0}^5(\Gamma_{1}^3+\Gamma_{1}^2\alpha_1
+2\alpha\Gamma_{2}(28\Gamma_{2}+15\alpha_1^{\prime})+\Gamma_{1}\Gamma_{2}(89\Gamma_{2}+77\alpha_1^{\prime}))).
\end{split}
\end{equation}

The RG equations on the green line can be written as
\begin{equation}
\frac{d\Gamma_{1}}{dl}=-\frac{(\Gamma_{1}-\alpha)^4\alpha\Lambda_7(\Gamma_{0},\Gamma_{1},\Gamma_{2})}{\Lambda_8(\Gamma_{0},\Gamma_{1},\Gamma_{2})},
\end{equation}
\begin{equation}
\frac{d\Gamma_{2}}{dl}=-\frac{(\Gamma_{1}-\alpha)^4\alpha\alpha_1^{\prime}\Lambda_7(\Gamma_{0},\Gamma_{1},\Gamma_{2})}{\Lambda_8^{\prime}(\Gamma_{0},\Gamma_{1},\Gamma_{2})},
\end{equation}
where the $\Lambda$s are obtainedas
\begin{equation}
\begin{split}
\Lambda_7(\Gamma_{0},\Gamma_{1},\Gamma_{2})&= (6\Gamma_{0}^2+4\Gamma_{1}(\Gamma_{1}-\alpha)-6\Gamma_{0}\Gamma_{2}-\Gamma_{2}
(\Gamma_{2}+\alpha_1^{\prime}))^{1/2}/(\Gamma_{1}-\alpha)(1/2(6\Gamma_{0}+\Gamma_{1}^2-\Gamma_{1}\alpha-6\Gamma_{0}\Gamma_{2}+\Gamma_{2}\\
&(\Gamma_{2}-\alpha_1^{\prime}))(2\Gamma_{0}\Gamma_{1}+(-7\Gamma_{1}+\alpha)\Gamma_{2}+(12\Gamma_{0}(\Gamma_{0}-\Gamma_{2})(\Gamma_{2}
+\alpha_1^{\prime}))/(-\Gamma_{1}+\alpha))+2(\Gamma_{1}\Gamma_{2}-(2\Gamma_{0}(\Gamma_{0}\\
&-\Gamma_{2}(\Gamma_{2}+\alpha_1^{\prime})))/
(-\Gamma_{1}+\alpha))(6\Gamma_{0}^2+3\Gamma_{1}^2-\Gamma_{1}\alpha-8\Gamma_{0}\Gamma_{2}+4\Gamma_{2}^2-(2\Gamma_{0}\Gamma_{1}+(-7\Gamma_{1}
+\alpha)\Gamma_{2})(\Gamma_{2}\\
&+\alpha_1^{\prime})/(2(-\Gamma_{1}+\alpha)))+6\Gamma_{0}^2(\Gamma_{0}-\Gamma_{2})\cos[2\sin^{-1}[(\Gamma_{2}+\alpha_1^{\prime})/(2\Gamma_{1}-2\alpha)]]),
\end{split}
\end{equation}
\begin{equation}
\begin{split}
\Lambda_8(\Gamma_{0},\Gamma_{1},\Gamma_{2})&=8\sqrt{2}\Gamma_{0}(24\Gamma_{0}^7+\Gamma_{1}^5(-\Gamma_{1}+\alpha)(\Gamma_{2}+\alpha_1^{\prime})
+2\Gamma_{0}^6(-43\Gamma_{2}+3\alpha_1^{\prime})+\Gamma_{0}^4(\Gamma_{1}\alpha(59\Gamma_{2}-3\alpha_1^{\prime})+\Gamma_{1}^2(-133\Gamma_{2}\\
&+\alpha_1^{\prime})+24\Gamma_{2}^2(-3\Gamma_{2}+\alpha_1^{\prime}))+\Gamma_{0}\Gamma_{1}(4\Gamma_{1}^5-4\Gamma_{1}^4\alpha+\Gamma_{1}\Gamma_{2}^3(\Gamma_{2}-\alpha_1^{\prime})
+\alpha\Gamma_{2}^3\alpha_1+4\Gamma_{1}^3\Gamma_{2}(2\Gamma_{2}+\alpha_1^{\prime})\\
&-2\Gamma_{1}^2\alpha\Gamma_{2}(3\Gamma_{2}+\alpha_1^{\prime}))+\Gamma_{0}^2(\Gamma_{1}\alpha\Gamma_{2}^2
(13\Gamma_{2}-5\alpha_1^{\prime})+2\Gamma_{2}^4\alpha_1-4\Gamma_{1}^4(9\Gamma_{2}+\alpha_1^{\prime})+2\Gamma_{1}^3\alpha(13\Gamma_{2}+\alpha_1^{\prime})\\
&+\Gamma_{1}^2\Gamma_{2}^2(-23\Gamma_{2}+3\alpha_1^{\prime}))-4\Gamma_{0}^5(-14\Gamma_{1}^2+6\Gamma_{1}\alpha+\Gamma_{2}(-29\Gamma_{2}
+5\alpha_1^{\prime}))+\Gamma_{0}^3(28\Gamma_{1}^4-20\Gamma_{1}^3\alpha+4\Gamma_{2}^3\\
&(5\Gamma_{2}-3\alpha_1^{\prime})-3\Gamma_{1}^2\Gamma_{2}(-33\Gamma_{2}
+\alpha_1^{\prime})+\Gamma_{1}\alpha\Gamma_{2}(-47\Gamma_{2}+7\alpha_1^{\prime}))),
\end{split}
\end{equation}

\begin{equation}
\begin{split}
\Lambda_8^{\prime}(\Gamma_{0},\Gamma_{1},\Gamma_{2})&=4\sqrt{2}\Gamma_{0}(8\Gamma_{0}^7(-4\Gamma_{1}+3\alpha)-\Gamma_{1}^6(-\Gamma_{1}+\alpha)
(\Gamma_{1}+\alpha_1^{\prime})+2\Gamma_{0}\Gamma_{1}^4(-\Gamma_{1}^3+\Gamma_{1}^2\alpha-3\Gamma_{1}\Gamma_{2}(\Gamma_{2}+\alpha_1^{\prime})\\
&+2\alpha\Gamma_{2}(\Gamma_{2}+\alpha_1^{\prime}))
+4\Gamma_{0}^6(31\Gamma_{1}\Gamma_{2}-17\Gamma_{1}\alpha_1^{\prime}+3\alpha(-7\Gamma_{2}+2\alpha_1^{\prime}))+\Gamma_{0}^2(-3\Gamma_{1}^4\alpha(3\Gamma_{2}
-\alpha_1^{\prime})-\Gamma_{1}^5\\
&(-15\Gamma_{2}+\alpha_1^{\prime})+\Gamma_{1}^2\alpha\Gamma_{2}^2(-7\Gamma_{2}+\alpha_1^{\prime})-2\Gamma_{1}\Gamma_{2}^4\alpha_1
+2\alpha\Gamma_{2}^4\alpha_1+\Gamma_{1}^3\Gamma_{2}^2(13\Gamma_{2}+5\alpha_1^{\prime}))+2\Gamma_{0}^3(-3\Gamma_{1}^5+\Gamma_{1}^4\alpha\\
&+\Gamma_{1}^2\alpha\Gamma_{2}(11\Gamma_{2}
-14\alpha_1^{\prime})+2\alpha\Gamma_{2}^3(5\Gamma_{2}-4\alpha_1^{\prime})+5\Gamma_{1}^3\Gamma_{2}(-4\Gamma_{2}+3\alpha_1^{\prime})+\Gamma_{1}\Gamma_{2}^3(-15\Gamma_{2}+11\alpha_1^{\prime}))+\Gamma_{0}^4\\
&(\Gamma_{1}^3(29\Gamma_{2}-35\alpha_1^{\prime})+10\alpha\Gamma_{2}^2(-7\Gamma_{2}+5\alpha_1^{\prime})\Gamma_{1}^2\alpha(-17\Gamma_{2}
+27\alpha_1^{\prime})-2\Gamma_{1}\Gamma_{2}^2(-57\Gamma_{2}+53\alpha_1^{\prime}))+2\Gamma_{0}^5(-\Gamma_{1}^3\\
&+\Gamma_{1}^2\alpha+2\alpha\Gamma_{2}(28\Gamma_{2}
-15\alpha_1^{\prime})+\Gamma_{1}\Gamma_{2}(-89\Gamma_{2}+77\alpha_1^{\prime}))).
\end{split}
\end{equation}
In all the equations $\alpha=\sqrt{\Gamma_{1}^2+4\Gamma_{0}(\Gamma_{0}-\Gamma_{2})}$ and $\alpha_1^{\prime}=(\Gamma_{2}-2\Gamma_{0})$.

\section{Numerical Results of Edge Modes at non-HS Criticalities}
\begin{figure}[t]
	\includegraphics[width=4.4cm,height=3.2cm]{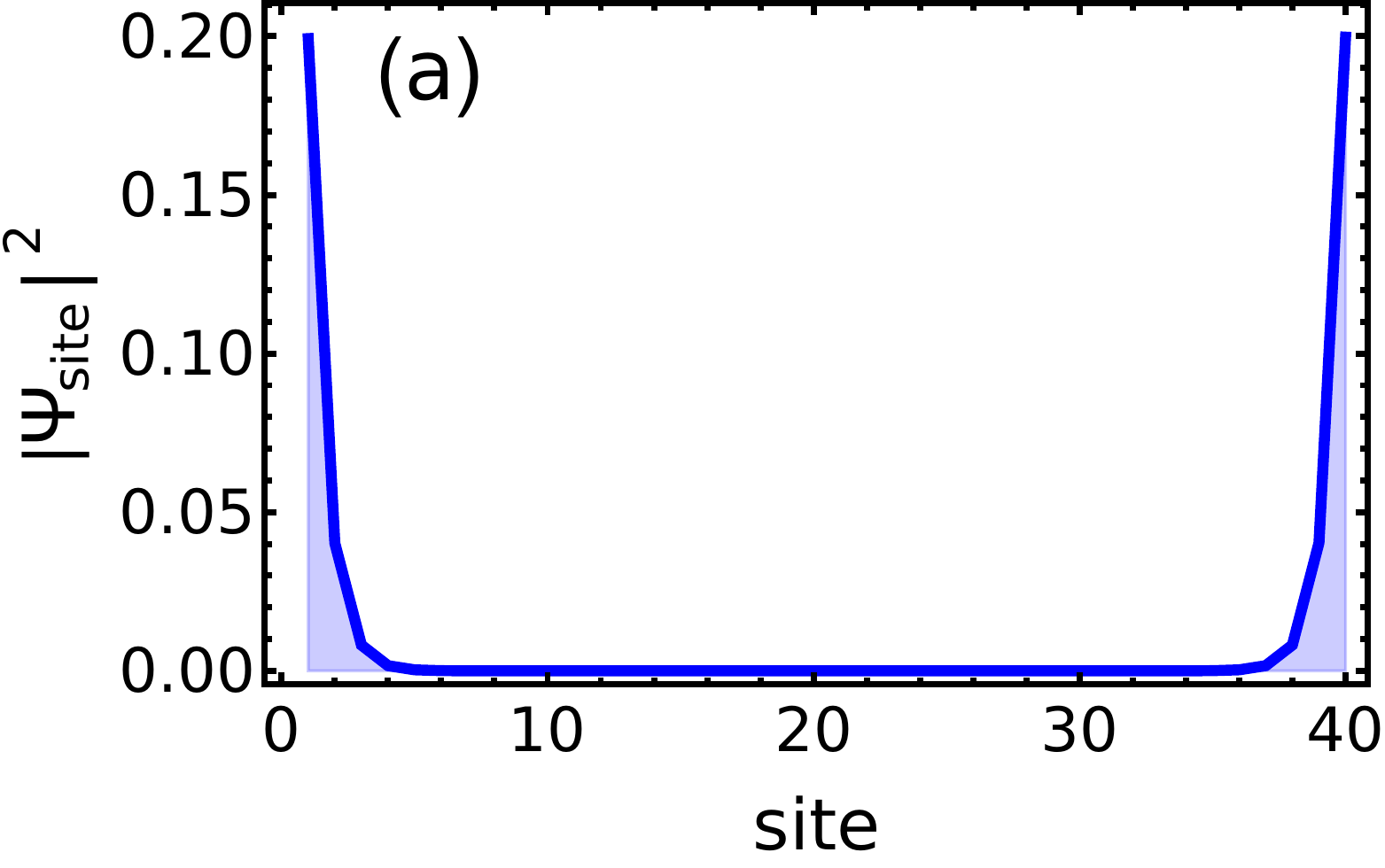}  
	\hspace{0.5cm} 
	\includegraphics[width=4.4cm,height=3.2cm]{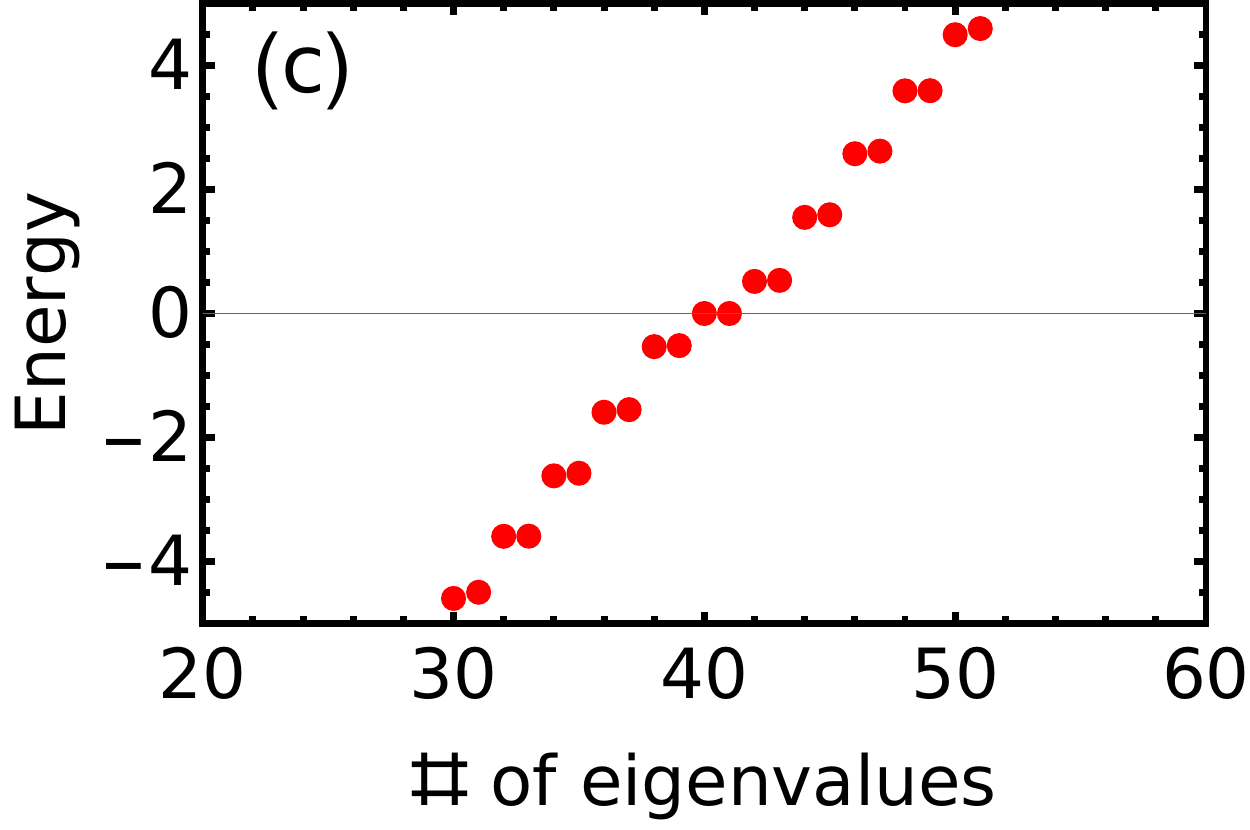}  
	\hspace{0.5cm}
	\includegraphics[width=4.4cm,height=3.2cm]{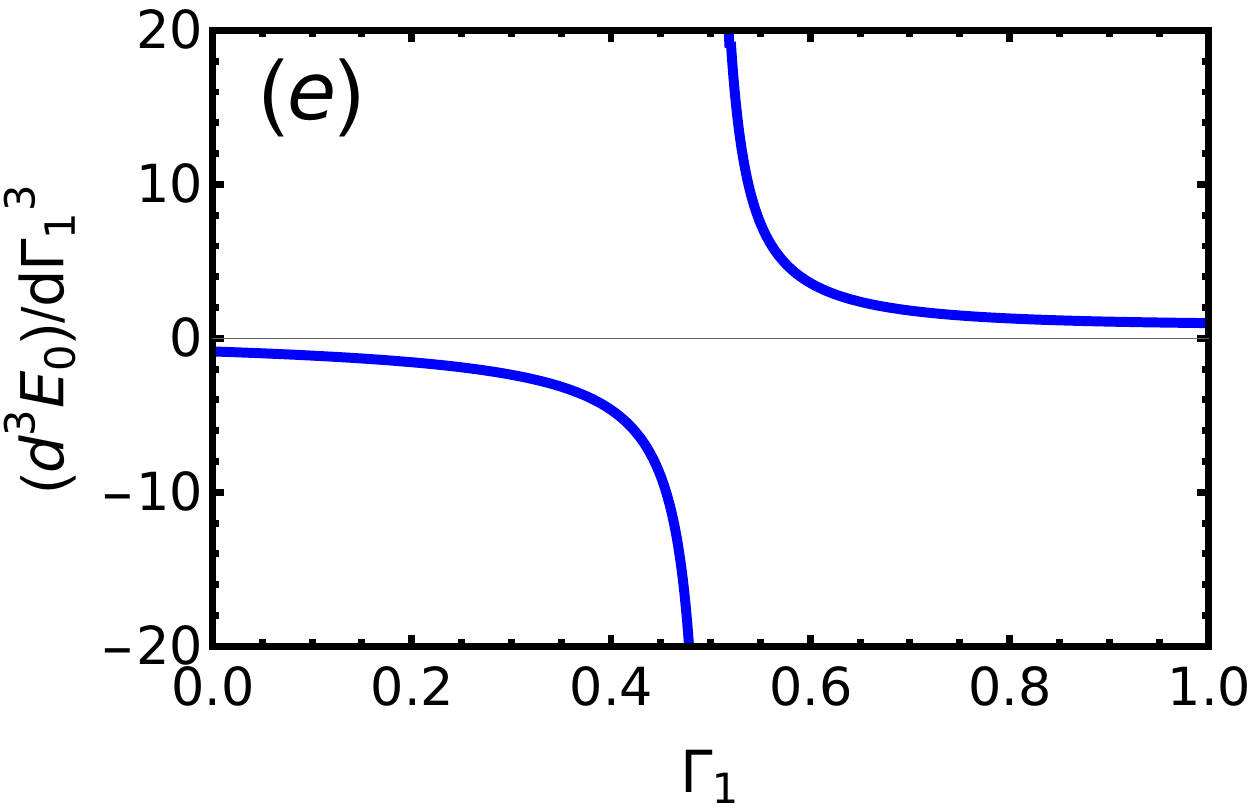} \\
	\includegraphics[width=4.4cm,height=3.2cm]{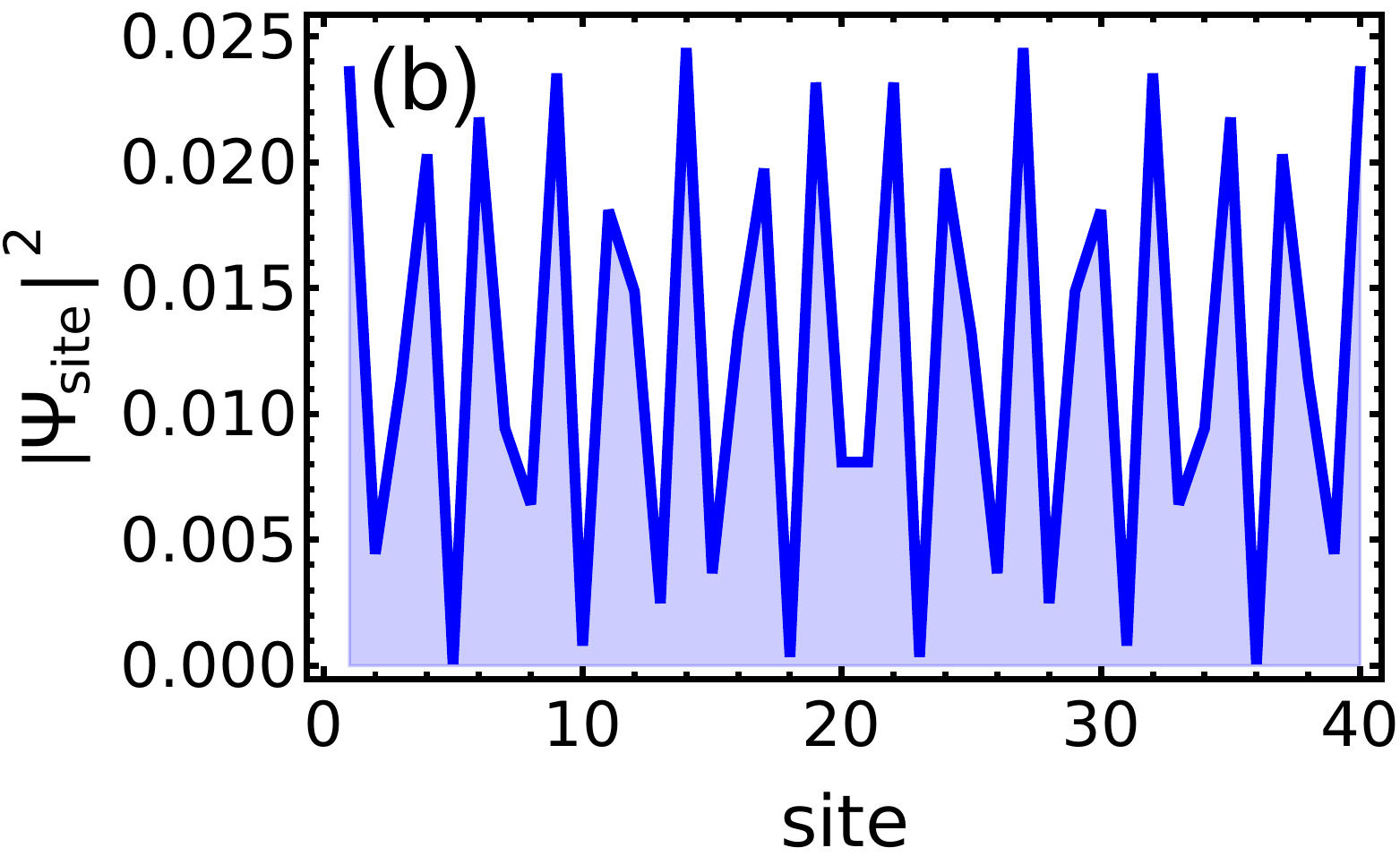}  
	\hspace{0.5cm} 
	\includegraphics[width=4.4cm,height=3.2cm]{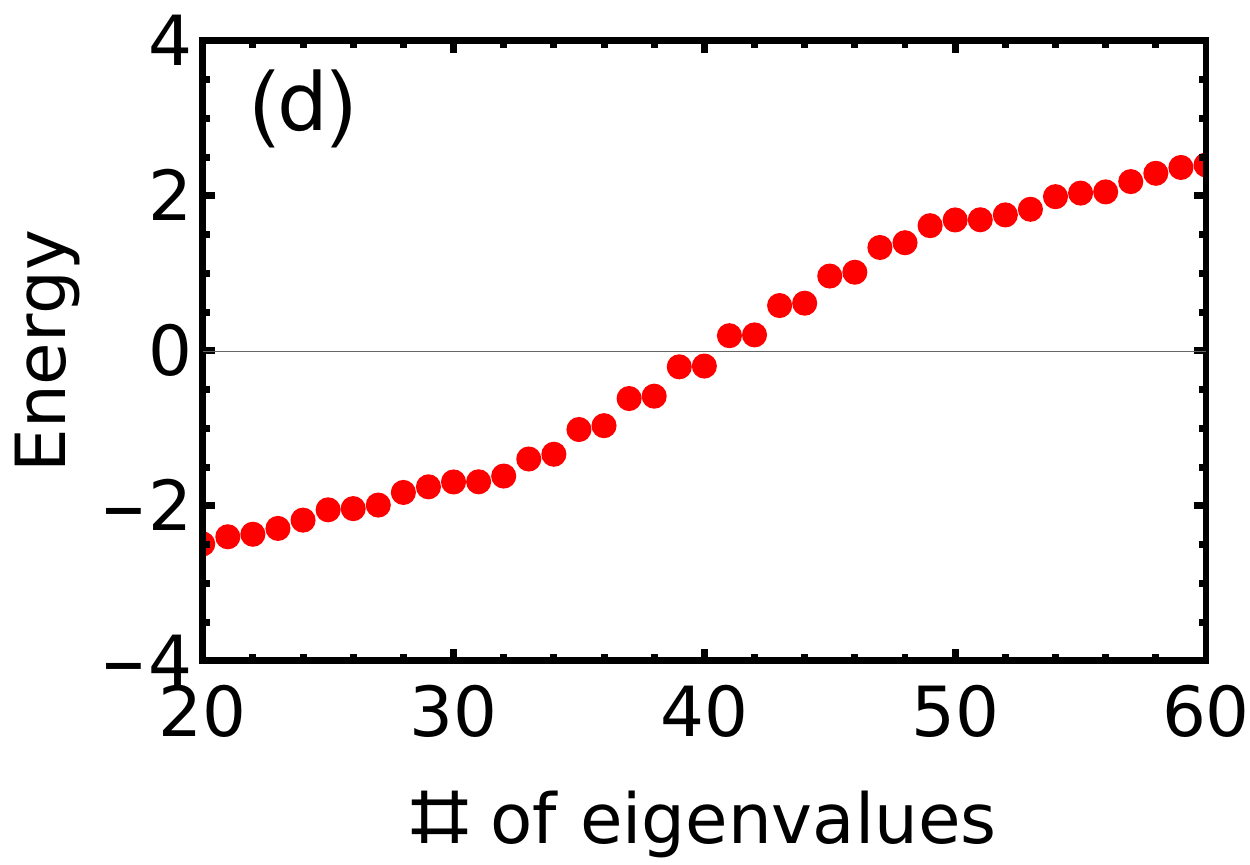}  
	\hspace{0.5cm} 
	\includegraphics[width=4.4cm,height=3.2cm]{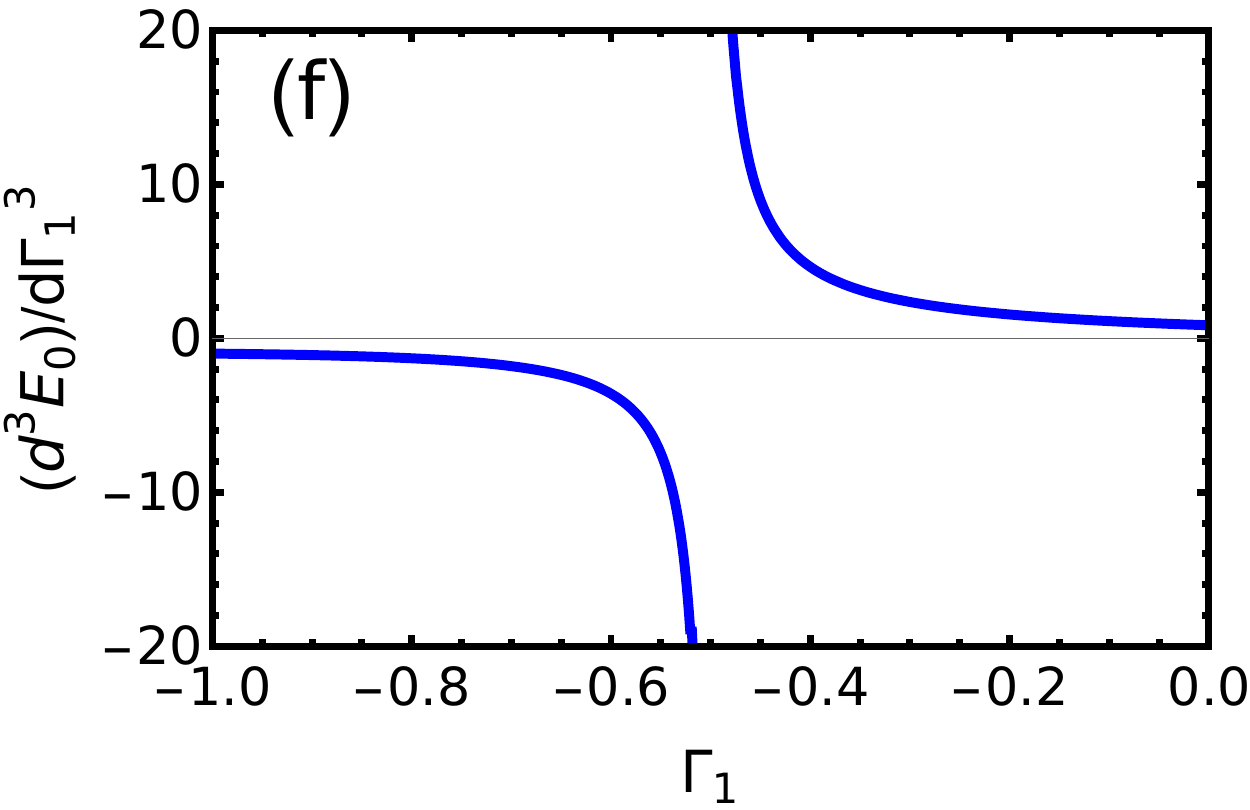} 
	\caption{\label{nume} Numerical analysis of the model. For all the plots $\Gamma_{0}=1$ and $\Gamma_{2}=0.5$. (a) Probability distribution at topological non-HS critical phase for the parameter space $\Gamma_{1}=\pm 3$ and $\Gamma_{3}=\pm 3.15$. (b) Probability distribution at trivial non-HS critical phase for the parameter space $\Gamma_{1}=0$ and $\Gamma_{3}=\pm 0.7$. (c) Eigenvalue distribution at topological non-HS critical phase for the parameter space $\Gamma_{1}=\pm 3$ and $\Gamma_{3}=\pm 3.15$. (b) Eigenvalue distribution at trivial non-HS critical phase for the parameter space $\Gamma_{1}=0$ and $\Gamma_{3}=\pm 0.7$. Third derivative of ground state energy density with respect to $\Gamma_{1}$ for (e) $M_1$ and (f) $M_2$.}
\end{figure}

\hspace{0.3cm}The model in the main text can be studied numerically under open boundary condition to identify the trivial and non-trivial critical phases. The model Hamiltonian can be written in the BdG form as~\cite{niu2012majorana}
\begin{equation}
H_{BdG} = \begin{pmatrix}
\chi_{x} & \chi_{y}\\
\chi_{y}^{\dagger} & -\chi_{x}
\end{pmatrix},
\end{equation}
with the submatrices
\begin{align}
\chi_{x}(i,j) &= - \Gamma_{0} \delta_{j,i} - \frac{\Gamma_{1}}{2} (\delta_{j,i+1}+\delta_{j,i-1})\\
&-\frac{\Gamma_{2}}{2} (\delta_{j,i+2}+\delta_{j,i-2})
-\frac{\Gamma_{3}}{2} (\delta_{j,i+3}+\delta_{j,i-3}), \nonumber\\
\chi_{y}(i,j)&= - \frac{\Gamma_{1}}{2} (\delta_{j,i+1}-\delta_{j,i-1})
-\frac{\Gamma_{2}}{2} (\delta_{j,i+2}-\delta_{j,i-2})\nonumber\\
&-\frac{\Gamma_{3}}{2} (\delta_{j,i+3}-\delta_{j,i-3}).
\end{align}
Diagonalizing the BdG Hamiltonian enable one to identify the trivial and non-trivial critical phase at both the non-HS criticalities. This can be achieved by analyzing the probability and eigenvalue distribution, as shown in Fig.\ref{nume}. The probability distribution of wave function of an open chain with the size $n=40$, shows distribution at the edges. These represents non-trivial non-HS critical phases with stable localized edge modes at each end of the chain. Fig.\ref{nume}(a), is obtained for the parameters $\Gamma_{0}=1$, $\Gamma_{1}=\pm 3$, $\Gamma_{2}=0.5$ and $\Gamma_{3}=\pm 3.15$, which represents topological non-HS critical phase at magenta and green lines respectively. 
%Similarly Fig.\ref{label}(b), is obtained for the parameters $\Gamma_{0}=1$, $\Gamma_{1}=3$, $\Gamma_{2}=0.5$ and $\Gamma_{3}=-3.15$, which represents topological non-HS critical phase at Green line. 
The the solutions for the trivial critical phases in Fig.\ref{nume}(b) is for $\Gamma_{0}=1$, $\Gamma_{1}=0$, $\Gamma_{2}=0.5$ and $\Gamma_{3}=\pm 0.7$ at both magenta and green lines respectively, where the probability distributes significantly at the bulk as a consequence of the delocalization of edge modes into the bulk. 

\hspace{0.3cm}This observation can be verified with the analysis of the eigenvalue distribution with the system size, as shown in Fig.\ref{nume}. The distribution of the eigenvalues at zero energy represents the topological non-HS critical phases while the absence of the eigenvalues at zero energy represents the trivial critical phase. For the parameter space $\Gamma_{0}=1$, $\Gamma_{1}=\pm3$, $\Gamma_{2}=0.5$ and $\Gamma_{3}=\pm3.15$, a pair of eigenvalue exists at zero energy represents the topological critical phase at both the magenta and green lines respectively, as shown in Fig.\ref{nume}(c). The parameter space $\Gamma_{0}=1$, $\Gamma_{1}=0$, $\Gamma_{2}=0.5$ and $\Gamma_{3}=\pm0.7$ is thus represents trivial critical phases of both the criticalities with no zero energy eigenvalues, as shown in Fig.\ref{nume}(d).

\hspace{0.3cm}These numerical evidences clearly shows the presence of trivial and non-trivial non-HS critical phases separated by multicritical points $M_{1,2}$, on both magenta and green critical lines. The $M_{1,2}$ can be identified with the divergence in the derivatives of ground state energy density with respect to $\Gamma_{1}$,
\begin{equation}
E_0=-\frac{1}{2\pi}\int_{-\pi}^{\pi} \dfrac{\partial^2}{\partial \Gamma_1^2} \left( \sqrt{\chi_{x}^2+\chi_{y}^2}\right)  dk.
\end{equation}
As shown in Fig.\ref{nume}(e,f), both the multicritical points are identified with the third order discontinuity in the ground state.